\documentclass[fleqn,11pt]{article}
\usepackage[headings]{espcrc1}
\readRCS
$Id: espcrc1.tex,v 1.2 2004/02/24 11:22:11 spepping Exp $
\usepackage{graphicx}
\usepackage{graphics}
\usepackage{epsfig}
\usepackage[figuresright]{rotating}

\usepackage{hyperref}
\usepackage{ae}
\usepackage{aecompl}
\usepackage{epsf}
\usepackage{makeidx}      
\usepackage{helvet}

\newcommand{\AmS}{{\protect\the\textfont2
  A\kern-.1667em\lower.5ex\hbox{M}\kern-.125emS}}

\title{Biological and synthetic membranes: What can be learned from a coarse-grained description?}
\author{
    Marcus M{\"u}ller
    \address{Institut f{\"u}r Theoretische Physik, Georg-August Universit{\"a}t, Friedrich Hund Platz 1, D37077 G{\"o}ttingen, Germany}
    \thanks{mmueller@theorie.physik.uni-goettingen.de}
    Kirill Katsov
    \address{Materials Research Laboratory, University of California, Santa Barbara, CA 93106, USA}
    \thanks{katsov@mrl.ucsb.edu}
    and
    Michael Schick
    \address{Department of Physics, University of Washington, Box 351560, Seattle, WA 98195-1560, USA}
    \thanks{schick@phys.washington.edu}
}
\runtitle{Coarse-grained models for biological and synthetic membranes}
\runauthor{M. M{\"u}ller, K. Katsov, and M. Schick}

\begin{document}
\def\bibdata{./bibtex}
\maketitle
\begin{abstract}
We discuss the role coarse-grained models play in the investigation of the
structure and thermodynamics of bilayer membranes, and we place them in the
context of alternative approaches. Because they reduce the degrees of freedom
and employ simple and soft effective potentials, coarse-grained models can provide
rather direct insight into collective phenomena in membranes on large time and
length scales.  We present a summary of recent progress in this rapidly
evolving field, and pay special attention to model development and
computational techniques. Applications of coarse-grained models to changes of
the membrane topology are illustrated with studies of membrane
fusion utilizing simulations and self-consistent field theory.  
\end{abstract}

\tableofcontents

\newpage
\section{Introduction}
\label{sec1}
The incredible complexity of biological systems combined with their immediate
importance makes them the most recent subject for the application of the
coarse-grained models of soft condensed matter physics
\cite{Daoud95,Larson99,Chaikin95}. While developed earlier for the elucidation
of the statics and dynamics of melts and solutions of very long and uniform
polymers
\cite{deGennesBook,FreedBook,Cloizeaux,Schafer,Baschnagel00,MuellerEncy01,Muller-Plathe02,Kremer02},
coarse-grained models seem particularly well suited to the study of biological
constituents
\cite{SHELLEY01_1,Muller03b,MARRINK04,NIELSEN04_1,IZVEKOV05,BOEK05}, such as
the heteropolymers DNA and RNA, as well as relatively short-chain lipids which
comprise all biological membranes.

Systems of polymers and of lipids share many common features, and exhibit
universal collective phenomena, those which involve many molecules
\cite{Muller03b}. Examples of such phenomena include thermodynamic phase
transitions, e.g., the main chain transition in lipid bilayers from a fluid, liquid crystalline
to a gel phase \cite{Gennis89,Lipo95}, and the lateral phase separation
which appears to be implicated in ``raft'' formation in the plasma membrane
\cite{Simons97,Brown98b,Pralle00,Dietrich01}, as well as thermally activated processes
such as vesicle fusion and fission
\cite{Chernomordik95b,Monck96,Zimmerberg99,JAHN02,Mayer02,Tamm03,Blumenthal03},
important in endocytosis and exocytosis, and electroporation
\cite{Potter88,Tsong91,Barnett91,Chang92,Weaver96} used in the micro-encapsulation of 
drugs and drug-delivery systems\cite{Mouritsen98,Fahr05}.

Since collective phenomena involve many molecules and entail large length and
time scales -- 10-1000 nm and $\mu$s-ms, respectively -- details of the structure and dynamics on short, atomistic length
scales are often irrelevant, and the behavior is dictated by only a small
number of key properties, e.g., the amphiphilic nature of the molecule. This
imparts a large degree of universality to the collective phenomena. These terms
are borrowed from the theory of critical phenomena \cite{Kadanoff00}. However the clear
separation in length, time and energy scales assumed by this approach, is often
missing in membrane systems.  Thus the universality of collective phenomena, or
the ability of coarse-grained models to describe collective phenomena, cannot
be taken for granted. It is important, therefore, to compare the behavior of
different experimental realizations among each other and with the results of coarse-grained models.

In the following we shall highlight some recent developments in this active
research area in which many new models and computational techniques are being
developed. We do not attempt to provide a comprehensive overview of this
rapidly evolving field, but rather try to give an introduction both to the basic
concepts involved in creating a  coarse-grained model, and to the
simulation techniques specific to membranes
and interfaces.  We shall emphasize the connection to polymer science whenever
appropriate. In particular, we will also discuss application of
field-theoretic techniques to calculate membrane properties. These techniques
employ very similar coarse-grained models as do the particle-based simulation schemes,
and they permit the calculation of free energies, and free energy
barriers, which are often difficult to obtain in computer simulations.  An application
of coarse-grained models in the context of computer simulations and
field-theoretic techniques is illustrated by the study of membrane
fusion, a choice biased by our own research focus on this area.

Many important applications are not covered by this manuscript. Most notably we
do not discuss important progress in the study of collective phenomena
exhibited by single molecules, as in the folding of a protein
\cite{Dill95,Karplus95,Shea01} or the processes that ensue when a protein is
inserted into a membrane \cite{Mouritsen93,Gil98,Tieleman01,Maddox02,Lopez04},
or those exhibited by assemblies of a small number of molecules, as in the
formation and subsequent function of a channel
\cite{BGROOT01,Tajkhorshid02,Saiz04b}.  In our view, details of the specific
molecular architecture are very important for these processes, and they lack
the type of universality which undergirds the application of coarse-grained
models.

In the next section we place coarse-grained models in the context of atomistic
ones that deal with molecular details, and of
phenomenological Hamiltonians that do not retain the notion of individual
molecules. We then discuss briefly a selection of simulation and
self-consistent field techniques utilized for coarse-grained models of
membranes. We illustrate the combination of computer simulation
and field-theoretic approach with the example of membrane fusion. The paper
closes with an outlook on further exciting, and open, questions in this area.

\section{Atomistic modeling, coarse-grained models and phenomenological Hamiltonians}
\label{sec2}

\begin{figure}[tb]
\epsfig{file=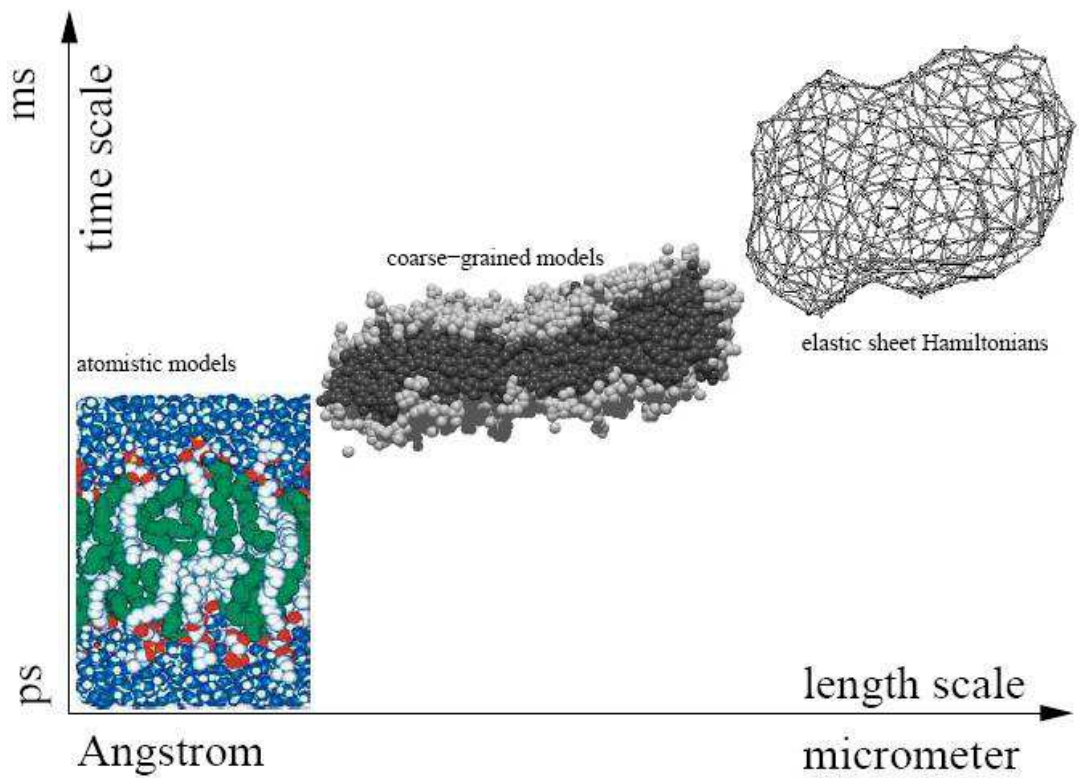,width=0.9\textwidth,clip=}
\caption{\label{fig:coarsegraining}
Illustration of different models used to tackle problems in membrane physics.
(left) snapshot of an atomistic simulation of a 1-stearoyl- 2-docosahexaenoyl-sn-glycero-3-phosphocholine (SDPC, 18:0/22:6 PC) lipid bilayer
from Ref.~\cite{Saiz01b}.
(middle) tensionless bilayer of a coarse-grained model from Ref.~\protect\cite{Muller03}
(right) snapshot of a randomly triangulated surface from Ref.~\cite{Gompper95b}.
}
\end{figure}

Processes in membranes evolve on vastly different scales of time, length
and energy. Consequently a variety of membrane models and computational techniques have been
devised to investigate specific questions at these different scales. We divide
them roughly into atomistic, coarse-grained, and phenomenological models as illustrated
in Fig.~\ref{fig:coarsegraining}.

\subsection{Atomistic molecular dynamics simulations}
\label{sec2.1}
 Atomistic models, which describe  bilayer membrane properties with chemical
accuracy,  have been successfully utilized to investigate the
detailed properties of particular membrane systems and
lipid-protein complexes.
Such models have a longstanding tradition.
The first Molecular Dynamics simulation of lipid bilayers, carried out in the early 1990's
\cite{DAMODARAN92,HELLER93}, were able to
simulate only a small patch of a bilayer,
one nanometer in extent, over a very short time interval, typically 0.1ns.
Since then, much effort has been
directed to the improvement of simulation algorithms,
(e.g., multiple time step algorithms
\cite{TUCKERMAN92}), and to the implementation of
the simulation code on parallel computers.
Consequently atomistic simulations have advanced significantly.
Today, there are many
complete simulation packages that comprise
standard routines for Molecular Dynamics simulations. Among them are
NAMD Molecular Dynamics Software (NAMD)
\cite{PHILLIPS05}, the Groningen Machine for Chemical Simulation (GROMACS)
\cite{LINDAHL01,SPOEL05}, Groningen Molecular Simulation (GROMOS)
\cite{CHRISTEN05,SCOTT99}, MDynaMix \cite{LYUBARTSEV00}, Assisted Model
Building with Energy Refinement (AMBER) \cite{CASE05}, NWChem
\cite{NWCHEM1,NWCHEM2}, Integrated Model Program, Applied Chemical Theory
(IMPACT) \cite{BANKS05}, Biochemical and Organic Simulation System (BOSS)/
Monte Carlo for Proteins (MCPRO) \cite{JORGENSEN05}, DL\_POLY
\cite{SMITH05DLPOLY,SMITH96}, Large-scale Atomic/Molecular Massively Parallel
Simulator (LAMMPS) \cite{PLIMPTON95}, and Extensible Simulation Package for
Research on Soft Matter Systems (ESPResSo) \cite{LIMBACH05}.
The first four packages in this list
are tailored to simulate lipid bilayers and proteins in atomistic
details, while the latter codes stem from polymer simulations and have also been
applied to coarse-grained simulations of soft matter. Most of the program
packages (NAMD, GROMACS, DL\_POLY, LAMMPS and ESPResSo) are freely available.
A nominal fee is charged for the use of GROMOS, while other packages,
(e.g., IMPACT), are commercial
products. Most of the programs run on parallel computing platforms using the
Message Passing Interface (MPI). Differences exist in the way the information
is distributed among the processors  (e.g.,
spatial domain decomposition vs.~particle
decomposition), the force fields that are implemented, and their ease of use
and the availability of tutorials.
In addition to ensuring that the code is error-free and
computational efficient, the developers make
efforts to keep it transparent and extendable.  Often the data formats
allow exchange of information with other software packages, such as high level
quantum chemistry packages (QM/MM methods), or visualization software (e.g., Visual
Molecular Dynamics (VMD) \cite{VMD} or PyMol \cite{PyMOL}).

In atomistic models the attempt is made to describe faithfully the
molecular architecture and the interactions between components.
The interactions include those specifically describing a chemical bond
(two particle, bond
angle potentials and dihedral, torsional, potentials), as well as
those between atoms not bonded to one another, such as electrostatic and
van der Waals interactions.
The quality of the interaction potentials is crucial for a
successful description. In lieu of a first-principles calculation, one would like  the interactions to fit the results of
quantum chemical calculations of small fragments of the molecules.  Often
parameters are additionally fitted to experimental observations of various quantities.  Many sets of potentials
have been devised for lipids in an aqueous environment, but  there is still a need
to refine the interactions and devise more accurate models.
The parameters of the interactions can be adjusted to standard force fields,
e.g., CHARMM \cite{BROOKS83}, GROMOS96 \cite{VANGUNSTEREN96}, OPLS-AA
\cite{JORGENSEN96,KAMINSKI01}, or Encad \cite{LEVITT90}.  Often there exists
the possibility of including customized, tabulated, potentials.

Atomistic force fields typically include Coulomb interactions. They arise from ionic groups
or  ``partial charges" that account for the ability of atomic
species to share charges on a common bond. For very small systems, the long-ranged Coulomb interactions
are handled by Ewald summation. Most modern
applications, however, employ Particle-Mesh-Ewald techniques
\cite{HOCKNEY88,DESERNO98_1,DESERNO98_2} which yield a scaling of the order
${\cal O}(N \ln N)$, with the number of charges, $N$,
or fast multipole expansions (e.g., IMPACT \cite{BANKS05})
which scale linearly with $N$.  ESPResSo \cite{LIMBACH05},
however, can additionally deal with Coulomb interactions via a strictly local,
field-theoretic, algorithm \cite{MAGGS02,ROTTLER04,DUNWEG04}, and provides
routines for simulating systems that are periodic in one or two
directions.

Typically, atomistic models are studied by Molecular Dynamics simulations \cite{FrenkelSmit}. The
advantage of this simulation scheme consists in its rather realistic description
of the microscopic dynamics of the constituents. It thereby permits the
investigation of kinetic processes, e.g., the transport of small molecules
across the bilayer, the lateral self-diffusion of lipid molecules in the
bilayer, the tumbling motion of the lipid tails, or the dynamics of the
hydrogen-bond network of water at the hydrophilic/hydrophobic interface.  The
simulation packages often use the time-reversible, symplectic Velocity Verlet or
Leapfrog algorithms \cite{RAPAPORT04,FRENKEL02}
and allow for multiple time step integration \cite{TUCKERMAN92}.

Most of the atomistic simulation packages also include a limited selection of
methods to calculate free energies. The most popular techniques comprise
thermodynamic integration and umbrella-sampling methods \cite{umbrella}.
Sometimes replica-exchange Monte Carlo, or parallel tempering methods, are
employed \cite{Hukushima96,Hansmann97,Sugita99,Mitsutake01}.

In the simplest case, the simulation follows the trajectory of the
particles by integrating  Newton's equations of motion in time.
The time step is
set by the shortest period in the system, e.g., that of the stretching mode of a
covalent bond. Since the time scale for these bond vibrations is orders of
magnitude smaller than the time scale of interest,  the
bond lengths are often constrained, thus eliminating the smallest periods and allowing for a larger
time step of the integrator. Common algorithms that incorporate these constraints into
the Molecular Dynamics scheme are SHAKE \cite{RYCKAERT77}, RATTLE
\cite{ANDERSEN83}, and LINCS \cite{Hess97}.  Moreover the
simulation packages also allow the possibility of  constraining the position of atoms,
or distances between constituents, a constraint often applied during the relaxation
period.

Newton's equations of motion lead to a microcanonical trajectory.
Unfortunately, most numerical integrations do not conserve the energy
on long time scales. Therefore  one often couples the system to a thermostat,
and uses the temperature, $T$, as a control variable. Analogously, it is often
desirable to simulate the system at constant pressure, or more usually, surface
tension,  using the Berendsen thermostat \cite{BERENDSEN84} or the
Anderson-Nos\'e-Hoover algorithm \cite{ANDERSON80,NOSE84,HOOVER85,MARTYNA92}.
This allows the volume, or area, to fluctuate.  Some
programs also permit the geometry of the simulation box to change in order to
equilibrate stresses in crystalline phases via the Parrinello-Rahman technique
\cite{PARRINELLO82}.  These simulation methodologies are very similar to what
is utilized in coarse-grained simulations.

In addition to interactions between atoms,  external forces have also been
included in recent simulation schemes. In these ``steered Molecular Dynamics
simulations'' one can mimic, for example,  the action of an AFM cantilever on a
biopolymer. This allows one to ``push'' the system over a free energy barrier.
If this is repeated often enough,  Jarzynski's equation
\cite{JARZYNSKI00_1,JARZYNSKI00_2,Park03,Park04,Speck05} can be used
to calculate free energies by integrating the corresponding Boltzmann factor.


While atomistic simulations in the early 1990s could simulate only an
extremely small patch of a pure lipid bilayer membrane on the order of a nanometer,
\cite{DAMODARAN92,HELLER93}, current Molecular Dynamics simulations study
membrane patches of a few tens of nanometers over time scales of a few
tens of nanoseconds. The time scale of observation has increased by two
orders of magnitude over the earliest attempts.
These simulations provide valuable information about
the molecular organization and dynamics of the lipids and of the water,
both in the bilayer and at the hydrophobic/hydrophilic interface
\cite{GIERULA99,AMAN03,LOPEZ04_2,SUITS05_2}. For example, recent atomistic
simulations revealed the molecular structure of the ripple phase in phosphatidylcholine
bilayers \cite{deVries05}. However, calculation of these basic
properties (e.g., free energy differences between different morphologies) still pose challenges to computer simulations because of the
large time scales which are introduced by the interactions between head
groups due to charge pairing or water bridging.

Another important aspect of atomistic simulations is that they are able
to correlate the
detailed lipid architecture (e.g., particular head group structure, length, and
degree of saturation of the tails) with the physical properties of
the bilayer (thickness, orientation of segments,  liquid-crystalline
ordering, and elastic moduli). Such physical properties can be obtained,
for example, by increasing sufficiently the length and
time scale of the simulation, so that one can observe the fluctuations of the
bilayer membrane. From an analysis of the undulations and peristaltic
fluctuations of the local
thickness, one obtains the
bending rigidity and area compressibility of the membrane,  as well as
the pertinent
relaxation times \cite{LINDAHL00,MARRINK01}.  For instance,
atomistic simulations of monoolein, which is an extremely simple lipid
with only one mono-unsaturated tail of eighteen carbon atoms and a very small
headgroup consisting solely of two hydroxyls, reveal
that fluctuations on the length scale of 20
nm have relaxation times of more than 5 ns. This time scale is expected
to be longer in the usual double-tailed phospholipids
\cite{MARRINK01}.

The combination of
sophisticated simulation codes and powerful parallel computers has permitted
atomistic simulations to investigate some collective phenomena in bilayer
membranes.
The
transformation from an inverted cubic phase to an inverted hexagonal phase
in the monoolein system has been studied \cite{MARRINK02}.
The transition, as a function of areal density, between liquid-condensed and
liquid-expanded phases of a {monolayer}
of DPPC at an air-water interface has been studied, as well as the monolayer's
eventual rupture \cite{KNECHT05}.  Recent
applications can even cope with the protracted time scales associated
with the spontaneous self-assembly of lipids into a bilayer
\cite{MARRINK01_2}, or the formation of small vesicles \cite{VRIES04}.


Another thrust of applications is the study of membranes that consist of
lipid mixtures, or membranes that have molecules adsorbed onto them.
The atomistic description
provides a detailed view of the role that small inclusions, like
cholesterol, or adsorbents, like sugars or polymers, play in modifying the
structure of the bilayer. In the following, we give several examples.

The addition of a surfactant \cite{BAND01} or cholesterol
\cite{TU98,HOFSASS03,FALCK04} to a lipid bilayer tends to
increase the ordering of the lipid tails, which causes them to lengthen.
Because the liquid-like interior of the bilayer is highly incompressible,
this results in a decrease of the area per head group.
In the
case of high concentrations of cholesterol \cite{HOFSASS03}, the decrease
of the area per head group is accompanied by  an increase of the bending
modulus and a decrease in
the lateral self-diffusion coefficient of the lipids. The change
of the lipid packing also affects the distribution of voids. They become
rarer, and those that remain elongate along the bilayer normal as the
concentration of cholesterol increases \cite{FALCK04}. Addition of
salicylate to a lipid bilayer also decreases the area per head group.
However in this case, the mechanical properties of the bilayer are
hardly affected
\cite{SONG05}. The effect of halothane, an anesthetic, is quite different.
This small molecule preferentially segregates to the upper part
of the lipid acyl chains, increases the area per head group, and decreases
the lipid chain order \cite{KOUBI00}.

Large sugar molecules do not penetrate into the hydrophobic interior of
the bilayer, but do impact the hydrogen bonding at the interface between the
head groups and the water.  An interesting example is provided by
trehalose, a disaccharide, which is
found in animals capable of enduring cold temperatures or dry environments. Experiments
indicate that it prevents leakage and fusion during drying and
freeze-drying, a property which has been exploited for practical applications \cite{Crowe84,Leslie95}.
Atomistic simulations \cite{SUM03,SUM03_2,Pereira04,VILLAREAL04,Doxastakis05b,Skibinsky05} show that the area
per head group remains unaffected.  In addition, the total number of
hydrogen bonds of
the lipid heads is conserved. However, hydrogen bonds with trehalose
now replace some of the bonds which had been made with water.
A single trehalose molecule can
interact with multiple lipids simultaneously.  This result suggests that,
at sufficiently high concentrations, disaccharides might serve as an
effective replacement for water.

The largest-scale simulations carried out
on the atomistic level are able to study
lipid-protein, or lipid-DNA, interactions \cite{BAN99}, and to investigate
channels \cite{BGROOT01,BGROOT02} through a bilayer lipid membrane.
The added complexity brought about by incorporating proteins into the
membrane poses huge
challenges to both the simulation techniques and computational
requirements due to the large number of additional interactions which
have to be accurately described. Some examples of the systems studied are
as follows.

A protein's conformations when it is
inserted into the membrane, and the distortion of the lipid bilayer in its
vicinity, can be studied by atomistic simulation.
The protein's interactions with the lipids are both strong,
compared to the thermal energy scale, $k_BT$, and specific.
They are difficult to simplify, with the result that the details of
the complex architecture on the molecular level have to be considered for
a quantitative description. Proteins can create interactions within the
bilayer due to the strain field generated by
a mismatch of its hydrophobic region with that of the
bilayer in which it is embedded. Large molecules like DNA \cite{BAN99}
and proteins
\cite{BGROOT01,BGROOT02,AKSIMENTIEV05} also give rise to significant
interactions outside of the bilayer.

Often proteins are not isolated in the lipid membrane,  but aggregate to
structures such as  pores, or ion channels \cite{Lopez04,BGROOT01,Tajkhorshid02}. The detailed
structure of these channels has attracted great interest in
understanding how they function to let some ions pass while stopping
others.
It has been possible to study the permeation
of water through an acquaporin pore \cite{BGROOT01,BGROOT02}. These
simulations reveal the motion of a single water molecule as it passes through
the channel. The trajectories provide insights into the specificity
mechanism by which the channel allows water, but not ions, to pass.
Recently, the permeability for water and ions
of the $\alpha$-hemolysin/lipid bilayer complex
has been studied by large-scale computer simulations
involving 300 000 atoms \cite{AKSIMENTIEV05}. The application of external
electrical fields permitted the ion permeability to be obtained
as a function of bias voltage,
and the selectivity of $\alpha$-hemolysin to chlorine ions to be elucidated.


The bilayer structure in almost all of these atomistic simulations
has to be pre-assembled because the time scale of
self-assembly from a homogeneous mixture of lipids and water typically
far exceeds the simulation time scale.  (For an exception, see
~Ref.~\cite{MARRINK01_2}.) This leaves unanswered the question of
the thermodynamic stability of the pre-assembled membrane. Even though the
atomistic potentials are parameterized from the interaction of atoms, the
manner in which these potentials determine the global stability of the
different lipid morphologies is subtle, and unknown.
Furthermore, the transitions between these lipid morphologies, and the formation of
out-of-plane structures as occurs in budding, are
beyond the scope of atomistic modeling.


\subsection{Coarse-grained models}
\label{sec2.2}
\subsubsection{Why are coarse-grained models useful?}

While atomistic simulations provide valuable, detailed, information about the
local structural properties of lipid membranes, they cannot access the time and
length scales involved in collective membrane phenomena, which are milliseconds and
micrometers respectively. One strategy to overcome this difficulty is to
eliminate some of the detail by coarse-graining the description.
Coarse-grained models do not attempt to describe the large scale phenomena by
starting from the smallest atomic length scale, but rather by lumping a small number
of atoms into an effective particle
\cite{SHELLEY01_1,MARRINK04,NIELSEN04_1,LARSON85,Haas96,GOETZ98,GOETZ99,Carmesin88,DOTERA96,NOGUCHI01,Soddemann01,SHILLCOCK02,GUO02,GUO03}.
These particles then interact via coarse-grained, simplified, interactions, ones which typically do not
include
electrostatic and torsional potentials,
for example. The reduced number of degrees of freedom, and the softer
interactions on the coarsened scale lead to a significant computational
speed-up with the consequence that  larger systems and longer time scales are
accessible. This makes possible the study of collective phenomena in membranes,
a study not possible via ab-initio methods now, or in the foreseeable future.
However the loss of chemical detail limits some of  the predictive power of
coarse-grained models.

The objectives of mesoscopic modeling are twofold: on the one hand, to
help to identify those interactions which are necessary to bring about
collective phenomena on a mesoscopic scale, such as self-assembly, and on the other
to elucidate the universal behavior on the
mesoscopic scale itself. These include the role of thermal fluctuations, or the
existence of phase transitions between self-assembled morphologies. Mesoscopic
models are also an ideal testing ground for phenomenological concepts.

The obvious question which presents itself is how the coarse-graining is to be achieved.
What are the
relevant degrees of freedom and interactions to be retained at the
coarse-grained scale in order to incorporate the essential physics of the
system? This is a fundamental question of {\em any} model-building procedure which must be addressed when one abandons ab-initio calculations.

One can respond that, due to the experimentally observed universality of
self-assembled structures, any coarse-grained model that includes the
relevant interactions will capture the qualitative features. Consequently  one
should use the simplest possible model in order to take maximum advantage  of
the computational benefits of coarse-graining. This is the strategy of
minimal models which were the first to study self-assembly. They  are
still very popular.
The question that remains  is just what are the ``relevant" interactions
necessary to bring about the collective phenomena observed in experimental
systems. Within the framework of minimal models, one can start with a
simple model and successively augment it with additional interactions
until the phenomenon of interest is captured. While this
method appears to be rather crude, it does provide  insight into
which interactions, on the length scales of a few atomic units, are
necessary to bring about collective behavior in membranes. It also
contributes to identifying those mechanisms that underly the phenomena and
the degree of universality.  Alternatively one can try to ``derive''
coarse-grained models from a specific atomistic system,  a procedure which
is termed ``systematic coarse-graining''. We shall discuss both techniques
in turn.

\subsubsection{Minimal models}
The idea of successively eliminating degrees of freedom from a specific mixture
of lipid and water to ``derive'' a coarse-grained model is a beautiful and
potentially powerful concept. This concept of coarse-grained models has a
long-standing tradition in polymer physics \cite{deGennesBook}, and during the last three years
much progress has been made in the area of  biological and synthetic membranes.
Unfortunately, the coarse-graining procedure is often impractical to implement
explicitly.  Notable exceptions are dilute and semi-dilute polymer solutions
for which the concept of coarse-graining can be formulated in terms of a
consistent theory, one which has been extensively exploited \cite{FreedBook,Cloizeaux,Schafer}.

The configurations of long, flexible,
linear polymers in dilute or semi-dilute solutions are characterized by a
self-similar, fractal structure.
This self-similarity extends from the structure of a few monomeric repeat units
to the size of the entire molecule, which is comprised of hundreds or thousands
of monomers. For long chain molecules, there is a clear separation
between the structure on the length scale of a monomeric unit, which strongly
depends on the chemical structure and details of the interactions on the atomic
scale, and the mesoscopic structure of the entire molecule.  Clearly the chain
dimensions depend on the chemical identity of the monomeric units in a
very subtle manner, but for the description of the large scale properties a
single, coarse-grained, parameter, the end-to-end distance, suffices.
The background of this statement is
the observation of de Gennes, in 1972, that the behavior of a long,
self-avoiding, walk is intimately related to the properties of a critical point
in a $n$-component field theory in the limit $n \to 0$ \cite{Degennes72}.  This opened the way
for the use of the machinery of the Renormalization Group for the description
of polymer solutions, and placed the heuristic observation of the universality
of the behavior of long chain molecules within a rigorous theoretical framework.
The inverse chain length plays the role of the distance to the critical point.
The behavior at the critical point is universal, i.e., it does not depend on
the microscopic interactions but only on a few, relevant, features that
characterize a universality class. For polymer solutions the relevant
properties are the connectivity of the monomeric units along the backbone and
the excluded volume interaction between monomeric units.  By virtue of
universality, any model characterized by these two properties will capture the
behavior of polymer solutions in the limit of long chain lengths.  The theory
has provided detailed insights into the large scale chain conformations in
dilute and semi-dilute solutions,  and has been utilized to describe
quantitatively the screening of the excluded volume interactions, and the
cross-over from dilute to semi-dilute solutions \cite{FreedBook,Cloizeaux,Schafer}.

Biological systems  do not exhibit the sort of scale invariance that lies at
the heart of the Renormalization Group approach to polymer systems. In
particular, there is no parameter, like the chain length,  that tunes the
separation between the microscopic scale of the atomic interactions and the
mesoscopic structure. Another practical complication is that, in contrast to
polymer systems in which one considers systems of very few components and with
simple interactions between them,  biological systems are composed of many
different, complex, structural units which are connected by means of several
different interactions.  As a consequence, the development of coarse-grained
models for membranes is more an art than a science. It is often guided by
physical intuition, computational constraints, and a large degree of
trial-and-error.  The underlying assumption is that, just as in polymer
solutions,  the qualitative behavior of the membrane depends only on a few
coarse-grained parameters that characterize the relevant interactions of the
mesoscopic model.  This assumption is not justified by a rigorous formal
theory. Consequently  it is {\em a priori} unknown what the relevant
interactions are that have to be incorporated in order for a coarse-grained
model to faithfully capture the behavior on mesoscopic length scales. The
answer to this crucial question  depends on the specific system, and on the
properties in which one is interested.  For example, the experimental fact that
systems which differ chemically a great deal, such as biologically relevant
lipids in aqueous solution and  amphiphilic water-soluble polymers, do exhibit
a great number  of common morphologies implies that the existence of these
morphologies can  be traced back to a small number of simple interactions.

A key ingredient is the connectivity of two strongly repelling entities within a single molecule.  
Since these two parts are joined together they
cannot separate and form macroscopic domains, but organize instead into supermolecular
structures so as to minimize the unfavorable contacts between their parts. The
particular physical mechanisms that cause the repulsion between the two
entities appear to be less important.

Another significant experimental observation is the correlation that exists
between the gross
amphiphilic architecture of the components of the system and the system's
phase behavior. Not only does the size of the
molecule set the scale of the self-assembled structures, such as the bilayer
thickness, but also its ``architecture", the relative volumes of the
two antagonistic molecular parts,  can be directly correlated with the various
morphologies. This correlation has been stressed by Israelachvili \cite{Israel,Israelachvili80}. Lipids in
which the head and tail groups are of similar volumes tend to form bilayers. If
the lipid tails volume is enlarged or the headgroup reduced (e.g. by replacing a
phosphatidylcholine with a phosphatidylethanolamine) then the lipids are said
to be ``cone-shaped" and they tend to form inverted hexagonal phases. In this
phase, the lipids assemble into tubes with the heads directed inward and the tails
outward, and the tubes form a periodic hexagonal lattice.  This concept also carries over
to $AB$ diblock copolymers which consist of two blocks composed of $N_A$
and $N_B$, monomeric units. In this case, the fraction, $f \equiv
N_A/(N_A+N_B)$, of one block is  employed to parameterize the molecular
architecture, and it also correlates with the observed phase behavior.  From these
observations, one can conclude that it is crucial to conserve the basic geometry
of the molecules during the mapping onto a coarse-grained model.

Notwithstanding these important universal aspects, the details of the molecular
architecture, interactions and the mechanisms of self-assembly, do vary from
system to system.  In block copolymers,  for example, the geometrical conformations of polymers
strongly fluctuate and, therefore, the average geometrical shape of a diblock
copolymer is strongly affected by its environment. The balance between the
repulsive interaction energy of the two components  and the
conformational entropy that describes the change of available molecular
conformations dictates
the self-assembled morphology.  In lipid systems, however, the molecules are
short and rigid. The reduced number of molecular conformations severely limits
their ability to alter their average geometric shape to adapt to the
environment. Thus, the concept of packing  rigid, wedge-shaped, objects is useful,
{i.e.}  the ``shape" of the molecules does not depend significantly  on the
environment. A mismatch between the molecular geometry and packing constraints
cannot be completely accommodated by changes of molecular orientation,
so that this mismatch also alters the local fluid-like packing. It is this
interplay between universal and specific aspects that make the development of
coarse-grained models in biological systems a
challenging one.

The amphiphilic nature of the molecules and the important
molecular geometry are characteristic of self-assembling systems. These two
relevant properties must be captured by a
coarse-grained model. They differ in detail as to how these
properties are incorporated, and they have been augmented by additional
interactions to provide a more detailed description of specific systems.

One of the simplest self-assembling system is that of oil and water and
amphiphile, and many simple lattice models were introduced to study it
\cite{LARSON85,WIDOM,GOMPPERSCHICK}.  Larson was one of the first to
ask how some of the simplest specific features of the amphiphile, such as the
presence of a multi-atom hydrophobic tail and the relative volume of head and
tail units, would affect the phase structure.  While water and oil were
represented by a single site on a lattice, amphiphiles were modeled as a
linear string of nearest or next-nearest sites. Interactions between
hydrophilic units, water or lipid heads, and hydrophobic units, oil or lipid
tails, were described by square-well potentials that  extended over the nearest and
next-nearest neighbors. Like units attracted each other while unlike units repelled
each other.  Monte Carlo simulations of this model yielded information about
possible phase diagrams of ternary water, oil, and amphiphile solutions
\cite{LARSON85,LARSON94,LARSON96}. This simple lattice model was even able to
reproduce the complex gyroid morphology that has been observed both in
amphiphile solutions and block copolymers.  Not surprisingly the Larson model
shares many features with simple lattice models that have been utilized to
study the universal characteristics of polymer solutions and melts. In the
latter context, simple lattice models have been employed to problems ranging
from the scaling properties of isolated chains in good solvent \cite{li95}, the equation of
state of solutions and mixtures \cite{flory41,huggins41}, to the ordering of diblock copolymers \cite{Kremer88,Lattice}.

To study further how microscopic details affect macroscopic behavior, one must
flesh out these schematic models by various structural details. Unfortunately,
it is difficult to capture details of the geometric shape of the amphiphiles in
simple lattice models. The restricted number of angles between bonds that
connect neighboring amphiphilic units makes very difficult a realistic
description of the rather stiff tails. Further, in lattice models, the head and tail
segments invariably occupy identical volumes. Some of these difficulties can be
overcome by complex lattice models, such as in the bond fluctuation model of
Carmesin and Kremer \cite{Carmesin88,Deutsch91,Muller99}. In this model,  each segment occupies the eight corners of
a unit cell of a simple cubic lattice. Monomers along the amphiphile are
connected by one of 108 bond vectors that are allowed to take lengths,
$2,\sqrt{5}, \sqrt{6},3$ or $\sqrt{10}$ in units of the lattice spacing. This
set of bond vectors is chosen such that the excluded volume constraint
guarantees that bonds do not cross in the course of local random monomer
displacements by one unit in one of the lattice axis. Thus, effects due to
entanglements are captured.  The large number of bond vectors and the extended
shape of the monomers yields a rather good description of continuum space. For
instance, the eighty-seven different bond angles permit a rather realistic
description of stiffness. Artifacts due to lattice discretization are strongly
reduced, yet the computational advantages of a lattice model (e.g., early
rejection of trial moves) are retained \cite{Kremer88,Lattice}. Moreover, sophisticated Monte Carlo
simulation techniques have been implemented for lattice models that allow for a
very efficient relaxation of the molecular conformations and the calculation of
free energies. The model can be quantitatively mapped onto the
standard Gaussian chain model that is used in self-consistent field (SCF)
calculations (cf.~Sec.~\ref{sec:SCFT}). This allows for a computationally efficient way to
explore a wide parameter range as well as to calculate corresponding free energies. Amphiphiles have
been modeled as flexible chains consisting of a hydrophilic and a hydrophobic
block. The solvent can be described by a homopolymer chain that consists of hydrophilic
segments. Like segments attract each other via a square well potential that
extends over the nearest fifty-four lattice sites, while hydrophilic and
hydrophobic segments within this range of interaction repel each other. The
strength of the interaction between the segments controls the free energy of the
hydrophilic/hydrophobic interface, where as the relative length of the hydrophilic
block, $f$, tunes the spontaneous curvature of a monolayer.  The model has been
successfully employed to study self-assembly in diblock copolymers and their
mixtures with homopolymers \cite{Muller96}, and pore formation of bilayers under tension \cite{Mueller96}. 
The bending rigidity of a monolayer and tension of a bilayer have been measured via the
spectrum of interface fluctuations and bilayer undulations \cite{Muller96,Mueller96},  and the fusion of membranes has also been studied \cite{Mueller02,Muller03} within this framework.

Although this lattice model includes only the bare essentials necessary to
bring about self-assembly, it is sufficient to describe its universal properties. A
mapping of length scale between lattice model and experimental realizations can be established
by comparing an experimental bilayer thickness in nanometers with the bilayer
thickness of the model expressed in lattice constants.  Similarly the model's
energy scale can be deduced by comparing the experimental and calculated free
energy of the hydrophilic/hydrophobic interface.  Additional characteristics, such as
the bending rigidity, or the area compressibility modulus, then can be combined
in dimensionless ratios. A comparison of such dimensionless ratios between liposomes, polymersomes, and
the bond fluctuation model is presented in Table~\ref{tab:bfm}. One observes that these
mesoscopic characteristics do not strongly differ between membranes formed by long
amphiphilic diblock copolymers and biological lipids in aqueous solution and that the
lattice model is able to reproduce the order of magnitude estimate of these properties. Therefore, this table
quantifies the universality of amphiphilic systems and justifies the use of
highly simplified models \cite{Muller03b}.

\begin{table}
\begin{tabular}{|l|lcl|}
\hline
          & polymersomes       & liposomes      & bond fluctuation model \\  \hline
$d_c$     & 80\AA              & 30\AA (DOPE$^{(a)}$), 25\AA (DOPC$^{(b)}$)  & 21u        \\
$f$       & 0.39               & $0.35\pm0.10$  & 0.34375    \\
$C_0 d_c$ & no data            & -1.1 (DOPE$^{(d)}$), -0.29 (DOPC$^{(c)}$)   & -0.68      \\
$\Delta A/A_0$
          & 0.19               & 0.05           & 0.19       \\
$\kappa_a/\gamma_{\rm int}$
          & 2.4                & 4.4 (DOPE$^{(b)}$), 2.9 (DOPC$^{(b)}$)   & 4.1        \\
$\kappa_b/\gamma_{\rm int} d_c^2$
          & 0.044              & 0.10 (DOPE$^{(c)}$), 0.12 (DOPC$^{(d)}$)  & 0.048      \\
\hline
\end{tabular}
\caption{Structural and elastic properties of bilayer membranes formed by amphiphilic diblock copolymers, biological lipids and a coarse-grained lattice model.
$d_c$ - thickness of membrane hydrophobic core in the tensionless state,
$f$ - hydrophilic fraction,
$C_0$ - monolayer spontaneous curvature,
$\Delta A/A_0$ - bilayer area expansion (critical value for the
experimental systems, and the actual strain used in simulations),
$\kappa_a$ - bilayer area compressibility modulus,
$\kappa_b$ - monolayer bending modulus,
$\gamma_{\rm int}$ - hydrophilic/hydrophobic interface tension
(oil/water tension of 50pN/nm for the experimental systems,
and A/B homopolymer tension for the simulations).
Data on EO7 polymersomes is taken from \cite{Discher99}; and on lipids
from (a): \cite{Rand89},  (b): \cite{Rand90},
(c): \cite{Chen97}, and (d): \cite{Leikin96}
(see also http://aqueous.labs.brocku.ca/lipid/).
Values of $d_c$, $C_0$ and $\kappa_a$ for DOPE were obtained by linear extrapolation
from the results on DOPE/DOPC(3:1) mixture.
Values of $\kappa_b$, $\gamma_{\rm int}$, and $C_0$ for the simulated
model were calculated by us using the method of \cite{Muller02f}. From Ref.~\cite{Muller03b}.
\label{tab:bfm}
}
\end{table}

An alternative procedure to include molecular detail is to use off-lattice
models. Clearly these models allow for much flexibility in describing the molecular
geometry and they can be studied by Molecular Dynamics. A generic off-lattice
model has been utilized by Smit and co-workers \cite{SMIT93} to elucidate micelle
formation. Water and oil molecules are modeled by Lennard-Jones particles while
the amphiphile is represented by a collection of particles bonded together via
harmonic springs. The hydrophobic beads form a linear chain tail, while the hydrophilic head beads
are all bonded to a single, central bead which, in
turn, is attached to the tail. This mimics the bulkiness of the lipid head.

Goetz and Lipowsky \cite{GOETZ98} employed a model in which the amphiphiles are
comprised of a single head bead and four tail segments. Water is modeled by a
single bead.  The interactions between the like beads (head-head, water-water,
head-water and tail-tail) are of Lennard-Jones type with a cut-off at $r_c
=2.5\sigma$. The energy, $\epsilon$, and range, $\sigma$, of the Lennard-Jones
potential set the scales.  The hydrophobic interaction between water and tail or
head and tail is a purely repulsive soft potential,

\begin{equation}
V_{\rm sc}(r)=4 \epsilon(\sigma_{\rm sc}/r)^{9},
\end{equation} with
$\sigma_{\rm rep}=1.05\sigma$, that is  cut-off
at $r_c=2.5\sigma$. The potentials are truncated and shifted
so that both the potential and the force are continuous at the
cut-off:
\begin{equation}
V_{\rm trunc}(r) = \left \{ \begin{array}{ll}
                           V(r)-V(r_c)-\left.\frac{dV}{dr}\right|_{r_c}(r-r_c) & \mbox{for}\; r\leq r_c  \\
                0 & \mbox{for}\; r> r_c.
                            \end{array}\right.
\end{equation}
Beads along the amphiphile are bonded
together via harmonic springs
\begin{equation}
V_{\rm bond}(r) = k_{\rm bond} (|{\bf r}|-\sigma)^2.
\end{equation}
The rather large value $k_{\rm bond}\sigma^2/\epsilon=5000$ is chosen to constrain the
average bond length to a value very close to $\sigma$.  Additionally  a bending potential of the form

\begin{equation}
V_{\rm bend}=k_{\rm bend}(1-\cos(\theta)),
\end{equation}
where $\theta$ denotes the bond angle, is included. By increasing the bending stiffness $k_{\rm bend} \leq
5 \epsilon$, one can change the conformations from those typical of a very flexible molecule to those characteristic of a rigid one.

A further step in the coarse-graining procedure is to eliminate the solvent
particles while preserving their effects implicitly.  Since the two-dimensional membrane is
embedded in a three-dimensional volume filled with solvent, the number of
solvent particles increases much faster than the number of amphiphiles as one
studies ever larger systems sizes. Yet the role of the solvent often is only to
stabilize the bilayer membrane whose properties are the focus of attention. Therefore,
if the solvent could be eliminated, it would result in a huge computational
speed-up. Typically, one can assume that the amphiphile-solvent mixture is
incompressible on a coarse scale.  Then the system configuration is completely
described by the configuration of the amphiphiles, and the interaction between
solvent and amphiphiles can be integrated out giving rise to an effective
interaction between the amphiphilic units.

The resultant {\em implicit} solvent models have enjoyed  long-standing popularity in
polymer physics where the behavior of polymers in solvents of different
qualities often is described by  polymers in vacuum with effective interactions
between the monomeric units.  Attractive interactions correspond to a bad
solvent and result in a collapse of the polymer, while  repulsive
interactions correspond to a good solvent because the isolated polymer adopts a
swollen, self-avoiding-walk like shape \cite{Cloizeaux,Doi}.  While there exists a formal one-to-one
correspondence between the thermodynamic properties of an incompressible
polymer-solvent mixture and the corresponding compressible polymer model with
effective interactions, these effective interactions might not be well
represented by {\em density-independent pair potentials}. For instance, by
replacing the repulsion between solvent and polymer by an effective attraction
between the polymer segments, one might observe a much higher local polymer
density than in the original incompressible mixture where the maximal value of
the local polymer density is limited by the incompressibility constraint.
The differences between incompressible mixtures and effective compressible
systems comprised of only amphiphiles are even more pronounced when one regards
dynamical properties because (i) the density variations in the implicit solvent
model results in variations in the local mobility of the amphiphilic units that
are absent in the original incompressible system and (ii) the solvent carries
momentum, and the concomitant hydrodynamic flow might promote cooperative
re-arrangements. These considerations illustrate that integrating out the
solvent degrees of freedom, though conceptually straightforward, does involve
some practical subtleties.

Initial attempts to construct solvent-free membrane models using simple
pairwise interactions were rather unsuccessful. The model of Drouffe et al.~\cite{Drouffe91} 
represented amphiphiles by single beads interacting via a spherical hard-core repulsion and 
an orientation-dependent short-ranged attraction. They found
that increasing the attraction between the lipid tails resulted in the
formation of membranes. These membranes consisted of a {\em single} layer of particles and 
the membranes were crystalline (gel), i.e. the lipids
laterally condensed onto a triangular lattice. This solid phase was
characterized by the pronounced reduction of lateral lipid diffusion.  When the
temperature was raised the membrane did not form a fluid membrane, but simply
disassembled. To overcome this difficulty, a many-body interaction was
introduced to mimic the hydrophobic effect and to stabilize a fluid membrane.
Additionally these interactions limited the number of neighbors and thereby
suppressed
three-dimensional aggregation in favor of sheet-like structures.
These multi-body, or density-dependent, interactions made the calculation of
thermodynamic quantities rather subtle (see below). 

Noguchi and
Takasu \cite{NOGUCHI01_2} modeled the amphiphiles by rigid rods comprised of three
interaction centers, a head and two tail beads. These beads interact via a
rotationally symmetric potential but the multi-body character of the attraction
of the hydrophobic tail beads is used to stabilize the membrane. Particles repel
each other via a soft core potential which defines the energy scale, $\epsilon,$
and the monomeric length scale, $\sigma$. The potential is of the form

\begin{equation}
V_{\rm rep}(r) = \epsilon e^{-20(r/\sigma-1)},
\end{equation}
and it is truncated and shifted at a cut-off $1.3\sigma$. The multi-body potential
takes the form
\begin{equation}
V_{\rm multi}[\bar \rho] = \left\{
            \begin{array}{ll}
            -0.5 \bar \rho & \mbox{for}\; \bar \rho < \rho^*-1 \\
            0.25(\bar \rho - \rho^*)^2-c & \mbox{for}\; \rho^*-1 \leq \bar \rho < \rho^* \\
            -c &  \mbox{for}\; \bar \rho^* \leq \rho
            \end{array}
                           \right.
\end{equation}
with parameters $\rho^*=10$ and $c=4.75$ for the tail bead nearest to the head and $\rho^*=14$ and $c=6.75$ for the tail bead at the end.
The smoothed density, $\bar \rho$, quantifies the local number of hydrophobic particles in a small sphere around
the reference particle at position, ${\bf r}$, according to
\begin{equation}
\bar \rho = \sum_{{\bf r}'} h(|{\bf r}-{\bf r}'|) \qquad \mbox{with} \qquad
h(r) = \left\{
        \begin{array}{ll}
        1 & \mbox{for}\; r<1.6\sigma, \\
        \frac{1}{\exp[20(r/\sigma-1.9)]+1} & \mbox{for}\; 1.6\sigma \leq r < 2.2\sigma,\\
        0 & \mbox{for}\; 2.2\sigma \leq r.
        \end{array}
       \right.
\end{equation}
where the sum over ${\bf r}'$ includes all hydrophobic segments on other amphiphiles.
At small smoothed densities, $\bar \rho < \rho^*-1$, the multi-body potential is linear
in the density and, thus,  represents a pairwise attraction between neighboring
hydrophobic beads on different molecules. At higher densities, the attractive strength
levels off and adopts a constant value independent of the local density. This feature
avoids the collapse of the hydrophobic tails into extremely dense structures and thus prevents
crystallization. In contrast to the previous model of Drouffe and co-workers \cite{Drouffe91} the membranes
in Noguchi's model are bilayer, i.e., they are comprised to two layers of amphiphilic molecules.

Wang and Frenkel \cite{WANG05}  described another variant of solvent-free models with
multi-body interactions, where amphiphiles were modeled as flexible chains consisting
of three coarse-grained beads. A bending potential along the backbone was
utilized to tune the molecular flexibility. They employed a qualitatively similar
density dependence of the multi-body term, but used an anisotropic weighting function to
construct the smoothed density, $\bar \rho$.

The first solvent-free model that resulted in the formation of  a fluid bilayer
from particles that interact via simple pairwise isotropic interactions was
devised by Farago \cite{FARAGO03}. In this model, amphiphiles consist of rigid, linear trimers
comprising one head-bead and two tail-beads. The interactions were tuned by a rather
lengthy ``trial and error'' process to make the attraction between molecules
sufficiently strong to support the stability of the membrane, but still weak enough
so that the membrane would not crystallize. They are as follows.  Let (1) denote the hydrophilic head bead and (2) and
(3) the hydrophobic beads along the rigid amphiphile that are spaced at a distance,
$\sigma$. Beads of the same type interact via a Lennard-Jones potential
\begin{equation}
V_{ii}(r) = 4 \epsilon_{ii} \left[ \left(\frac{\sigma_{ii}}{r}\right)^{12} - \left(\frac{\sigma_{ii}}{r}\right)^6\right].
\label{eqn:farago1}
\end{equation}
Interactions between the head and the first hydrophobic bead are repulsive
\begin{equation}
V_{12}(r) = 4 \epsilon_{12}  \left(\frac{\sigma_{12}}{r}\right)^{12},
\label{eqn:farago2}
\end{equation}
and the repulsion between the head and the end tail bead is even harsher
\begin{equation}
V_{13}(r) = 4 \epsilon_{13}  \left(\frac{\sigma_{13}}{r}\right)^{18}.
\label{eqn:farago3}
\end{equation}
The attraction between different hydrophobic beads, however, has a broad attractive minimum
\begin{equation}
V_{23}(r) = 4 \epsilon_{23}  \left[ \left(\frac{\sigma_{23}}{r}\right)^{2} - \frac{\sigma_{23}}{r} \right].
\label{eqn:farago4}
\end{equation}
All potentials are truncated and shifted at a cut-off distance $2.5\sigma_{33}$. The potential
parameters are detailed in Table~\ref{tab:farago}.

\begin{table}
\begin{tabular}{|l|llll|l|lll|}
\hline
$\sigma_{ij}/\sigma_{33}$ & head (1) & tail (2) & tail (3) & & $\epsilon_{ij}/k_BT$ & 1 & 2 & 3                   \\
\hline
1           & 1.1      & 1.15     & 1.4      & &                    & 0.1875 & 1.1375 & 200           \\
2           &          & 1.05     & 0.525    & &                    &        & 1.75   & 375           \\
3           &          &          & 1        & &                    &        &        & 1.875         \\
\hline
\end{tabular}
\caption{\label{tab:farago} Parameters of the interaction potentials in Equations (\ref{eqn:farago1})-(\ref{eqn:farago4})
of Farago's solvent-free model. From Ref.~\cite{FARAGO03}.
}
\end{table}

\begin{figure}
\includegraphics[scale=0.45]{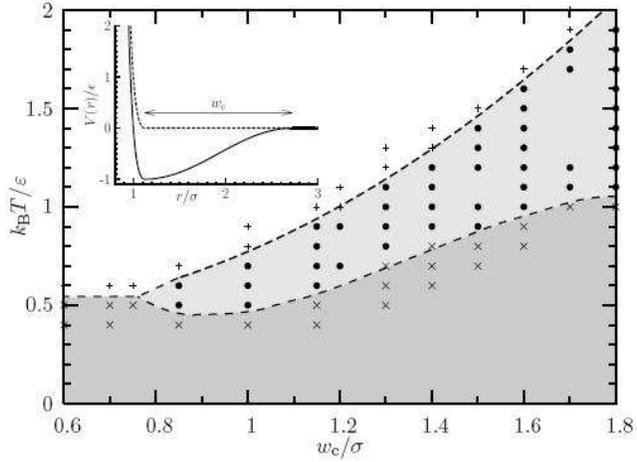}
\caption{
Phase diagram resulting from Deserno's model as a function of rescaled
temperature, $k_BT/\epsilon$ and range, $w_c/\sigma_{33}$ of the attraction
between tail beads according to Eq.~(\ref{eqn:deserno1}).  The areal
density corresponds to zero tension.  Each symbol corresponds to one simulation
and identifies different bilayer phases: $\times$ -- gel/crystalline, $\bullet$ -- fluid,
{\footnotesize $+$} -- unstable.  Lines are merely guides to the eye.  The inset
shows the pair-potential between tail lipids (solid line) and the purely
repulsive head-head and head-tail interaction (dashed
line). From Ref.~\cite{COOKE05b}.\label{fig:deserno1}}
\end{figure}

Deserno and co-workers \cite{COOKE05b,Cooke05} argued that an increase of the range of the interaction is
crucial for stabilizing fluid bilayers.  They
represented amphiphiles as flexible trimers. All beads repel each other via a Lennard-Jones potential
of the type of Eq.~(\ref{eqn:farago1}) which is truncated and shifted at the minimum,
$r_{\rm min}=\sqrt[6]{2}\sigma_{ii}$, resulting in a purely repulsive potential. The size
of the tails defines the length scale, $\sigma_{33}=\sigma_{22}$ while the
heads are smaller, $\sigma_{11}=0.95\sigma_{33}$ and the repulsive interactions
between head and tails are non-additive,
$\sigma_{12}=\sigma_{13}=0.95\sigma_{33}$. In addition to this purely repulsive
interaction, hydrophobic tail beads interact with each other via an attraction
with tunable range, $w_c$ (see inset of Fig.~\ref{fig:deserno1}):

\begin{equation}
V_{\rm att}(r) = \left\{
                 \begin{array}{ll}
         -\epsilon & \mbox{for}\; r < r_{\rm min}, \\
         -\epsilon \cos^2\left(\frac{\pi(r-r_{\rm min})}{2w_c} \right) & \mbox{for}\; r_{\rm min} \leq r < r_{\rm min}+w_c,\\
         0  & \mbox{for}\;r_{\rm min}+w_c<r.
         \end{array}
         \right.
         \label{eqn:deserno1}
\end{equation}
The beads are linked together via a FENE potential
\begin{equation}
V_{\rm bond} = - \frac{1}{2}k_{\rm bond} r_0^2 \ln \left[ 1-\left(\frac{r}{r_0}\right)^2\right],
\end{equation}
with $k_{\rm bond}=30\epsilon/\sigma_{33}^2$ and maximal bond length,
$r_0=1.5\sigma_{33}$. There is no bond angle potential, but the
flexibility is tuned by applying a harmonic spring between the head and the
last tail bead

\begin{equation}
V_{\rm bend}(r_{13})= \frac{1}{2}k_{\rm bend}\left( r_{13}-4\sigma_{33} \right)^2.
\end{equation}
with $k_{\rm bend}=10\epsilon/\sigma_{33}^2$.

Figure \ref{fig:deserno1} presents the phase diagram at zero lateral tension as
a function of rescaled temperature and the range of the attractive interaction,
$w_c$.  If the range of the attraction, $w_c$, is small compared to the
effective hard core diameter, $\sigma_{33}$, the membrane assembles into a
solid sheet upon cooling.  Only if the range of the attraction is sufficiently
large does one encounter two transitions upon cooling. As the system is cooled, the
amphiphiles first assemble into a fluid membrane.  Upon further cooling,  the
membrane crystallizes. In this solid phase the lateral mobility of the
amphiphiles is strongly reduced. The temperature interval in which the fluid
membrane is stable increases with the range of the attraction and extends to
higher temperatures.

The role of the range of the attractive interactions in stabilizing fluid,
self-assembled membranes qualitatively resembles the role it plays in
stabilizing a fluid phase of simple molecules. If such molecules interact
via a hard-core repulsion and a weak, but longer-ranged, attraction a fluid
phase exists only if the range of the attraction is sufficiently large,
roughly greater than $20\%$ of the hard core diameter. Otherwise the fluid
directly condenses into a solid \cite{Hagen94,Dijkstra98,Likos01}.

Another solvent-free model has been devised by Brannigan and co-workers
\cite{Brannigan04,Brannigan04b,Brannigan05,Brannigan05c,Brannigan06b}. The
amphiphiles consist of five beads. The first bead (h) corresponds to the
hydrophilic head, the second bead (i) is associated with the interface between
hydrophilic and hydrophobic entities, and the other three beads constitute the
hydrophobic tail (t). The distance between neighboring beads along the
amphiphile are constrained to a distance $\sigma$, which defines the length scale. A
bending potential of the form $V_{\rm bend}(\theta)=k_{\rm bend}\cos\theta,$
with $5\epsilon \leq k_{\rm bend} \leq 10\epsilon,$ tunes the geometrical shape
of the molecules.
A repulsive interaction
\begin{equation}
V_{\rm rep}(r) = c_{\rm rep} \left(\frac{\sigma}{r}\right)^{12} \qquad \mbox{with} \qquad c_{\rm rep}= 0.4\epsilon,
\end{equation}
is applied between all beads
except intramolecular pairs separated by less than
three bonds.  Tail beads, and a tail and an interface bead, attract
each other via a standard van-der-Waals attraction:

\begin{equation}
V_{\rm att}(r) = - \epsilon \left(\frac{\sigma}{r}\right)^{6}.
\end{equation}
Both, repulsion and attraction, are truncated at a distance, $2\sigma$.
The interface beads, however, interact among each other via a longer-ranged
potential

\begin{equation}
V_{\rm int}(r) = - c_{\rm int} \left(\frac{\sigma}{r}\right)^{2} \qquad \mbox{with} \qquad c_{\rm int}=3 \epsilon,
\end{equation}
which is cut off at $3\sigma$. This longer-ranged attraction, which acts only between  the
interface beads, is sufficient to stabilize fluid bilayer membranes
at reduced temperature, $k_BT/\epsilon=0.9$.

A different path to the development of  generic coarse-grained models has been
pursued by Groot \cite{Groot00,GROOT01}, Smit
\cite{KRANENBURG03_1,KRANENBURG04_1,KRANENBURG04_2,KRANENBURG04_3,KRANENBURG04_4,Kranenburg05},
Shillcock \cite{SHILLCOCK02,IMPARATO03,ILLYA05,Ortiz05} and Mouritsen \cite{Jakobsen05b,Jakobsen05} with their co-workers.
These
coarse-grained models utilize ultra-soft interactions in conjunction with a
dissipative particle dynamics (DPD) thermostat \cite{DPD1,DPD2,DPD2b,DPD3b,DPD4b}.
Unlike the Langevin thermostat that adds random noise and friction to each
particle, the DPD thermostat adds random noise and friction to each neighboring
pair of particles. Thus, momentum is locally conserved and hydrodynamic
flow can be described. The use of ultra-soft potentials allows for rather large
time steps for integrating the equation of motions (see Refs.~\cite{BESOLD00,VATTULAINEN02,Jakobsen05,Allen06} for a detailed discussion). Their use can be justified
by recognizing that the center of mass positions of the group of atoms that
constitute a coarse-grained segment can overlap and their interaction is much
softer than the harsh repulsions on the atomistic scale (cf.~Sec.~\ref{sec:syscg}). This is a generic
feature of coarse-grained models: the larger the length scale the weaker the
interactions. By the same token, the density of the soft beads exceeds unity
when measured in units of the particles' interaction radius.

In DPD simulations, particles of type $i$ and $j$ (denoting water (w), head (h),
glycerol-linking (e) and tail (t) ) interact via a very simplistic soft force of the form:

\begin{equation}
{\bf F}_{ij}({\bf r})= \left\{ \begin{array}{ll}
        -a_{ij} \left(1-\frac{|{\bf r}|}{r_c}\right)\frac{{\bf r}}{|{\bf r}|} & \mbox{for}\; r \leq r_c \\
        0 & \mbox{for}\; r>r_c
              \end{array}\right.
\end{equation}
where ${\bf r}$ is the distance vector between the particles' positions. The range of the
interactions, $r_c$, between these soft beads sets the length scale.

Originally, the DPD simulation scheme had been devised to simulate fluid flow,
and a soft bead was thought of as a fluid volume comprising many molecules but
still being macroscopically small.  In the context of membrane simulations, a
soft bead represents a rather small fluid volume that consists only of several
molecular groups comprising the amphiphile. Often one identifies the range of
interaction, $r_c$, with 1 nm, i.e., one tail bead corresponds to three or four
methyl units. By the same token, a solvent bead corresponds to a small number
of water molecules.  (Attempts to map a single methyl unit onto a soft bead
were rather unsuccessful \cite{KRANENBURG04_2} in reproducing the internal
bilayer structure and resulted in a significant interdigitation of the apposed
monolayers.)
Typically the amphiphilic molecules consist of only a very small number of beads
-- one to three hydrophilic head beads and four to ten hydrophobic tail beads.
The longer the amphiphiles the more stable and rigid the bilayer is.

The strength of the interaction simultaneously describes  the repulsion between unlike species, and the excluded volume of
the coarse-grained beads which imparts a small compressibility to the liquid.
The parameters of the model are
tailored to reproduce key characteristics of the amphiphiles in solution (e.g.,
the compressibility of the solvent and the bilayer compressibility).

A typical set of interaction strength is:
\begin{equation}
\begin{array}{l|lll}
a_{ij}/k_BT & w & t & h \\
\hline
w      & 15(25) & 80 & 15 \\
t      & 80     & 25 & 80 \\
h      & 15     & 80 & 35 \\
\end{array}
\end{equation}
These values were originally proposed by Groot to parameterize the interactions
of ionic surfactants \cite{GROOT01}.  Smit and co-workers \cite{KRANENBURG04_2} increased $a_{ww}$
from 15 to 25 to avoid very high densities in the bilayer's hydrophobic core.
Different sets of parameters have been devised for double-tail lipids.
Shillcock and Lipowsky \cite{SHILLCOCK02} used the set

\begin{equation}
\begin{array}{l|lll}
a_{ij}/k_BT & w & t & h \\
\hline
w      & 25 & 75 & 35 \\
t      & 75 & 25 & 50 \\
h      & 35 & 50 & 25 \\
\end{array}
\end{equation}

Interactions along the amphiphile determine the molecular shape of the amphiphile. They
include a harmonic bonding potential of the form \cite{GROOT01}
\begin{equation}
V_{\rm bond}(r) = k_{\rm bond} {\bf r}^2,
\end{equation}
or \cite{KRANENBURG03_1}
\begin{equation}
V_{\rm bond}(r) = k_{\rm bond} (|{\bf r}|-r_0)^2.
\end{equation}
Additionally, bending potentials of the form \cite{KRANENBURG03_1}
\begin{equation}
V_{\rm bend} = \frac{1}{2} k_{\rm bend} (\theta - \theta_0)^2,
\end{equation}
or \cite{SHILLCOCK02}
\begin{equation}
V_{\rm bend} = k_{\rm bend} (1-\cos(\theta-\theta_0)),
\end{equation}
have been applied.
Shillcock and Lipowsky \cite{SHILLCOCK02} highlight the role of the bending potential on the internal
structure of the bilayer membrane. The earlier
DPD model by Groot \cite{GROOT01}, and also self-consistent field models \cite{Leermakers88}, describe the lipid tails as
completely flexible which leads to a broad distribution of the tail ends throughout the
bilayer. This interdigitation of the two apposing monolayers that form the membrane typically is
not observed in biological membranes. A very large incompatibility between hydrophobic
and hydrophilic entities would be required to stretch the monolayers sufficiently to
reproduce the structure of a biological membrane.

\begin{figure}
\begin{minipage}{0.55\textwidth}
\epsfig{file=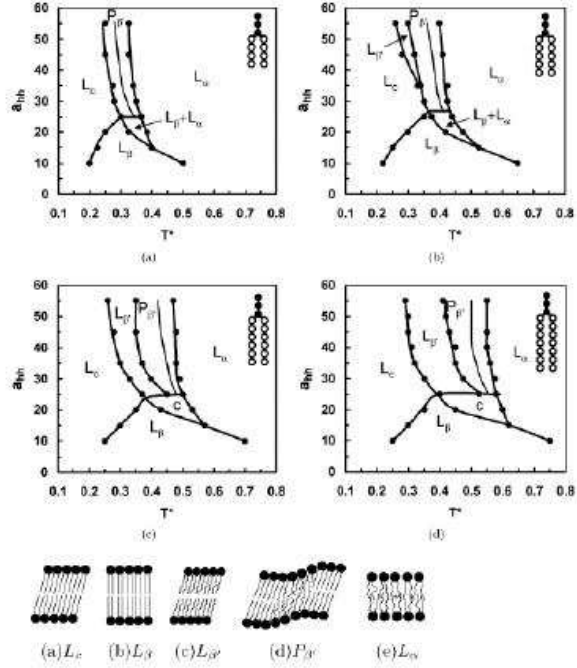,width=0.9\textwidth,clip=}
\end{minipage}
\begin{minipage}{0.45\textwidth}
\caption{
Phase diagrams of model lipids as a function of head-head repulsion, $a_{hh}$, and temperature, $T^*$ for various lipid architectures as indicated in the key. The different bilayer phases are sketch in the bottom row: $L_c$ -- tilted subgel phase, $L_\beta$ -- gel phase, $L_\beta'$ -- tilted gel phase, $P_\beta'$ -- ripple phase, $L_\alpha$ -- fluid phase. In the two lower phase diagrams ``c'' denotes a coexistence region, of which the exact structure was difficult to determine. 
From Ref.~\cite{Kranenburg05}.\label{fig:DPD}}
\end{minipage}
\end{figure}

This example illustrates that bilayer structure and properties do sensitively depend on the
model parameters of the minimal coarse-grained models. Since the potentials of these models are not
directly related to molecular interactions, the parameters have to be adjusted so as  to reproduce
macroscopic observables. Although the minimal coarse-grained models are quite simple, they have been shown to
form several of the known bilayer phases: fluid, gel and crystalline \cite{KRANENBURG04_1,Brannigan04b,Lenz05,Marrink04b,Marrink05,Kranenburg05,Cooke05}, as well as more exotic ones,
like the ripple phase \cite{KRANENBURG04_1,Kranenburg05}. 
The different phases that can
be obtained by the DPD model of Smit and co-workers are illustrated in Fig.~\ref{fig:DPD}.
The phase diagram as well as the area compressibility modulus, bending stiffness and spontaneous curvature \cite{ILLYA05,Brannigan06b,Cooke06} have been explored 
as a function of model parameters. This ensures that one can choose parameters that make the coarse-grained
model mimic, qualitatively, the behavior of an experimental system. 

\subsubsection{Systematic coarse-graining: potential and limitations}
\label{sec:syscg}
While the above minimal coarse-grained models are able to explore the generic
features of the behavior of amphiphiles in solution, much recent interest has
focused on ``deriving'' coarse-grained models for a specific chemical
substance \cite{SHELLEY01_1,AYTON02,MARRINK04,NIELSEN04_1,IZVEKOV05,BOEK05}.  The general feature of these systematic coarse-graining schemes is
the attempt to utilize information about the microscopic structure, obtained by atomistic
simulations, for instance, in order to parameterize the interactions between the
coarse-grained entities.

These techniques have originally been developed for polymer systems, and they
have been extensively employed to describe quantitatively the structure and
dynamics of polymer melts and solutions \cite{Baschnagel00,Muller-Plathe02,Kremer02,Paul91,Murat98b,Tschop98,Bolhuis01b,Doi03}.  While the universal properties of
amphiphilic systems only require a coarse-grained model to capture a few relevant interactions, the
{systematic  construction} of such a model that quantitatively
reproduces the behavior of the underlying microscopic system is a very
ambitious task.  From the onset, it is obvious that decimating the degrees of
freedom will result in a loss of information, and will generate very
complicated multi-body interactions \cite{Ma85}. The latter imparts additional
complications on extracting thermodynamic information from coarse-grained
models. Commonly the multi-body interactions are, in turn, approximated by
pairwise interactions in order to retain the computational efficiency of the
coarse-grained representation, and these effective pairwise interactions
depend on the thermodynamic state of the system (i.e., they depend on density or
temperature, and they are different in different thermodynamic phases).

The general principles of systematic coarse-graining consist in (i) choosing a
set of key characteristics that the coarse-grained model shall reproduce, and
(ii) determining the interactions between the coarse-grained degrees of freedom so
as to reproduce these characteristics.  The first step is the most crucial, and
is guided by insight into the physics of the phenomena that one wants to
investigate.  Three qualitatively different properties can be distinguished --
structural quantities, thermodynamic properties, and dynamic characteristics.

Structural quantities are related to the geometry of the molecules on different
length scales. In order to capture the specific details of the molecular
architecture, a coarse-grained model should not only reproduce the overall size
of the molecule (e.g., the end-to-end distance), but it should also include finer
details of the molecular architecture (e.g., the stiffness along the molecular
backbone, the bulkiness of the lipid's head, the location of double bonds in the
hydrocarbon tails). It is essential to capture the rough features of the
molecular geometry and its fluctuations, or ability to deform in response to its
environment.

In the ideal case the parameterization of a coarse-grained model is based on
the  detailed atomistic information about the molecular conformations under the
same thermodynamic conditions, i.e., the same temperature and density.
Then, one explicitly defines a mapping from a configuration of the atomistic
model, $\{{\bf r}\}$, onto a configuration of the coarse-grained model, $\{
{\bf R}\}$.  Utilizing a large equilibrated sample of atomistic
configurations, one can obtain the probability distribution of distances between
coarse-grained entities.

\begin{figure}
\includegraphics[width=0.45\textwidth]{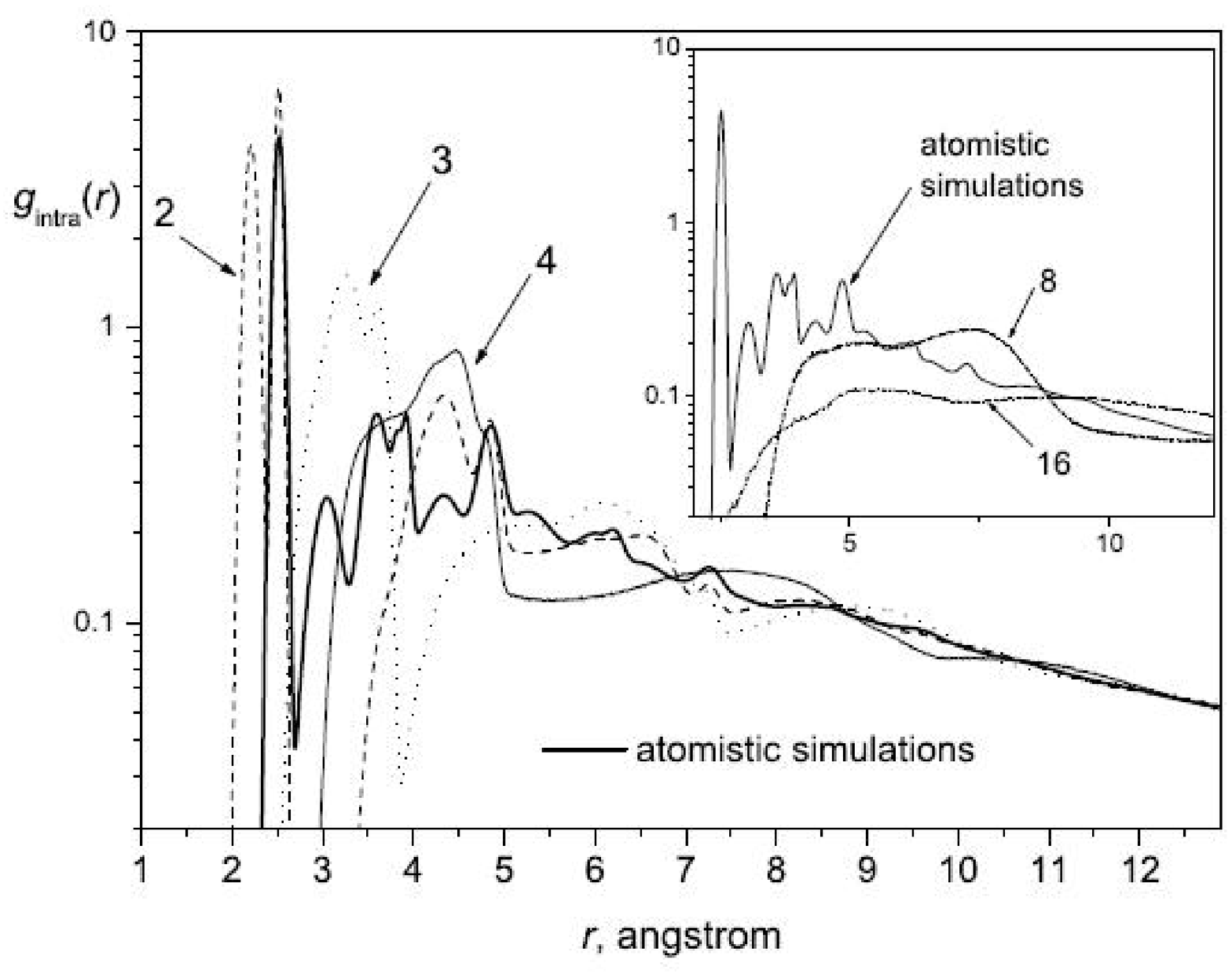}
\includegraphics[width=0.45\textwidth]{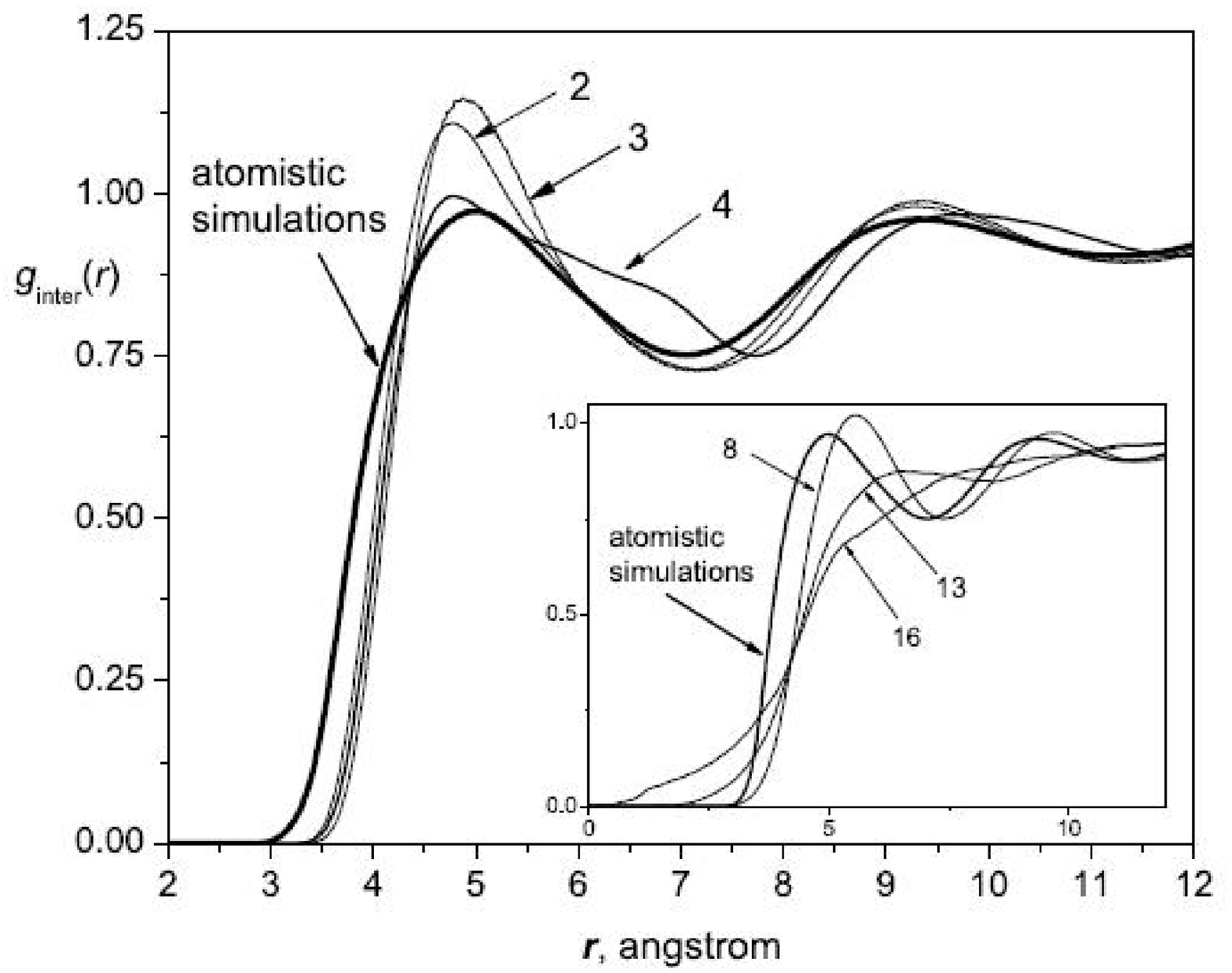}
\caption{
(a) Coarse-graining of the intramolecular segment-segment correlation function
of polybutadiene ($240$ K).  The numbers mark data with different degrees of the
coarse-graining.  For example, the coarse-graining degree $n=4$ yields a chain
molecule for which four united atoms are approximated by one effective segment.
The bold curve is the correlation function calculated using the united-atom model
($\delta$-type peaks at $r=1.34$~{\AA}, $1.5$~{\AA}, $1.53$ {\AA} arising from
the CH$=$CH, CH$_2-$CH and CH$_2-$CH$_2$ bonding distances are omitted).
(b) Coarse-graining of the intermolecular segment-segment correlation function
of polybutadiene at $240$ K.  For $n\le 8$, the correlation function changes
quantitatively.  For large degree of the coarse-graining (e.g., $n > 13$), the
correlation peaks and minima disappear and the correlation hole at $r<3$A
shrinks. The bold curve is the correlation function calculated using the united-atom
model. From Ref.~\protect\cite{Yelash06}.
\label{fig:leonid}}
\end{figure}

This procedure is illustrated in Fig.~\ref{fig:leonid} for the case of a
polybutadiene melt at 240 K and atmospheric pressure. The rich structure of local
correlations is specific to the chosen system, but the qualitative features
discussed in the following are born out by a wide class of systems and have been
observed in a large variety of studies \cite{Muller-Plathe02,Tschop98,Baschnagel91,Doruker97,FALLER99,Meyer00,Abrams03,Abrams03b,Faller04b,MARRINK04,MURTOLA04,NIELSEN04_1,IZVEKOV05_2,BOEK05,Sun05,Milano05,Praprotnik05,Sun06b,Sun06,Silbermann06,Izvekov06,Bedrov06,Prampolini06}.
The two panels of the
figure present the intramolecular and intermolecular pair correlation function
as obtained from a simulation within a united atom model \cite{Smith98,Smith99,Smith01,Krushev02}.  The
{\em intramolecular} correlations exhibit a rich structure on short length scales
that mirrors both the correlations due to bond length and torsional
interactions, as well as the delicate packing of the dense polymeric fluid. At
large distances, those correlations have died away, and the intramolecular pair
correlation function smoothly decays like $g_{\rm intra}(r) \sim 1/r$ as
expected for a Gaussian polymer in a melt \cite{deGennesBook}.  The {\em intermolecular} pair correlation
function describes the probability of finding two coarse-grained segments
on different molecules at a distance, $r$. It exhibits qualitatively a form that
one also expects for a simple liquid. At small distances the correlation
function vanishes -- this distance characterizes the ``thickness'' of the
polymer. There is a broad nearest-neighbor peak around 5 A, and there are a
few further oscillations at larger distances. The long-range approach of the
intermolecular correlation function to unity is dictated by the polymeric correlation hole effect and it is identical to the decay of the
intramolecular correlations because the two cancel one another at  length
scales larger than the correlation length of density fluctuations due to the
near incompressibility of the dense liquid \cite{deGennesBook}.

One can use the explicit configurations of an atomistic simulation, and lump
$n$ successive carbon atoms along the backbone of the polymer into an effective
coarse-grained segment. In this specific example, the location of the coarse-grained bead is taken to be
the center of mass of its constituents, without accounting for the differences
in molecular mass of CH and CH$_2$ units.  The correlations between the
 coarse-grained beads constructed in this manner for different levels, $n$, of
coarse-graining are also depicted in Fig.~\ref{fig:leonid}. Two general
characteristics can be observed: First, the larger the degree of
coarse-graining, the smoother are the intra- and intermolecular pair correlation
functions. This behavior stems from the fact that the coarse-grained beads can
partially overlap. This softening increases with the degree of
coarse-graining, $n$. Second, the local structure is very sensitive to $n$,
while the large distance behavior is not. In the specific example, constructing
a coarse-grained bead from $n=3$ segments yields a rather poor representation
of the intramolecular and intermolecular correlations: The first peak of
$g_{\rm intra}$ coincides with the minimum of the data for the united atom
model and the first peak of $g_{\rm inter}$ is too high.  The value $n=4$
yields a better description. Increasing the value of $n$ even further, one finds that the
local structure gradually fades away and the beads become so soft that there is
a significant probability that the coarse-grained beads overlap (c.f., inset
of Fig.~\ref{fig:leonid} b).  Thus there is an optimal degree of coarse-graining.
Values of $n$ which are too small lead to comparably harsh coarse-grained potentials which, in turn, give
rise to packing effects that are not related to the structure of the underlying
atomistic systems. Large values of $n$ result in a loss of local structure
and thereby of chemical specificity.

The thermodynamic equilibrium properties of a coarse-grained model can be constructed formally via a partial trace,
i.e., a summation over all microscopic configurations compatible with a fixed set of coarse-grained degrees of freedom. This
procedure is in complete analogy with Renormalization Group calculations which have
successfully been employed in polymer physics and critical phenomena \cite{FreedBook,Cloizeaux,Schafer}. Let $\{
{\bf r}\}$ denote the coordinates of the detailed microscopic model (e.g., the atom
positions in a chemically realistic model) and let $\{ {\bf R}\}$ denote the
coarse-grained degrees of freedom. There is a mapping from the detailed degrees
of freedom, $\{ {\bf r}\}$, onto the coarse-grained ones, ${\bf R}[\{{\bf
r}\}]$.  For instance, $\{ {\bf R}\}$, might be the center of mass of a small
group of atoms or the location of a particular group of a lipid molecule. Let
$E[\{{\bf r}\}]$ denote the pairwise interactions of the microscopic system,
so one can calculate an effective interaction, $U[\{{\bf R}\}]$ between the
coarse-grained degrees of freedom via \cite{Ma85}

\begin{equation}
\exp\left(-\frac{U[\{{\bf R}\}]}{k_BT}\right) = \int {\cal D}[\{{\bf r}\}] \exp\left(-\frac{E[\{{\bf r}\}]}{k_BT}\right)
\delta\left({\bf R}-{\bf R}[\{{\bf r}\}]\right)
\label{eqn:cg}
\end{equation}
With this definition of the effective interaction, $U$, the partition function
of the microscopic system ${\cal Z}$ can be obtained according to

\begin{equation}
{\cal Z} = \int {\cal D}[\{{\bf R}\}] \exp\left(-\frac{U[\{{\bf R}\}]}{k_BT}\right)
\end{equation}
and the probability of the coarse-grained degrees of freedom is identical to
that generated by equilibrium configurations of the microscopic model. There
are, however, two caveats:

(i) $U$ does not possess the typical characteristics
of an interaction but rather those of a free energy.  In particular, $U$
depends on the thermodynamic state characterized by temperature, pressure, etc.
Due account of this state dependence has to be taken if thermodynamic
quantities are extracted. For example, the internal energy is calculated
according to \cite{Louis02}:

\begin{eqnarray}
\langle E \rangle & \equiv & \frac{\int {\cal D}[\{{\bf r}\}] E[\{{\bf r}\}]
\exp\left(-\frac{E[\{{\bf r}\}]}{k_BT}\right)}{\int {\cal D}[\{{\bf r}\}] \exp\left(-\frac{E[\{{\bf r}\}]}{k_BT}\right)} \\
                  & = & k_BT^2 \frac{\partial \ln {\cal Z}}{\partial T} \\
          & = & \frac{\int {\cal D}[\{{\bf R}\}] \left( U[\{{\bf R}\}]-T\frac{\partial}{\partial T} U[\{{\bf R}\}]\right)\exp\left(-\frac{U[\{{\bf r}\}]}{k_BT}\right)}{\int {\cal D}[\{{\bf R}\}] \exp\left(-\frac{U[\{{\bf r}\}]}{k_BT}\right)} \\
          & = & \left \langle U[\{{\bf R}\}]-T\frac{\partial}{\partial T} U[\{{\bf R}\}] \right\rangle_{|\{{\bf R}\}}
\end{eqnarray}
By the same token, the effective potential depends on density, $\rho$, and care has to
be taken when calculating derivatives of thermodynamic quantities with respect to the
number of particles or volume. For density-dependent central pair potentials, $U(R,\rho,T,\cdots)$ the pressure is given by \cite{Ascarelli69}
\begin{eqnarray}
p & \equiv & - \frac{\partial F}{\partial V} \\
  & = & p_{\rm ideal} + p_{\rm virial} + \frac{\rho^3}{2} \int {\rm d}^3R\;  \frac{\partial U(R,\rho,T,\cdots)}{\partial \rho} g(R,\rho,T,\cdots) \\
  \mbox{with} && p_{\rm ideal} = k_BT \rho \qquad \mbox{and} \qquad
  p_{\rm virial} = - \frac{\rho^2}{6} \int {\rm d}^3R\; R \frac{\partial U(R)}{\partial R}  g(R)
  \label{eqn:virial}
\end{eqnarray}
where the first terms are the familiar ideal gas term and the viral expression for the pressure which
ignore the density dependence of the effective potential. The additional
second term arises due to the dependence of the effective interaction, $U$, on the thermodynamic conditions
under which it has been obtained \cite{Ascarelli69,Louis02}.  $g(R)$ is the pair correlation function of the coarse-grained fluid (cf.~Eq.~(\ref{eqn:cgpair})).

(ii) Generally, $U[\{{\bf R}\}]$ cannot be decomposed into pairwise
interactions but it consists of complicated many-body interactions. It is
neither feasible to  construct these many-body interactions numerically through the formal
coarse-graining procedure outlined by Eq.~(\ref{eqn:cg}) (see Refs \cite{Ma76,Swendsen79,Reynolds80,Pawley84} for
explicit schemes for simple lattice models), nor can they be utilized in an
efficient computational model.

In order to
obtain a computationally tractable model, one has to approximate the effective,
many-body interaction by pairwise potentials, and one often ignores the state
dependence of the effective interaction.
To this end, one seeks an approximation of the form

\begin{equation}
U[\{{\bf R}\},\rho,T, \cdots] \approx \sum_i \sum_{j<i} V({\bf R}_i-{\bf R}_j)
\end{equation}
The detailed form of $V$ is unknown. Often one utilizes a generic form that contains several
free parameters. These parameters are adjusted to the specific system (see below).
Moreover, one requires that even after ignoring the state dependence of the interactions, the
model still reproduces thermodynamic properties such as the equation of state, the interfacial
tension, etc. It is this uncontrolled approximation in the construction of coarse-grained models
that requires physical insight.

Once the general type of the interactions to be included and the key characteristics the
coarse-grained model should reproduce have been chosen, the second step of
constructing the detailed form of the coarse-grained interactions, $V$, is conceptually
straightforward but tedious. As a first approximation one utilizes the {\em
potential of mean force}, $V_{\rm mf}$ obtained via the pair correlation function of
coarse-grained degrees of freedom:

\begin{equation}
g({\bf R}_{ij}) \equiv \exp \left(-\frac{ V_{\rm mf} ({\bf R}_{ij})}{k_BT} \right)
\label{eqn:MF}
\end{equation}
with ${\bf R}_{ij} = {\bf R}_i-{\bf R}_j,$ and
where the pair correlation function is given by
\begin{equation}
g({\bf R}) = V\left\langle \delta\left({\bf R}-(R_i[\{{\bf r}\}]-R_j[\{{\bf r}\}])\right)\right\rangle
\label{eqn:cgpair}
\end{equation}
through the microscopic configurations.

By definition this construction procedure works well if the coarse-grained
degrees of freedom are dilute, but  fails if higher order correlations become
important (condensed phase effect). Nevertheless this potential of mean force serves as an initial guess
for more sophisticated procedures. One strategy consists in using the
Ornstein-Zernike equation

\begin{equation}
h({\bf R}_{12}) = c({\bf R}_{12})+\rho \int {\rm d}{\bf R}_3 \; h({\bf R}_{23}) c({\bf R}_{13})
\label{eqn:OZ}
\end{equation}
to obtain the direct correlation function, $c({\bf R})$ from the total
correlation function, $h({\bf R})=g({\bf R})-1$. Then, one can use an approximate
closure relation to express the direct correlation function in terms of the interparticle
potential. Liquid state theory provides guidance in choosing accurate closures.
An example is the hypernetted chain closure (HNC) \cite{Silbermann06}

\begin{equation}
\frac{V({\bf R})}{k_BT}= - \ln g({\bf R}) + h({\bf R}) - c({\bf R})
\end{equation}
which differs from the potential of mean force in Eq.~(\ref{eqn:MF}) by the last two terms.

This scheme provides a systematic, yet approximate, approach for constructing the
effective pair interactions.  In polymer physics, one often employs a numerical
method -- iterative Boltzmann inversion \cite{Reith03,Faller03d,Sun06,Sun06b,Silbermann06}. One starts with an initial guess for
the pair interactions, $V_0$. This pair interaction is used in a computer
simulation of the coarse-grained system, and the correlation function, $g_0$, is
monitored. In the next step of the iteration, $i=1$, the potential is improved
according to

\begin{equation}
V_i({\bf R}) = V_{i-1}({\bf R}) + k_BT \ln \frac{g_{i-1}({\bf R})}{g({\bf R})}
\end{equation}
where $g({\bf R})$ is the  reference pair correlation function obtained from
the microscopic system.  The procedure is continued until the coarse-grained
simulations with the potential $V_i$ reproduce the reference pair correlation
function with sufficient accuracy. Other iterative scheme, e.g., self-adjusting
Wang-Landau-type methods \cite{WangLandau1}, can be envisioned as well.

In addition to these structural quantities the potential parameters of the coarse-grained
representation are tailored to reproduce thermodynamic properties such as pressure, interface
tension or elastic moduli. Some quantities (like pressure or membrane tension) can be included
in the iterative Boltzmann inversion scheme utilizing constant pressure or tension simulations \cite{Sun06b}. Alternatively,
non-bonded interactions that do not strongly impact
the molecular structure are adjusted to reproduce the desired thermodynamic properties.

While the mapping of equilibrium properties between atomistic and
coarse-grained models is conceptually well understood, the mapping of dynamical
properties might prove an even bigger challenge. The softer interactions in
coarse-grained models typically speed up the dynamics on large length scales.
This is a major computational benefit of coarse-graining. However the speed up is
not uniform, but rather depends upon the specific dynamic property.
For instance, the ratio of
time constants between lateral diffusion of the lipids' center of mass, the
structural relaxation of the lipid conformations and the rotational diffusion
around the bilayer's normal differ between the coarse-grained model and
atomistic simulations \cite{NIELSEN04_1}. Several factors contribute to the
non-uniform speed-up:

The mobility of lipid head groups is coupled to the structural rearrangement of
water in their vicinity, but the local structure (e.g., hydrogen bonding) is not
faithfully reproduced by coarse-grained models.  Likewise, the elimination of
degrees of freedom of the hydrophobic tails (C--C bond stretching, bond angle and torsional potentials) 
in the course of the
coarse-graining procedure speeds up the structural relaxation of the lipid
tails. Moreover, the neglect of topological constraints in the molecules'
dynamics (non-crossability) by very soft potentials (e.g., DPD models) will
selectively speed up the dynamics of extended molecules compared to e.g., small
solvent molecules.  These effects have quite different origins and, in general,
will result in different speed-up factors. Moreover, electrostatic interactions
omitted in coarse-grained models might be not essential for structural
properties, but still may significantly alter the dynamics \cite{LOPEZ02}.
Likewise, integrating out the solvent molecules results in strong attractions
between like units of the amphiphilic molecules. This might dramatically
influence the dynamics (see e.g., Ref.~\cite{Chang01} for a careful discussion of the role of
explicit solvent on the dynamics of the collapse of a polymer chain in a bad
solvent).  Therefore, one cannot expect that a {\em single} scale factor
relates the characteristic times of different motions and dynamic processes
between atomistic and coarse-grained models \cite{Muller03b,NIELSEN04_1,Chang05b}.
Thus particular care has to be exerted
when non-equilibrium process are to be described by coarse-grained models.
Nevertheless, a natural and common identification of the time scale in coarse-grained models consists
in matching the characteristic time
of the slowest process of relevance, e.g., the lateral diffusion of a
amphiphilic molecule in the bilayer.

\subsection{Coarse-grained field-theoretic models and molecular field theories}
\label{sec:SCFT} An alternative approach to a particle-based
description has its roots in statistical field theory
\cite{Doi-Edwards-Book,Fredrickson-Book}. In this framework one
operates with collective variables, such as density and interaction
fields instead of the coordinates of the individual molecules. The
formal advantage of this approach is that it allows one to
decouple many-body (particle-particle) interactions and to replace
them with one-body (particle-field) interactions. This
``simplification'' comes at a price -- to calculate the partition
function one has to perform thermodynamic averages over the field
variables. Nevertheless, under suitable conditions, one can
introduce controlled approximations that make the field-based
approaches tractable.

Formally, the particle-based canonical partition function of a
system of $n$ particles, occupying volume $V$ and interacting
through a (multi-body) potential $U(\{{\bf{r}}\})$
\begin{equation}\label{PFparticle}
    \mathcal{Z}\propto \prod_{j=1}^{n}\int_{V} {\rm d}^3{\bf{r}}_j\ e^{- U(\{{\bf{r}}\})/k_BT},
\end{equation}
can be transformed into the following field-theoretic partition
function:
\begin{equation}\label{PFfield}
    \mathcal{Z}\propto \int\mathcal{D}w_\alpha \int\mathcal{D}\phi_\beta\;
    e^{-{\cal H}[w_\alpha,\phi_\beta]/k_BT},
\end{equation}
where ${\cal H}[w_\alpha,\phi_\beta]$ is usually called an effective action
or Hamiltonian of the system, and it depends on
density and interaction
field variables, $\phi_\beta({\bf r})$ and $w_\alpha({\bf r})$. Here, the indices $\alpha$ and $\beta$ run over
different {\em types} of microscopic units (monomers, segments), and the
functional integration $\int\mathcal{D}f$ is performed over all
spatial realizations of the field variables.

Despite the fact that the field-theoretic transformation replaces
one very hard problem, Eq.~(\ref{PFparticle}), with another,
Eq.~(\ref{PFfield}), in this representation it is simple to implement
the {\em mean-field} approximation which, in polymer systems, is quite accurate \cite{Muller05f,SCFT2}.
Physically, the mean-field approximation amounts to the assumption
that the statistics of a system is dominated by a particular { field}
configuration, so that the partition function is also dominated by a
single term and takes the simple, approximate form
\begin{equation}\label{PFfield2}
    \mathcal{Z}\propto  e^{-{\cal H}[w_\alpha^*,\phi_\beta^*]/k_BT},
\end{equation}
where the dominant field configurations, $w_\alpha^*$ and
$\phi_\beta^*$, are the stationary (and, actually, saddle point)
configurations of the effective Hamiltonian:
\begin{equation}
 \left. \frac{\delta {\cal H}}{\delta w_\alpha({\bf r})}\right|_{w^*,\phi^*} =
 \left. \frac{\delta {\cal H}}{\delta \phi_\beta({\bf r})}\right|_{w^*,\phi^*}=0
\end{equation}
This approximation is usually not very accurate for systems composed
of low-molecular weight species with short-ranged interactions, but it
becomes more so for particles with long-ranged
interactions and for high-molecular weight polymers at high
concentrations. Depending on the system of interest, one can
frequently quantify deviations away from the mean-field solution
caused by long-wavelength fluctuations, and define the corresponding Ginzburg
parameter \cite{Muller05f,SCFT2}. In polymer systems with short-ranged segment interactions,
it is usually the dimensionless overlap parameter, or invariant degree of polymerization, $\bar{\cal N} \equiv (\rho
R_e^3/N)^2$, which characterizes the number of polymer chains interacting directly
with a given chain. Here $R_e$ is the molecules' end-to-end distance,
$N$ denotes the number of monomeric units (segments) per molecule and, $\rho$ is
the segment number density. This
parameter can be compared to that in the case of simple lattice
models. In the latter systems, the mean-field approximation usually
becomes accurate at large enough site coordination number, commonly
achieved either at high spatial dimensionality or for long-ranged
interaction potential. In polymer systems, however, high effective
coordination numbers can be achieved by increasing the
polymerization index at a constant segment density.

There are two basic issues that have to be addressed within any
mean-field approach: (i) calculation of the average molecular-field
acting on a given molecule due to a density distribution of all
other molecules present in the system, (ii) evaluation of a
single-molecule partition function in the presence of this field,
which can be formally written as:
\begin{equation}
Z_1[w]=\sum_{\{s\}} g_s e^{-w(s)/k_BT}.
\end{equation}
Here $w(s)$ is the mean-molecular field felt by the chain in
microscopic state $s$. These states have {\em a priori} statistical weights $g_s$, that
are solely determined by intra-chain degrees of freedom and characterize the molecular architecture in a spatially homogeneous system. Ultimately,
the two ingredients of a mean-field approach have to be solved
self-consistently for all the species, therefore the name {\em
self-consistent field (SCF) theory}. The variety of the SCF
approaches in the literature can be classified by the types of
approximations used for solving the two problems above.
Below, we describe a number of such approaches as applied to modeling of bilayer membranes.

\subsubsection{Anchored chain models} One of the first
molecular-field theories of a lipid bilayer was proposed by Marcelja
\cite{Marcelja73,Marcelja74b}. In this model, lipid head groups and
solvent molecules are modeled by a simple boundary to which lipid
tails are anchored at a given areal density. Intra-chain degrees of
freedom are sampled using Flory's Rotational Isomeric State (RIS)
model \cite{flory54,Mattice94} and the segments interact through an
anisotropic aligning potential of Maier-Saupe kind:
\begin{equation}
w(\Theta_{12}) = V_0
\left[\frac{3}{2}\cos^2(\Theta_{12})-\frac{1}{2}\right],
\label{EQ:Maier-Saupe}
\end{equation}
where $\Theta_{12}$ is the angle between segment axes. This
potential is borrowed from the theory of nematic liquid crystals,
and in the context of lipid packing the aligning tendency between
segments stems from the fact that all-{\em trans} chains pack at a
higher density and, hence, experience stronger dispersive van der
Waals attractions. Further, being better aligned, they are subject to fewer hard
core repulsions. In the mean-molecular field approximation the two-body 
interactions in Eq.~(\ref{EQ:Maier-Saupe}) are replaced by a
one-body interaction with the scalar order parameter field:
\begin{equation}
\eta = \left<\frac{3}{2}\cos^2(\Theta)-\frac{1}{2}\right>,
\label{EQ:Maier-Saupe-OP}
\end{equation}
where $\Theta$ denotes now the angle between the segment and the average
axis of orientation. This relatively simple theory is capable of a
quantitative (albeit phenomenological)  description of experimental
data on the liquid-gel ordering transition thermodynamics.

Gruen \cite{Gruen85,Gruen80,Gruen81,Gruen81b} extended Marcelja's
model by replacing the phenomenological Maier-Saupe ordering
interaction with a local incompressibility condition in the lipid tail
region imposed through an inhomogeneous Lagrange multiplier field,
$\pi(z)$, coupled to the local microscopic segmental density
$\hat\phi(z)=\sum_{i=1}^N \delta(z-z_i)$, where $z$ is the
coordinate across the laterally homogeneous bilayer and the
sum runs over all the positions $z_i$ of the $N$ units per molecule.
Results derived
with this model show semi-quantitative agreement with results of
molecular simulations both for thermodynamics as well as for the
structural properties of bilayers. Minor modifications of this
approach allow the study of properties of cylindrical and spherical
micelles \cite{Gruen81c,Gruen85b,Gruen85c}.

Similar ideas of imposing packing constraints on anchored chain
systems have been explored by a number of researchers. Differences
between these methods are mostly technical. Dill et al.~\cite{Dill84,Dill88} and Scheutjens and co-workers \cite{Scheutjens79} use a
lattice to describe both chain conformations as well as molecular
field configurations, which can be treated with efficient transfer
matrix methods originally proposed by Kramers and Wannier for
solving the Ising model. Ben-Shaul et al.~\cite{Ben-Shaul85,Ben-Shaul95,Ben-Shaul85b,Ben-Shaul85c} enumerate a
very large set of the most probable chain configurations off-lattice
and use iterative techniques to reach self-consistency. For a more
detailed exposition of these ideas we refer the reader to excellent
reviews of these works \cite{Dill88,Ben-Shaul85}.

Recently Elliott et al.~\cite{Elliott05,Elliott06} has extended the mean-field
approach to study mixtures of saturated and unsaturated lipids and
cholesterol. It has been shown that the incompressibility (or
packing) constraint is not sufficient to provide a quantitative
description of both liquid and gel lipid membranes and the
corresponding main-chain transition. This deficiency can be
eliminated by including additional explicit orientational
interaction, similar in spirit to the original ideas of Marcelja and
Gruen. The following form of pairwise interaction potential has been
proposed:
\begin{equation}
V({\bf u},{\bf u'})\rightarrow V({\bf u\cdot c},{\bf u'\cdot
c})\approx -J f({\bf u\cdot c})f({\bf u'\cdot c}),
\end{equation}
where ${\bf u}$ and ${\bf u'}$ are the local orientations of CH$_2$
groups, ${\bf c}$ is the bilayer normal, and $J$ characterizes the coupling
strength. In the absence of head group orientational interactions,
the chain tails on average will align parallel to the bilayer
normal, which implies that $f(x)$ should be a monotonically
decreasing function of its argument. Elliott et al.~\cite{Elliott05} utilize the generic 
form
\begin{equation}
f(\cos\Theta)=\frac{2m+1}{2}(\cos^2\Theta)^m\approx m
\exp(-m\Theta^2),
\end{equation}
where $\Theta$ is the angle between the local chain orientation ${\bf
u}$ and the bilayer normal ${\bf c}$. The parameter $m$ controls the
width of the orientational interaction and the case $m=1$ is comparable
to the original Maier-Saupe interaction function used by Marcelja
\cite{Marcelja74}. Elliot et al.~\cite{Elliott05} take $m=18$.
The higher power dependence was first proposed by
Gruen \cite{Gruen81,Gruen81b}.

The experimental value of the main-chain transition temperature of a
single component DPPC system is used to fit the strength, $J$, of the
orientational interaction whereas the power, $m$, is estimated from
the average alignment of an acyl tail in the bilayer. 
The same set of parameters can be used to describe
membranes containing unsaturated dioleoylphosphatidylcholine (DOPC) lipids. This theory
successfully predicts the absence of the liquid-gel transition in DOPC
membranes due to disordering effect of the unsaturated double bond.
Binary dipalmitoyl phosphatidylcholine (DPPC)/DOPC model bilayers exhibit phase coexistence between a
saturated-lipid-rich gel phase and an unsaturated-lipid-rich liquid
crystalline phase with the transition temperature below that of the
pure saturated-lipid system. An extended model of a ternary system
containing saturated and unsaturated lipids and cholesterol has been
used to provide strong evidence for the existence of
a liquid-ordered to liquid-disordered phase transition
\cite{Elliott06}.

Despite the fact that the anchored chain models provide a detailed
description of chain-packing in both gel and liquid lipid bilayers,
their description of the interfacial polar-head region is very
rudimentary. It is either completely ignored in favor of fixing the
lipid areal density or treated essentially phenomenologically. The
inequivalence of tail, head and solvent segments effectively
makes the bilayers {\em pre-assembled} structures, as opposed to
{\em self-assembled} ones. Since this limitation does not come from
the mean-field approximation, it can be completely eliminated within
a more general self-consistent field-theoretic framework.

\subsubsection{Self-assembled membrane models}
The self-consistent theory of polymer systems, both for solutions and
melts of long-chain molecules, was formulated by Edwards
\cite{Edwards1965}. Helfand and Tagami
\cite{Helfand71,Helfand72b,Helfand72c} pioneered the use of this
approach for the study of multicomponent polymer systems with diffuse interfaces.
Since then the self-consistent field theory has been utilized for
a wide range of polymeric systems \cite{Scheutjens79,Scheutjens80,Hong81b,Hong81,Matsen94,Fredrickson-Book,SCFT2,Matsen06}, notably it has been highly successful
in studying self-assembly of block copolymers \cite{Matsen96,Shi96,matsen02b} and
copolymer-homopolymer mixtures \cite{Matsen95b,Janert97,Janert97b,Janert98,Matsen99}.

Leermakers et al.~used similar ideas to study self-assembled
bilayers composed of polymer chains with random-walk and RIS (Rotational Isomeric State)
statistics \cite{Leermakers88}. For the sake of computational
convenience, chain configurations are restricted to a tetragonal
lattice aligned with the bilayer interface. The segment-segment
dispersive interactions are modeled by a set of Flory-Huggins
parameters: $\chi_{AW}$ for tail segment/water, $\chi_{BW}$ for head
group/water and $\chi_{AB}(\approx \chi_{AW})$ tail segment/head
group interactions. The harshly repulsive interactions at short distances
are taken into
account by imposing the common incompressibility condition
$\phi_A(z)+\phi_B(z)+\phi_W(z)=1$, where $\phi_{A,B,W}(z)$ are local
volume fractions of the tail, head, and water segments, respectively. The authors use two
types of chain statistics to describe the intramolecular
conformations. The first one is a random walk on the tetragonal
lattice, which allows neighboring bonds to overlap and does not
distinguish between the {\em trans} and {\em gauche} conformations.
The second more elaborate model, based on the RIS
scheme, takes into account the small energy difference between the
local bond configurations and also forbids direct backfolding of the
chain onto itself. In both approaches the chain partition function and
density distributions can be evaluated using straightforward matrix
algebra. Nevertheless, for computational tractability only laterally
homogeneous structures are considered.

This fully self-consistent framework is capable of describing stable,
tensionless, self-assembled bilayers, provided the interaction
parameters are within certain limits. The random-chain and the
RIS-chain models both result in membranes with qualitatively similar
segment distributions and thermodynamic properties. Quantitatively, however,
this approach underestimates the experimentally measured membrane
thickness by about $50\%$, suffers from an excessive chain-disorder
and fails to predict the gel-liquid main-chain transition. Despite
these deficiencies, this simple framework is very flexible and
efficient and it has been extended to studies of bilayer vesicles,
membrane elasticity, bilayer-bilayer interactions and effects of
surfactants on bilayer structure and stability
\cite{Leermakers89}.

The absence of the main-chain transition in this model has been
traced  to the inability of the simple incompressibility constraint
to properly describe {\em anisotropic} packing of the chains. To
cure this problem, Leermakers et al.~formulated an {
anisotropic} SCF theory \cite{Leermakers88b}. The idea is based on
a generalization of Flory's and Di Marzio's treatment of rigid rod
statistics in anisotropic states and results in an effective
entropic aligning interaction between chains (cf.~Elliott et al.~\cite{Elliott05}).
The modified theory predicts a first order main-chain transition and
the existence of two gel phases: one with intercalated monolayers in
the dilute regime, and the other with almost non-overlapping
monolayers in the concentrated regime.

With straightforward modifications, this theory was used to
investigate mechanical properties of the lecithin bilayer membranes.
To calculate the bending moduli of the lecithin vesicles, Leermakers
et al.~\cite{Leermakers89} proposed to use spherically curved
lattices that can accommodate either lattice random-walks or RIS
chains. The fact that the underlying lattice is ``commensurate'' with
the segment structure of the chains, leads to strong lattice
artifacts. These numerical artifacts give rise to strong variations of
both structural and thermodynamic properties as a function of
vesicle radius. To resolve this problem, the authors propose to
consider only  vesicles with radii that are commensurate with the
underlying lattice. Despite the fact that the model then provides
a reasonable account of the stability and structure of the curved,
pure lecithin bilayers, this approach has been criticized in a later
work (see below). Bending moduli of multicomponent vesicles are
predicted to be smaller than those of pure vesicles composed of
either of the components. This prediction is in agreement with the
earlier studies on anchored chains grafted to curved interfaces
\cite{Szleifer88}.

Oversteegen et al.~\cite{Oversteegen00} studied in detail the
dependence of elastic constants on microscopic interaction
parameters. In particular, they found that the bending modulus,
$\kappa_b$, strongly depends on the strength of interactions between the
head groups and the solvent. As the strength of head group hydration
decreases, so does the bending modulus, apparently vanishing for
sufficiently weak head group/solvent interaction parameters.
Interestingly, under the same conditions the saddle splay modulus,
$\bar{\kappa}$, progressively becomes less negative and eventually adopts 
positive values. This favors formation of saddle-like structures and should
lead to the instability of a flat bilayer. The authors also carefully
investigate the spherical lattice artifacts mentioned above and
conclude that there is actually no completely consistent way to eliminate
them, despite earlier claims to the contrary. Nevertheless, the
authors argue that the lattice model may still prove to be a
valuable tool to to make qualitative predictions. The same approach
was used \cite{Kik05b} to study elastic properties of surfactant
monolayers. It was found that the monolayer spontaneous curvature
does not vanish, as in the case of the bilayers, and strongly
depends on molecular architecture. The monolayer bending modulus was
found to be essentially half of that of the corresponding bilayer,
in agreement with previous studies which used an anchored chain model
\cite{Szleifer90}.

Meijer et al.~\cite{Meijer99} used extensions of the lattice
mean-field model to study inclusions of foreign molecules in bilayer
membranes utilizing a detailed description of both rigid and flexible
molecules with, and without, electrostatic interactions. To handle the
rigid inclusions whose segment structure is not necessarily
commensurate with the underlying tetragonal lattice (used to
generate RIS hydrocarbon chain configurations), the authors must
make several approximations to preserve efficiency of the propagator
scheme. This approach is implemented in a computer program called
GOLIATH. Partitioning coefficients for a wide range of molecules
into bilayer membranes were predicted in reasonable agreement with
experimental data. Detailed theoretical information on the density
and orientational distributions of these inclusions allowed the
authors to conclude that it is possible to design isomeric molecules
that can have roughly the same partitioning coefficient, but would
differ greatly with respect to the position and orientation of the
inclusion in the membrane. This  could be potentially of use
in designing membrane-incorporated drug systems.

In a more recent work, Leermakers et al.~refined their SCF approach
by including several new features \cite{Leermakers03}. First,
the introduction of different segment volumes for CH$_2$ and CH$_3$
groups counteracted the interdigitation of hydrocarbon
chains from apposing monolayers. Second, by allowing the formation of clusters of
water molecules \cite{Suresh96b}, they were able to prevent
excessive penetration of the water molecules into the membrane's
hydrophobic region. Third, polarizability of chain segments was
included for thermodynamic consistency. The predictions of this
detailed approach compare favorably with all-atom MD simulations
by the same authors \cite{Rabinovich03} of SOPC and SDPC bilayers.
This validates the SCF parameter set, which can then be used to
investigate other membrane properties.

Whitmore et al.~extended the lattice, self-consistent, anisotropic
field theory to compressible, fully hydrated, phospholipid membranes
\cite{Whitmore98}. The authors assumed that the lipid headgroups are
strongly confined both positionally and orientationally throughout
the temperature range of interest. This approximation 
considerably simplifies the numerical solution of the set of
self-consistent equations, and emphasizes the bond density and
orientation in the hydrocarbon region of the bilayer. The assumption, however,
is in disagreement with the the rather wide interfaces observed by
Leermakers et al.~in a similar model bilayer system
\cite{Leermakers03}. Similarly, the solvent molecules (water) were
excluded from the hydrocarbon layer. To determine the microscopic
interaction parameters the authors used the main-chain transition
temperature for DPPE, its dependence on the chain length, and the
change in thickness of the hydrocarbon layer at the transition. The so-parameterized
model exhibits a first order phase transition between the highly
ordered gel phase, $L_\beta$, to the partially disordered 
liquid crystalline phase, $L_\alpha$. It also provides a reasonable account of
equilibrium changes in the bilayer thickness and hydrocarbon density
across the main chain transition. The predicted increase of the
transition temperature with the hydrocarbon chain length is in
qualitative agreement with experimental observation, but is still
weaker by a factor of $3$. The authors attribute this inaccuracy to
the neglect of orientational effects of water on the gel-to-liquid
transition.

In an alternative approach, M\"uller and Schick \cite{Muller98c} used
an off-lattice representation of the field theory, and obtained the
single-chain partition function via a partial enumeration \cite{Szleifer96} over a
large set of molecular conformations of a lipid chain with the RIS
statistics. The model takes into account an explicit solvent, which
permits the study of thermotropic, as well as lyotropic, phase transitions
between lamellar, $L_\alpha$, and inverted hexagonal, $H_{\rm II}$,
phases. In agreement with experiment, it was found that the
transition from $H_{\rm II}$ to $L_\alpha$ occurs upon increasing the
solvent content or decreasing the temperature. It was also confirmed
that an increase in the length of the hydrocarbon tail or decrease
in the head-group volume stabilizes the inverted hexagonal phase.

In a series of papers, Schick and coworkers used a flexible chain model to
study bilayer membranes. This model is known to be incapable of producing the
main-chain transition to the gel phase, $L_\beta$, due to the excessive entropy
of a flexible chain.  Nevertheless, in the liquid crystalline phase,
$L_\alpha$, it is expected to account reasonably well for the competition between
hydrophobic forces and chain stretching. In particular, Netz and Schick
\cite{NETZ96} used a standard flexible diblock copolymer model in the presence of a
solvent, modeled by a flexible hydrophilic homopolymer, to study structure and
stability of the fluid bilayer phase. It was found that under external compressive
stress the bilayer thickness decreases until the bilayer is in thermodynamic
coexistence with either the disordered or catenoid lamellar (hexagonally
perforated lamellar) phases.  This result was used to describe membrane
rupture. It was found that the relative areal expansion at the onset
of rupture is about $2\%$, which is in agreement with experiments on lipid
systems. Pores, which appear in  the hexagonally-perforated lamellar phase, formed reversibly with stress and were slightly hydrophobic.

Li and Schick \cite{Li00} used a more detailed microscopic model of a
lipid with a charged headgroup, two flexible hydrophobic tails, and
a neutral solvent with counter ions. The hydrophobic interactions
are described through the usual contact interactions. First, the
model is applied to a neutral, anhydrous, lipid and is shown to describe reasonably
well the transitions between lamellar, hexagonal and cubic phases.
With the addition of a solvent, the system exhibits a re-entrant
$H_{\rm II}\to L_\alpha \to H_{\rm II}$ transition, a peculiar feature
observed experimentally in the  dioleoylphosphatidylethanolamine (DOPE)-water system. Addition of a negative
charge on the headgroup leads to an effective increase of the
head-group volume and drives the transition from the $H_{\rm II}$ to the
$L_\alpha$ phase, also in agreement with experiments. In a second
paper, Li and Schick \cite{Li01} extend their approach to study
mixtures of cationic, anionic and neutral lipids. This model
provides a detailed description of the $pH$-tunable
$L_\alpha-H_{\rm II}$ phase transition in mixed lipid systems, which
could play an important role in liposomal drug delivery systems \cite{Hafez00}.

We used the SCF approach with the Gaussian chain model to elucidate the
evolution and free energy costs of intermediates involved in the fusion of bilayer
membranes \cite{Katsov04,Katsov06}. This study goes beyond the
previous work on laterally homogeneous bilayers and provides a
description of complicated structural rearrangements that 
occur during the fusion reaction. Detailed results of this approach
are presented in Sec~\ref{sec:FSCF}. Here we mention important
modifications necessary to study such complicated problems within
the field-theoretic framework.

In practice, the SCF theory in its common implementation is only capable
of identifying thermodynamic, locally stable
configurations of a system. Such structures can correspond to stable
configurations of a system or metastable {\em intermediate states}
encountered along the transformation of the system towards a stable
structure. In studying activated processes (e.g., rupture, fusion or
fission of membranes), however, one should also be able to describe 
properties of {\em transition states} which correspond to saddle
points of the SCF free energy functional. Unfortunately, finding a
saddle point of a functional poses a serious numerical problem. To
address this issue we propose the use of {\em reaction coordinate
constraints} \cite{Matsen99}. In some cases, using either physical intuition,
qualitative input from numerical simulations, or SCF fluctuation calculations \cite{Shi96}, one can identify the
unstable directions of a saddle point and stabilize them by
applying a suitable constraint. Such a constraint, for instance, can be a restriction on the
symmetry of the segment density distribution, or a direct
constraint on the field variables.

As an example, consider nucleation of a hole in a bilayer membrane.
First, we confine the possible solutions to be axially symmetric, and
to have a mirror symmetry with respect to the bilayer mid-plane. We
further constrain the possible segmental density field
configurations by requiring that in the bilayer mid-plane, the
interface between the hydrophilic and hydrophobic segments be
localized along a circle of some prescribed radius, which plays the
role of a reaction coordinate. This constraint can be imposed via a
Lagrange multiplier field and results in a minimal modification of
the usual set of SCF equations. In general, for an arbitrary
value of the constraint, the value of the corresponding
Lagrange multiplier field is non-zero, which means that the
corresponding configuration of the system is not a solution of the
original unconstrained problem. Therefore the details as to how the constraint is
applied matters. Nevertheless, exactly at the points where
the Lagrange multiplier field vanishes, the solutions coincide with
those of the unconstrained system, and it is an extremum, a minimum, of the free
energy along the reaction coordinate path, i.e., these points are intermediate states. 
Furthermore, an extremum which is a maximum of the free energy along the reaction coordinate path, corresponds to a saddle point of the
unconstrained problem, i.e. to the transition states of the system
along the considered reaction coordinate. This approach provides a
wealth of additional information that cannot be accessed by the
standard, non-constrained, calculations.

Yet another approach that uses the collective density degrees of
freedom as the basic degrees of freedom is the Density Functional
Theory (DFT). This formalism has its roots in the quantum many-body
theory, where it is widely used for calculating electronic structure
of many-atomic systems. Similar ideas have been applied to the
description of simple classical fluids, in particular to the studies
of vapor-liquid and liquid-solid interfaces, nucleation of  new
phases (see Ref.~\cite{Talanquer03} for an application to hole nucleation in amphiphilic bilayers), and solvation of extended solutes. One of the major advantages of
this approach is the systematic incorporation of liquid state packing effects 
on the level of segments. 
Chandler and co-workers \cite{Ch86}
have developed an approach to study properties of inhomogeneous
molecular systems, which combines molecular simulations and DFT.
Despite the fact that the DFT is formally an exact framework it is
limited by the accuracy of the approximate density functional. The
common approach to construct an approximate density functional
relies on  mean-molecular field ideas, similar to those utilized
in the SCF framework \cite{SCFT2}. In particular, the interacting many-particle
problem is reduced to that of a reference system consisting of
non-interacting particles in an effective external field chosen so
that it results in distributions equivalent to those in the
original interacting system. Not surprisingly, in complete analogy
with the SCF theory, the problem can be divided into two
coupled sub-problems. First, one determines the mean-molecular
fields produced by a given density distributions. Second, one
evaluates the thermal averages of the density operators in the
presence of these fields. The major difference between the DFT and
SCF theory approaches lies in the microscopic details utilized to obtain the
expressions for the mean-molecular fields. In particular, in the DFT
framework, the fields are customarily obtained within a linear
response theory of a bulk uniform fluid to a weak perturbation. For
example \cite{Frischknecht02,Frischknecht02b}, the mean-molecular
field $w_\alpha({\bf r})$ acting on a segment of type $\alpha$ can
be approximated by
\begin{equation}
w_\alpha({\bf r})=V_\alpha({\bf r}) - \sum_\beta \int d{\bf r'}
c_{\alpha\beta}^{\rm bulk}({\bf r-r'})\left[\rho_\beta({\bf r'}) -
\rho_\beta^{\rm bulk}\right],
\end{equation}
where $\rho_\beta({\bf r'})$ and $\rho_\beta^{\rm bulk}$ are the local density and the bulk density of site type
$\beta$, respectively, $V_\alpha({\bf r})$ denotes the ``bare'' external fields (e.g., interactions with confining surfaces) and
$c_{\alpha\beta}^{\rm bulk}({\bf r-r'})$ is the bulk direct correlation
function between segments of the types $\alpha$ and $\beta$.
The direct correlation function provides information about microscopic pair correlations in the system.
In the mesoscopic SCF approaches discussed above, these microscopic
correlations are usually treated on more phenomenological grounds and replaced by
a simple contact interaction. The additional microscopic detail of
the DFT approaches comes with the additional cost of calculating the direct correlation functions
of a molecular fluid. This can be obtained either from simulations (see Eq.~\ref{eqn:OZ}) or from
various integral equation approaches. Despite the much more elaborate
microscopic input to the DFT framework, it is {\em a priori} not clear whether
the density functional itself, obtained in the linear response approximation,
is accurate in the strongly inhomogeneous systems such as self-assembled
bilayer membranes. More sophisticated relations between fields and densities
have been utilized, e.g., to capture the equation of state of polymers in bad solvents
\cite{Muller00f,Muller03d,Patra03} or the local structure of the segment fluid \cite{Yu02b,Wu06}.

Frink et al. applied the DFT approach to study coarse-grained lipid
bilayers \cite{FRINK05} using a freely jointed chain model for the
lipid tails. The direct correlation functions of the bulk disordered
system were obtained from the polymer reference interaction site
model (P-RISM) \cite{curro87,Schweizer88,yethiraj92b,Schweizer97} in conjunction with the Percus-Yevick closure.
Additional single chain Monte Carlo simulations had to be performed
to obtain the intramolecular structure factor, which is a necessary
input to the P-RISM approach. The bilayer solutions obtained with
this DFT framework have structural properties that are in a
reasonable agreement with experimental DPPC data. Interestingly, the
model even predicts a highly ordered bilayer phase at a sufficiently
low temperature, which has an apparent resemblance to the ordered
gel phase, $L_\beta$. Nevertheless, the disorder-order transition
appears to be continuous, in contradiction with the experimentally
observed first-order main chain transition. The reason for this
discrepancy  presumably lies in the high entropy of the flexible chain
model.  A more accurate treatment would require the inclusion of chain stiffness to properly
describe the main chain transition. Furthermore, in their comparison
of this DFT approach to the MD simulations of the same model system,
Frischknecht et al.~\cite{FRISCHKNECHT05} found no ordering
transition in the latter at all. Therefore, despite the overall agreement with the
MD simulations at relatively high temperatures, it seems that the
DFT approach can become qualitatively inaccurate at low
temperatures.

\subsection{Phenomenological Hamiltonians}
\label{sec2.3}
While the coarse-grained models discussed above retain the notion of lipid
molecules, phenomenological Hamiltonians take the coarse-graining idea one step
further and describe the configuration by the collective density of hydrophilic 
and hydrophobic entities or the spatial position of interfaces.
These approaches address the universal
properties of amphiphilic systems, and are stripped down to  essential
ingredients. This permits a study of  wide parameter ranges, and a systematic
exploration of universal properties and mechanisms. The form of the Hamiltonian
is usually dictated by symmetry considerations. The few parameters of the Hamiltonian
can be established through comparison of simple properties which result from the calculation, such as interface tension or fluctuation spectra, either with experiments
or with simulation results of more detailed models. Two major
approaches can be distinguished: (i) Ginzburg-Landau type models which describe
the system via collective densities of hydrophilic and hydrophobic entities,
(ii) Helfrich Hamiltonians which conceive of the system as an assembly of 
infinitely thin interfaces characterized by their elastic
constants. We briefly discuss both approaches in turn.

\subsubsection{Ginzburg-Landau type models}
Ginzburg-Landau type models resemble the field-theoretic approach
summarized above because they describe the system via collective densities. Their
staring point is a free energy functional ${\cal H}[\phi]$ that, in the
simplest case, describes the free energy of a system with a spatially varying
distribution of hydrophilic particles. Within the Random Phase Approximation \cite{deGennesBook}
one can derive such a Hamiltonian from the field-theoretic formulation of a
microscopic model. This technique is useful for rationalizing the general form of
the Hamiltonian, but the assumption of weak segregation between hydrophilic and
hydrophobic entities inherent in the Random Phase Approximation is not valid in
biological systems. Hence the approach is often quantitatively unreliable. Typically, the
Hamiltonian is constructed via an expansion in terms of an order parameter such as the 
difference between hydrophilic and hydrophobic densities, $\phi({\bf r})$,  and its
spatial derivatives, $\nabla \phi$. Only the lowest terms 
consistent with isotropy of space are retained;

\begin{equation}
{\cal H}[\phi] = \int {\rm d}^3{\bf r}\; \left[ f(\phi) + g(\phi)(\nabla \phi)^2 + c (\nabla^2 \phi)^2 \right]
\end{equation}
The function $f(\phi)$ determines the phases of the homogeneous system. Generically, it
exhibits two or more
minima. The function $f$ can take the form of a polynomial, as in the
original Ginzburg-Landau model of phase separation, or  the form of a piecewise
parabolic function of $\phi$ \cite{Gompper90c}. The second term determines the free energy costs
of interfaces. In phases characterized by an {\em extensive} amount of interface, such as a microemulsion,  the coefficient in this term must become
negative, at least for some composition range, in order to generate these
stable interfaces. The third term with $c>0$ guarantees  that spatial variation in the order parameter will not grow indefinitely, and therefore ensures the
stability of the system.

Although simple, this form of Hamiltonian suffices to reproduce many structures observed in
lipid-water mixtures including disordered structures like microemulsions and
complex periodic phases like the gyroid morphology \cite{Gozdz96}.  Typically, the spatial
derivatives are discretized on a regular cubic lattice.  Then, the densities at
the different lattice nodes are the degrees of freedom and the derivatives give
rise to nearest and next-nearest neighbor interactions. The statistical
mechanics of the model can be studied with Monte Carlo simulations \cite{Gompper93c} which includes thermal
fluctuations. The fluctuations are important, e.g., for the formation of microemulsions. Alternatively,
one seeks spatially modulated structures that minimize the Hamiltonian, and  thereby one
ignores fluctuations \cite{Gozdz96}.

The attractive features of this type of model is the computational ease with which
fluctuations or complex spatial structures can be investigated. Rather large systems
can be studied over a wide parameter range, and often one can obtain analytic
solutions in various asymptotic regimes. The connection between the model parameters
and biological systems often remains qualitative, and we shall not discuss this
approach further but rather refer to a comprehensive review \cite{GOMPPERSCHICK}.

\subsubsection{Helfrich's curvature Hamiltonian and its numerical implementation}

Another popular description consists of describing a membrane as an infinitely thin sheet that is characterized by a small number
of elastic properties: spontaneous curvature and bending moduli. These coarse-grained parameters encode the architecture of the
constituent lipids and the way they pack in the bilayer. In its simplest form the Hamiltonian is written as  an expansion in the
invariants that can be constructed from the local principal curvatures, $C_1$ and $C_2$ \cite{Canham70,Helfrich73,Evans74}:
\begin{equation}
{\cal H}=\int {\rm d}^2A\; \left( \gamma + \frac{\kappa_b}{2} (C_1+C_2-C_0)^2+\bar \kappa C_1C_2 \right)
\label{curvature}
\end{equation}
where $C_0$ is denoted the spontaneous curvature, and the integration is extended over the entire surface.  The coefficient
$\gamma$ is the lateral membrane tension, $\kappa_b$ and $\bar \kappa$ are the bending rigidity and saddle-splay modulus.
This expansion is appropriate if the curvature is small compared to the characteristic curvature set by the inverse
bilayer thickness or additional non-linear elastic effects.
For homogeneous membranes of fixed topology, the integral over the Gaussian curvature, $C_1C_2$, is an invariant
(Gauss-Bonnet theorem), and the term proportional to $\bar \kappa$ contributes only a constant. The saddle-splay modulus
is, however, important if the membrane topology changes, e.g., pore formation, fusion or fission. Membranes are often characterized as being tensionless, i.e., by $\gamma=0.$ In this case,
one can constrain the average membrane area to a fixed value, $A_0$, and utilize a term that is proportional to the square of the deviation of the membrane area, $A$,
from it,
\begin{equation}
\Delta {\cal H}_A = \frac{\kappa_a}{2} (A-A_0)^2
\end{equation}
to describe the small area compressibility of the membrane.
Such a model is able to describe successfully the large-length scale behavior of interfaces, membranes, and vesicles and their fluctuations.
It has been the starting point for analytical calculations, such as the renormalization of elastic constants
by fluctuations of the local membrane position \cite{Peliti85}. Furthermore, the model can very efficiently be studied by computer
simulations. Much effort has been focused on exploring, for instance, the shape of vesicles and their fluctuations \cite{Kroll92,Kroll92b,Gompper92,Morse94,Morse95,Gompper95b,Kraus96,Gompper97c,Dobereiner97,Seifert97,Gompper98,Gompper00b}.

There also have been attempts to successively incorporate more local details into the description by first applying the curvature Hamiltonian to
each monolayer of a bilayer membrane and then to augment it by terms that account for the tilting of hydrocarbon tails \cite{Hamm98,Fournier98}. These techniques
have been applied to estimate the free energy of localized, highly curved membrane structures as they occur at the edge of a bilayer \cite{May00} or in morphological
changes of the bilayer structure, e.g., in transitions from the micellar  to the inverted hexagonal phase \cite{Hamm98} or in the fusion of membranes \cite{Kozlovsky02,KOZLOVSKY02_2,SIEGEL99,May02}. The description of topological changes, e.g., pore formation, is difficult, however. 

One important simplification of the general curvature Hamiltonian (\ref{curvature}) is achieved when overhangs of the membrane can be neglected. In this case 
the sheet can be described by a single-valued function, $z=h(x,y)$.
In this Monge representation the mean and Gaussian curvature, $H$ and $K$,  are given by
\begin{equation}
H= \frac{C_1+C_2}{2} = \frac{(1+h_x^2)h_{yy}+(1+h_y^2)h_{xx}-2h_xh_yh_{xy}}{2(1+h_x^2+h_y^2)^{3/2}}
\end{equation}
and
\begin{equation}
K=\frac{h_{xx}h_{yy}-h_{xy}^2}{(1+h_x^2+h_y^2)^2}
\end{equation}
where $h_x \equiv \partial h/\partial x$ denote the derivative with respect to one of the two lateral coordinates.

If the amplitude of fluctuations are small, then one can approximate $2H \approx h_{xx}+h_{yy}=\triangle h$ and Eq.~(\ref{curvature})
takes the simple form (with $C_0=0$):
\begin{eqnarray}
{\cal H} &\approx & \int {\rm d}x{\rm d}y \sqrt{1+h_x^2+h_y^2} \left( \gamma + \frac{\kappa}{2} (\triangle h)^2 \right)\\
         &\approx & \gamma A + \int {\rm d}x{\rm d}y \left(\frac{\gamma}{2} |\nabla h|^2 + \frac{\kappa}{2} (\triangle h)^2 \right)
\label{eqn:Helfrich}
\end{eqnarray}

Often this approximation is utilized to extract the interface tension and bending rigidity from computer simulations of small membrane
patches \cite{Muller96b,GOETZ99,LINDAHL00,Otter03,COOKE05b,BOEK05,Brannigan05c,Otter05}.
To this end, one determines the local position of the membrane, $h(x,y)$, from the simulation.
Utilizing the Fourier spectrum of interface fluctuations, $h({\bf k })$,
\begin{eqnarray}
h({\bf x}) &=&  \frac{1}{L^2} \sum_{\bf k} h({\bf k})  \exp[i{\bf k}\cdot{\bf x}] = \frac{1}{(2\pi)^2} \int {\rm d}^2{\bf k}\; h({\bf k}) \exp[i{\bf k}\cdot{\bf x}] \\
h({\bf k}) &=& \int {\rm d}^2{\bf x}\;  h({\bf x}) \exp[-i{\bf k}\cdot{\bf x}]
\end{eqnarray}
one diagonalizes the Hamiltonian (\ref{eqn:Helfrich})
\begin{equation}
{\cal H}[h({\bf k})] = \frac{1}{(2\pi)^2} \int {\rm d}^2{\bf k}\; \left(  \frac{\gamma}{2} k^2 + \frac{\kappa}{2} k^4 \right)|h({\bf k})|^2
                     = \frac{1}{L^2} \sum_{\bf k} \left(  \frac{\gamma}{2} k^2 + \frac{\kappa}{2} k^4 \right)|h({\bf k})|^2
\end{equation}
which shows that  the different Fourier modes of the local interface position are independent.
The statistical mechanics of the interface position can be described by the partition function
\begin{equation}
{\cal Z}_{\rm int} = \int {\cal D}h \; \exp \left[ - \frac{{\cal H}[h]}{k_BT}\right]
\end{equation}
where the functional integral $\int {\cal D}h$ sums over all local positions of the interface.
Since the Hamiltonian is the sum of independent, harmonic degrees of freedom, $h({\bf k})$, the Fourier modes
are uncorrelated and Gaussian distributed around zero. Their variance is given by the equipartition theorem:
\begin{equation}
\frac{1}{L^2}\left(\gamma k^2 + \kappa k^4 \right)\langle |h({\bf k})|^2 \rangle = k_BT
\end{equation}
On short length scales, $k>(\gamma/\kappa)^{1/2}$, the bending stiffness dominates and $|h(k)|^2 \sim k^{-4}$.
The large scale fluctuations are controlled by the surface tension, $\gamma$, and obey $|h(k)|^2 \sim k^{-2}$. This
method has successfully been employed to extract the tension and bending rigidity from simulations of interfaces in polymer blends \cite{Muller96,Werner99b,Werner99}, polymer-solvent interfaces \cite{Muller00f}
liquid crystals \cite{Akino01,Wolfsheimer06}, and lipid bilayers \cite{Muller96b,GOETZ99,LINDAHL00,Otter03,COOKE05b,BOEK05,Brannigan05c,Otter05}. To do so, one numerically determines the position, $h$, of the interface, or bilayer, in sub-columns centered
around $(x_i,y_j)$ and with lateral size $\Delta$ (block analysis \cite{Werner97,Werner99b}). The distance between the grid points, $\Delta$, is a compromise: if
the discretization is too fine, the instantaneous composition profile along a column will strongly fluctuate. Composition
fluctuations may affect the estimate of the interface position. If $\Delta$ is too large, however,
the Fourier modes with large wavevector are averaged out and the spectrum is underestimated. This spectral damping is a function
of the product of the spatial resolution $\Delta$ and the wavevector $q$. It only becomes irrelevant in the limit $\Delta q \ll 1$. Cooke and Deserno \cite{Cooke05} explicitly calculated the strength of
the spectral damping assuming that the smooth interface location is averaged inside each column. In this case, the spectrum
of the grid-averaged fluctuations $h({\bf k})_{\rm grid}$ and the original spectrum, $h({\bf k})$ are related via
\begin{equation}
|h({\bf k})_{\rm grid}|^2 = |h({\bf k})|^2 \left[ \frac{\sin (\Delta k_x/2)}{(\Delta k_x/2) } \frac{\sin (\Delta k_y/2)}{(\Delta k_x/2)}\right]^2
\end{equation}
This discretization artifact becomes important when the fluctuation spectrum of very small membrane patches is studied by atomistic
simulations or when the crossover between the tension-dominated and bending rigidity-dominated part of the spectrum occurs at rather
large wave vectors (i.e., $\gamma>0$). A possible way to deal with these discretization artifacts is to divide out the {\em a priori},
$k$-dependent spectral damping factor.

\begin{figure}
\includegraphics[scale=0.7]{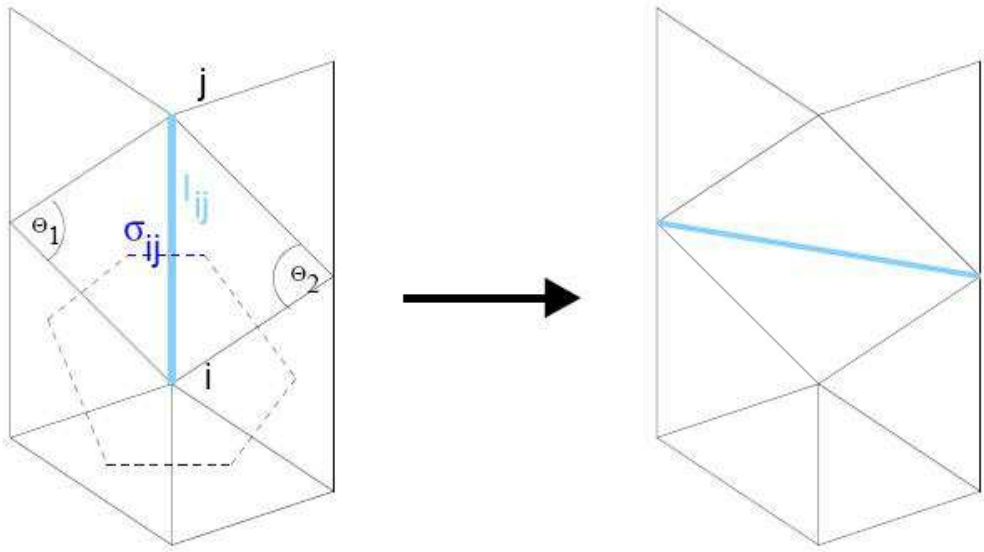}
\caption{Dynamic triangulation of fluid membranes. The bonds between vertices $i$ and $j$ has length $l_{ij}$, 
and the length of the dual tether, $\sigma_{ij}$ is indicated. The bond that connects vertices $i$ and $j$ is flipped.
In order to conserve the triangular nature of the network each vertex must initially be
connected to at least four neighbors.
\label{fig:dyntri}}
\end{figure}

The full Hamiltonian of Eq.~(\ref{curvature}) has to be used to describe large amplitude fluctuations of  microemulsions and of closed vesicles.
To this end one triangulates the surface. A typical snapshot is presented on the right hand side of Fig.~\ref{fig:coarsegraining}. 
The vertices are connected by tethers that define the internal topology of the sheet. If the
topology is fixed, the membrane will exhibit only elastic response that is characteristic, {\em inter alia}, of polymerized membranes. If one
allows for changes of the internal topology in the course of the simulation (dynamic triangulation) \cite{Ho90,Kroll92b,Boal92c}, one can mimic the in-plane fluidity.
The vertices diffuse in this fluid membrane. A common strategy consists in cutting and re-attaching tethers between the four beads of two neighboring
triangles. Such an elementary move is sketched in Fig.~\ref{fig:dyntri}.

The self-avoidance of the membrane is modeled by an excluded volume interaction between vertices at positions, ${\bf R}_i$. The strength and range of the excluded volume
interaction and the interactions along tethers can be chosen so as to avoid crossings. Let ${\bf n}_i$ denote the surface normal of a triangle, $i$.
The discretization of the mean curvature, $H={\bf n}\cdot \triangle {\bf R}$ at vertex $i$ takes the form
\begin{equation}
H_i = \frac{1}{\sigma_i}{\bf n}_i \cdot \sum_{j \in n.n.(i)} \frac{\sigma_{ij}}{l_{ij}}({\bf R}_i-{\bf R}_j)
\end{equation}
where the sum runs over all tethers $ij$ that are connected to the vertex $i$. The length of such a tether is $l_{ij}$, $\sigma_{ij}=l_{ij}(\cot \Theta_1+\cot \Theta_2)/2$ is the length
of the corresponding tether in the dual lattice which is created from the intersections of the perpendicular bisectors of the bonds.
 $\Theta_1$ and $\Theta_2$ are the angles opposite to the link $ij$ in the two triangles that border the link.
The quantity  $\sigma_i=\frac{1}{4} \sum_{j \in n.n.(i)} \sigma_{ij}l_{ij}$ represents the area of the virtual dual cell. Using this expression and the fact that the local normal {\bf n} is collinear to $\triangle {\bf R}$ in three dimensional space, one obtains
\begin{eqnarray}
{\cal H} &=& \frac{\kappa}{2} \int {\rm d}^2A\; H^2 = \frac{\kappa}{2} \int {\rm d}^2A\; (\triangle {\bf R})^2 \\
         &=&  \frac{\kappa}{2} \sum_i \frac{1}{\sigma_i} \left[\sum_{j \in n.n.(i)} \frac{\sigma_{ij}}{l_{ij}}({\bf R}_i-{\bf R}_j)\right]^2
\end{eqnarray}
for $\gamma=0$. We have omitted the integral over the Gaussian curvature which does not depend on the membrane configuration if the the topology is preserved.

If one assumes that all triangles are equilateral, $\sigma_{ij}=l_{ij}/\sqrt{3}$,  the equation above reduces to
\begin{equation}
{\cal H} = \frac{\sqrt{3}\kappa}{2} \sum_{\langle \alpha,\beta\rangle} ({\bf n}_\alpha - {\bf n}_\beta)^2 = \sqrt{3}\kappa \sum_{\langle \alpha,\beta\rangle} (1-{\bf n}_\alpha \cdot {\bf n}_\beta)
\label{eqn:simple}
\end{equation}
where ${\bf n}_\alpha$ denotes the normal vector of the triangle $\alpha$. In the general case of randomly triangulated surfaces, however, this
latter expression suffers from deficiencies. Most intriguingly, use of the expression (\ref{eqn:simple}) for a randomly triangulated sphere
recovers the expected result, while a random triangulation of the cylinder in conjunction with Eq.~(\ref{eqn:simple}) fails to yield the
result of the continuum description \cite{Gompper96}.

Generalizations of this class of models have be utilized to study freezing of vesicles \cite{Gompper97d,Gompper97} or mixed membranes \cite{Kumar98,Kumar01}. These models can also be employed in conjunction with a hydrodynamic description of the surrounding solvent \cite{Kraus96,Noguchi04,Noguchi05c,Noguchi05b}.

\section{An example of an integrated approach: fusion of membranes}
\label{sec5}

\subsection{Motivation and open questions}
One example of a collective phenomena in membranes is the fusion of two
apposing lipid bilayers.  It is a basic ingredient in a multitude
of important biological processes ranging from synaptic release, viral
infection, endo- and exocytosis,  trafficking within the cell, and  fertilization \cite{Chernomordik95b,Monck96,Zimmerberg99,JAHN02,Mayer02,Tamm03,Blumenthal03}. The fusion
process can be roughly divided into two steps \cite{Blumenthal03}: first the two membranes to be
fused are brought into close apposition. Fusion peptides embedded in the
membranes play an important role during this initial step. They ensure that only specific
membranes are brought into close proximity with one another.
One way to accomplish this is for a fusion protein, embedded in one of the membranes, to be is inserted  into  the
apposing membrane, followed by a conformational change of the protein. This active mechanism
imparts energy into the system. The specific role of  proteins in the fusion process,
their spatial arrangement and conformational changes, have attracted much
interest for they are of great importance in regulating fusion \cite{JAHN02,Chen01}. The second
step consists of the fusion event itself in which the topology changes from two
apposing bilayer of  two vesicles to a fusion pore in a now-single vesicle. There is
evidence that this second step is dictated by the amphiphilic nature of the
bilayer constituents \cite{Lee97,Zimmerberg99,Lentz00}, for fusion occurs in very different systems ranging from
tiny, highly curved, synaptic vesicles to whole cells. Moreover, 
sophisticated fusion peptides are not necessary to initiate fusion between laboratory vesicles.  The simple depletion force that arises on the addition of 
water-soluble polymer (polyethylene glycol, PEG) to a vesicle solution \cite{Cevc99,Evans02,Lentz97}, shear \cite{Zhou05d}, or sonication \cite{Zhou05d} serve equally well in inducing it. Another
important piece of evidence comes from synthetic, polymer-based, membranes made of
amphiphilic polymers \cite{Forster98,Discher99,Luo01,Discher02,Antonietti03,Ortiz06}. The behavior of these polymersomes resembles those of
the much smaller and more fragile vesicles comprised of biological lipids and,  in the absence of proteins, includes processes like fusion and fission \cite{Luo01,Zhou05d}. This all suggests that  fusion is a universal collective phenomenon.
Therefore coarse-grained models are well-suited to investigate its underlying
mechanism.

While many specific details are known about the first step of the fusion process,
much less is known about the second. Generally, fusion is
considered a ``messy'' process because of the drastic disruption of the bilayers' integrity.
The time and length scales exclude a direct experimental observation of an individual
fusion event. However even in the absence of direct information about the fusion intermediate,
much can be inferred from a systematic variation of parameters (e.g., composition of  mixed
bilayers, or tension), and careful experimental techniques (e.g., electrophysiological measurements
of membrane conductance). Some of the main experimental observations are:
\begin{enumerate}
\item Lipids that favor a large negative spontaneous curvature of a monolayer, such as DOPE, increase the fusion rate \cite{Chernomordik95b,Chernomordik96}.
\item Increasing the tension of the apposed membranes results in an increased fusion rate \cite{Monck90}.
\item During fusion, the lipids in the two apposing {\em cis} leaves mix \cite{Evans02,Lee97,Melikyan95}.
\item In some experiments, fusion is leaky. This implies that, correlated in space and time with the
      fusion event, there is a passage from the interior of the vesicle to the outside solution \cite{Shangguan96,Bonnafous00,Dunina00,Haque02,Frolov03}.
\item Some experiments report on the transient mixing of lipids between the {\em cis}, or most closely apposed, leaf
      and the {\em trans}, or farther leaf,  of the other membrane \cite{Lentz97,Evans02}. This process is much faster than the flip-flop tunneling of lipids from one
      leaf to the other in an intact membrane.
\end{enumerate}

\begin{figure}[htbp]
\epsfig{file=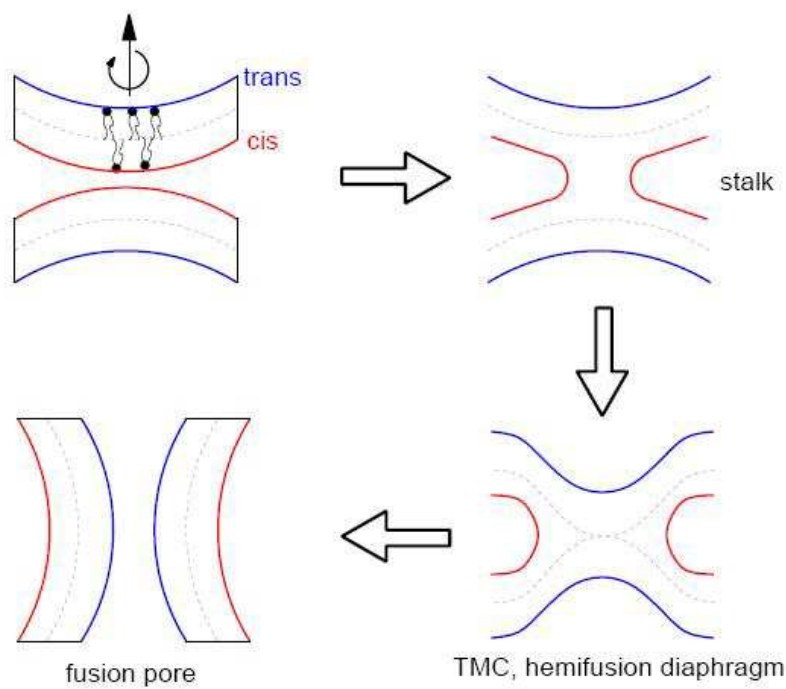,width=0.7\textwidth,clip=}
\caption{\label{fig:classical}
Sketch of the classical fusion path.
}
\end{figure}

The first three observations can be rationalized by the classical model of
membrane fusion proposed by Kozlov and Markin in 1983 \cite{Kozlov83,Markin83}. They {\em conjectured} a
fusion pathway and calculated the free energy barrier utilizing an effective
interface Hamiltonian. The monolayers of the membranes were modeled as thin
elastic sheets, and the description was augmented by a free energy penalty for
the packing difficulties that arise in the intermediate structures.

In their model the proper fusion process starts with the formation of a stalk (see Fig.~\ref{fig:classical}),
a rotationally symmetric connection between the two apposing {\em cis}
monolayers.
Once the stalk has formed, the two inner {\em cis} leaves
retract from it leaving  a transmembrane contact that consists of a small circular
membrane patch built from the two outer {\em trans} leaves. This transmembrane
contact can expand radially  to form an extended hemifusion diaphragm. The
rupture of this diaphragm creates a fusion pore. The expansion of the fusion
pore completes the process. The classical fusion model is able to rationalize
the first three observations. (1) Lipids that favor a large negative curvature of a monolayer
tend to form an inverted hexagonal phase, and this non-lamellar morphology
shares common local geometrical features with the stalk intermediate. (2) An
increase of the lateral membrane tension, or free energy per unit area makes more favorable the
decrease in membrane area which fusion brings about. (3) The
stalk and the outer rim of the hemifusion diaphragm establish a connection
between the two inner {\em cis} leaves along which lipids between the two {\em
cis} monolayers can mix by diffusion. The last two
experimental findings, however, cannot be explained. At no time during the
classical fusion scenario is there a path between the inside of either vesicle
and the outside solution. Also there is never any direct connection between the
inner and outer monolayers of a bilayer membrane.

The geometry of the stalk intermediate and the free energy penalty associated
with the packing frustration of the hydrocarbon tails is an input into the classical
model. Earlier calculations used a toroidal shape of the stalk, and  assumed that
the thickness of the monolayers remained constant as a void formed. An {\em
ad-hoc} upper limit of the free energy costs of chain packing was obtained
utilizing the macroscopic surface tension of hydrocarbons.  The total free
energy barrier associated with the formation of a stalk was estimated to be
$200\  k_BT$,  a value much too large for fusion to occur in soft matter
systems \cite{Siegel93}.  Subsequent refinements of the model utilizing a catenoid shape and
including tilt and splay of the hydrocarbon tails solved this
``energy crisis'' \cite{Kozlovsky02,May02} and yielded a significantly lower barrier of $30$ to $40\  k_BT$. The large variation of the
estimated barrier with the assumptions of the model suggests that the effective
Hamiltonian representing the monolayers as thin elastic sheets might not be able
to accurately describe the highly bent structures of the
intermediate. Models that retain the notion of amphiphilic molecules, that
incorporate the packing of the molecular conformations in non-lamellar
geometries, and that do not assume the structure of the fusion intermediates, are
better suited to provide direct insights.

\subsection{Model and techniques}
By virtue of the experimentally observed universality of the proper fusion
event, and due to the concomitant time and length scales that are unattainable
with atomistically detailed models, coarse-grained models are well-suited to
investigate the basic mechanism of membrane fusion. Ideally, such a coarse-grained
model would combine the following characteristics:
\begin{enumerate}
\item It describes the architecture of the amphiphilic molecules. The parameters of
      the model are directly related to experimentally measurable characteristics.
      The change of the molecular conformations and the associated loss of entropy
      in a non-planar environment can be calculated.
\item It can be used to  observe whether  the fusion event does, or does not, occur without prior assumptions about the pathway.
\item It can be solved with a computational technique which permits one to simulate a number of independent fusion events.
      Note
      that in experiment,  several different outcomes of an encounter between two vesicles can  be observed, e.g.,
      adhesion, complete fusion, or rupture.
\item Its various parameters, such as the lipid architecture and  thermodynamic conditions, can be explored, and the free energy of the fusion intermediates can be calculated.
\end{enumerate}
The first requirement is best met by a detailed model that mimics as
much of the local structure of the bilayer and the surrounding solvent as
possible. Such models, however, are computationally very expensive and do not
permit the systematic exploration of the fusion process or the calculation of
free energy barriers.

In our own investigation, we have chosen the bond fluctuation model
(cf.~Sec.~\ref{sec2.2}).  Coarse-grained models strike a balance between
specificity of description and  efficiency of computation. The bond fluctuation model, in which the amphiphiles are modeled as diblock copolymers on a cubic lattice and
the solvent consists of homopolymers of identical length, is certainly  one of the
coarser, and therefore more computationally efficient, models. Although the model  captures only
the universal amphiphilic characteristics of the membrane components, it provides a reasonable description
of bilayer properties (cf.~Tab.~\ref{tab:bfm}).  This efficient model allowed us
to study rather large systems that contain several thousand amphiphiles, and
to observe thirty-two fusion events for one set of parameters.  Besides computational
efficiency, the model has two additional advantages: (1) Much is known about the
properties of the model, e.g., interface tension, bilayer compressibility,
phase behavior, spectrum of interface fluctuations, etc. (2) The model can be
quantitatively compared to the standard Gaussian chain model of the SCF
theory without any adjustable parameter. We use Monte Carlo
simulations \cite{Muller03,Mueller02} to provide an unbiased insight into the fusion pathway. Once this is attained, we
perform extensive SCF calculations \cite{Katsov04,Katsov06} of the same model in order to obtain
the free energies of intermediates over a wide range of parameters.

\subsection{MC simulation}
\subsubsection{Separation of time scales}
Our Monte Carlo simulations are performed in the canonical ensemble \cite{Muller03,Mueller02}.  The
molecular conformations are updated by local segment displacements and
slithering-snake-like movements. These movements conserve the local densities
and thus lead to a diffusive behavior on large length scales.  Moreover, the
molecules cannot cross each other during their diffusive motion. In this sense
we have a slightly more realistic time evolution on local length scales than in
dissipative particle dynamics simulations \cite{SHILLCOCK02}, but Monte Carlo
simulations cannot include hydrodynamic flow, which might become important on
large length scales. We count one attempted local displacement per segment and
three slithering-snake-like attempts per molecule as four Monte Carlo steps
(MCS). This scheme relaxes the molecular conformation rather efficiently \cite{Muller95d}. The
time scale of the MC simulations can be compared to experiments by matching the
self-diffusion coefficient of the lipids in a single bilayer (see below). At any rate, we
do not expect the time sequence to differ qualitatively from that of a
simulation with a more realistic dynamics on time scales much larger than a
single Monte Carlo step. Most importantly, fusion is thought to be an activated
process. Therefore the rate of fusion is dominated by free energy barriers encountered
along the fusion pathway, while the details of the dynamics only set the absolute time
scale.

\begin{figure}[htbp]
\includegraphics[width=0.5\textwidth]{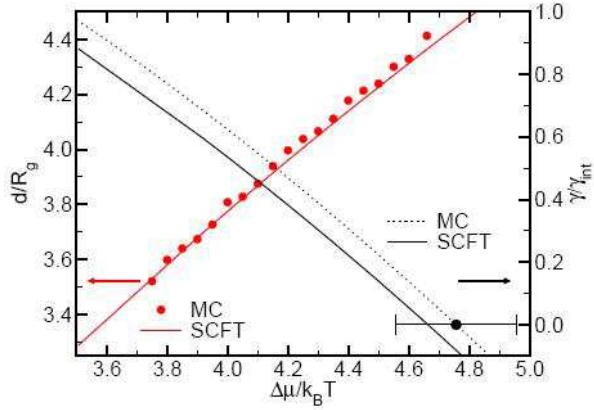}
\caption{Total thickness of the bilayer membrane, $d$, measured in units of the radius of gyration, $R_g \approx 7u$,
vs.~the exchange chemical potential $\Delta \mu$ between amphiphiles and solvent molecules. Membrane tension, $\gamma$,
as a function of exchange potential, $\Delta \mu$, is shown on the right scale.
}
\label{fig:tension}
\end{figure}

We begin our simulation by preassembling flat, tense, bilayers. The tension is
dictated by the areal density of amphiphiles, i.e., the thickness of the membrane (cf.~Fig.~\ref{fig:tension}). 
In the lattice MC simulations we cannot simulate at constant pressure or
surface tension (which is routinely used in off-lattice models \cite{Feller96,FELLER99}) because the size of the simulation box cannot be changed
continuously. The lattice also prevents us from measuring the pressure or
tension via the virial expression (cf.~Eq.~(\ref{eqn:virial})) or the anisotropy
of the virial of the forces. The thickness of a tensionless membrane, however, can be measured by
simulating a bilayer patch that spans the periodic boundary conditions of the
simulation cell in only one direction but exhibits a free edge in the other direction.
The area of the bilayer then will adjust so as to establish zero tension and the
thickness can readily be measured in the center of the bilayer patch. Such a tensionless
configuration is shown in the middle of Fig.~\ref{fig:coarsegraining}.

The dependence of the tension on the exchange potential, $\Delta \mu$, between
amphiphiles and solvent can be obtained by simulations in the grand-canonical
ensemble where the density is constant but, in addition to the MC moves that
renew the molecules' conformations, one ``mutates'' amphiphiles into solvent
molecules and vice versa \cite{Sariban87,Sariban88}. These moves change the composition of the system while the thermodynamically conjugate variable, the exchange potential $\Delta \mu$,
is controlled. 

The excess free energy per unit area, in the thermodynamic limit of infinite
area, defines the lateral membrane tension 
\begin{eqnarray}
 \gamma(T,\Delta\mu)&\equiv&\lim_{A\rightarrow\infty}
 \frac{\delta \Omega_m(T,\Delta\mu,A)}{ A }, \qquad \mbox{with}\nonumber\\
  \delta\Omega_m(T,\Delta\mu,{\cal A})&\equiv&
  \Omega_m(T,\Delta\mu,V,{\cal A})- \Omega_0(T,\Delta\mu,V)
\end{eqnarray}
where $A$ denotes the area of the membrane, and $\Omega_m$ and $\Omega_0$ the grand canonical free energy of the system with and without membrane, respectively. In an incompressible system the
tension $\gamma$ can be related to the temperature and chemical potential by
means of the Gibbs-Duhem equation

\begin{equation} 
{\rm d}\gamma(T,\Delta\mu)=-\delta s\ {\rm d}T-\delta \sigma_a {\rm d}(\Delta\mu), 
\end{equation}
where $\delta s$ is the excess entropy per unit area, and $\delta\sigma_a$ is
the excess number of amphiphilic molecules per unit area. This relation can be
exploited to integrate the thickness of the bilayer $d=\sigma_a N/\rho$ with
respect to the exchange potential $\Delta \mu$ at constant temperature and
calculate the change of the membrane tension \cite{Muller96b}. Using the
tensionless state as staring point we obtain the relation between the membrane
thickness, $d$ and the tension, $\gamma$ as shown in Fig.~\ref{fig:tension}.
 
\begin{figure}[htb]
\includegraphics[width=0.6\textwidth]{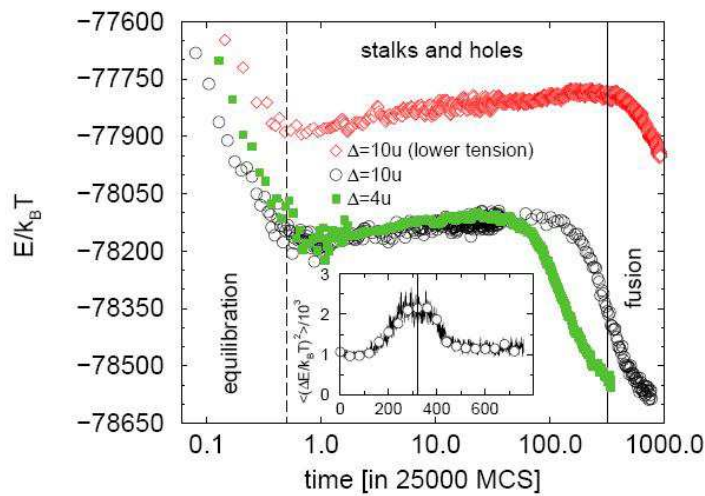}
\caption{
Evolution of internal energy in fusion simulations. The curves correspond to
different initial bilayer separations $\Delta=4u$ (squares) and $\Delta=10u$
(circles) and bilayer tensions, $\gamma/\gamma_{\rm int}=0.75$ and $0.375$,
respectively.  To reduce fluctuations, the data are averaged over all 32
configurations at equal time and additionally over small time windows. The
large negative value of the energy mirrors the attractive interactions in the
solvent.  The inset shows the sample-to-sample energy fluctuations as a
function of time (for the system with high tension and large membrane
separation).  Large fluctuations identify the onset of fusion.  }
\label{fig:energy} 
\end{figure}

After the bilayer has assembled without defects in a thin film geometry, we
stack two bilayers on top of each other with a distance $\Delta$ between them.
The system of two apposing bilayers is embedded into a three dimensional
simulation cell with periodic boundary conditions in all three directions.
Configuration bias MC moves \cite{Siepmann90,Laso92,siepmann92} are utilized to
fill the remaining volume with homopolymer solvent.  This starting
configuration of flat bilayers mimics the approach of two vesicles whose radii
of curvature are much larger than the patch of membrane needed for fusion. Then
we let the system evolve and observe the fusion process in the simulation.

Because fusion is a kinetic, irreversible process, the starting condition
might have a pronounced influence on the outcome. In Fig.~\ref{fig:energy} we present
the internal energy of the system as a function of MC steps on a logarithmic
scale. Immediately after the assembly of the two bilayer system, one observes a
rapid decrease of the energy.  This corresponds to the relaxation of the bilayer
structure after the insertion of the solvent. After $\tau_R \approx 25 000$
MCS ($\approx$ 9ns, see below), however, the local internal structure of the bilayer has equilibrated and
the energy starts to rise very slowly. This increase is compatible with a
logarithmic growth law and stems from thermally excited interface fluctuations
that increase the area of the hydrophilic-hydrophobic interface. At the end of
this stage, stalks and holes are formed in the simulations. (Here and in the
following, we denote a pore across a single bilayer as a ``hole" in contrast to the
``fusion pore" that spans both bilayers.) Finally, around
$\tau_F \approx 8 \cdot 10^6$ MCS, the energy suddenly drops indicating the
formation of a fusion pore that reduces the total bilayer area and, therefore, the
stored tension. The important point is that there is a difference of two orders of magnitudes
between the relaxation time of the local structure of an individual bilayer, and
the time scale on which fusion occurs. Due to this separation of time scales
between initial relaxation and fusion, we do not expect the preparation of the
starting configuration to affect the irreversible fusion process. Similarly we do not expect
our results to depend on our particular choice of relaxation moves, as other
choices would also lead to relaxation of the bilayers which takes place on a
much shorter time scale than does fusion.

The inset of Fig.~\ref{fig:energy} shows the fluctuations in the energy, i.e.,
the fluctuations between the thirty-two different runs at equal time.  Strong
fluctuations indicate energy differences between the independent runs. The peak
at around $\tau_F \approx 8 \cdot 10^6$ MCS (for the system with high tension
and large separation) indicates that some systems have already formed a fusion
pore, and therefore have a lower energy, while other systems have only stalks
and holes, and therefore have a higher energy.  The width of the peak provides
an estimate for the spread of the time during which a fusion pore appears.
Specifically, for the system with the lower tension and large initial
separation, at time $1.26 \cdot 10^7$ MCS we have observed the formation of
stalks in 23 systems while 9 runs have already formed a fusion pore. At time
$2.52 \cdot 10^7$ MCS, 20 runs have successfully completed fusion, in 3
configurations only stalks have formed and in 9 runs stalks and holes have
appeared but the fusion has not been successfully completed.
The figure shows data for two different bilayer tensions and two
different distances between the bilayers. Increasing the bilayer tension, and
reducing the bilayer distances accelerates the fusion process. Moreover, the
slower the fusion process, the clearer is the transition state we observe.
Fusion occurs around $8 \cdot 10^6$ MCS and $2 \cdot 10^7$ MCS for the systems
at large separation and with high and low tension respectively.

The lateral diffusion constant of a lipid in a single bilayer is $D \approx
10^{-4} u^2/$MCS.  If we identify the thickness $d_c=21u$ of hydrophobic core
in the tensionless state with $3$nm and utilize a typical value for the
self-diffusion coefficient of lipids in bilayer membranes, $D \approx 6 \cdot
10^6$nm$^2/$s, we estimate that one MCS corresponds to 0.36 ps. Thus, the time
scale on which we observe fusion is 3$\mu$s and 7$\mu$s for the tense and less
tense bilayers, respectively. This estimate is about an order of magnitude
larger than what is observed in the DPD simulations of Shillcock and Lipowsky
\cite{SHILLCOCK05} and about 2 orders of magnitude slower than the fusion process
in the Lennard-Jones-type model of Smeijers et al \cite{Smeijers06}. In the
latter study, however, the time scale was not estimated from the comparison of the
lipid self-diffusion coefficient in a bilayer but directly from the time of the
MD simulations without further rescaling which is likely to underestimate the duration
of the fusion process \cite{Hilbers06}.

Unless noted otherwise, we will discuss in the following the data for the
larger tension and larger initial bilayer separation. The data for the lower
tension are very similar.

An important property of the bilayer which we prepare is its tension, or free
energy per unit area. Thermodynamically, a tense bilayer is only metastable,
and  will eventually rupture. Hole formation reduces the bilayer's area and
lead to a stable, tensionless state. Hole formation is an activated process
and, indeed, we do observe the rupture of isolated tense bilayers in very long
simulation runs. Since vesicles have to be stable for long times in order to
fulfill their carrier and enclosing function, it is reasonable to require that
the time scale of hole formation and rupture, $\tau_H$, of an isolated membrane
be much larger than the times scale of fusion of two apposed bilayers. 
Yet in order to undergo fusion, just such long-lived holes must occur at some
point along the fusion pathway. It would seem that vesicles could either be
stable, or they could undergo fusion, but not both. The question that
immediately arises is how membranes actually manage to exhibit these two
conflicting properties. We return to this question in Sec.~\ref{sec7}.

\begin{figure}[htbp]
\begin{minipage}{0.55\textwidth}
\epsfig{file=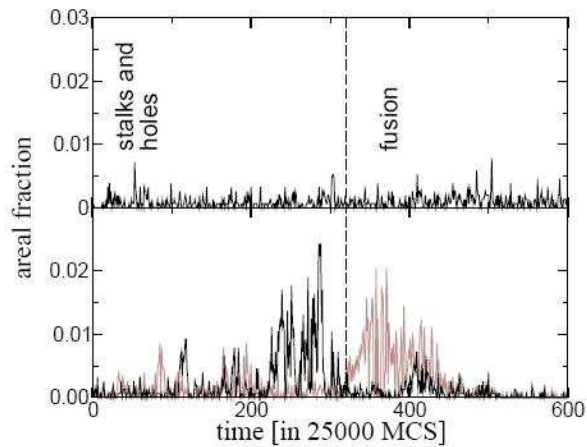,width=0.9\textwidth,clip=}
\end{minipage}
\begin{minipage}{0.45\textwidth}
\caption{Area of holes vs.~time in the system of two apposed bilayers
(gray for one bilayer and black for the other on the bottom panel) and in an isolated bilayer
(top panel). From Ref.~\protect\cite{Muller03}}
\label{fig:holes_2vs1}
\end{minipage}
\end{figure}

In Fig.~\ref{fig:holes_2vs1} we present time traces of the areal fraction of holes in
an isolated tense bilayer and in the two apposed bilayer system. In the
isolated bilayer, small holes form and close but the bilayer remains stable on
the time scale of the fusion process. In the two apposed bilayer system,
however, larger holes form more readily and their occurrence is correlated with the fusion event.

This additional MC data further supports the observation that in our coarse-grained model, there is
a clear separation of time scales, $\tau_R \ll \tau_F \ll \tau_H$, between the
local relaxation time of bilayers, $\tau_R$, the time scale, $\tau_F$, to
nucleate a fusion pore, and the lifetime of an isolated tense bilayer, $\tau_H$.

\subsubsection{Observed fusion pathways}
\begin{figure}[htbp]
\begin{minipage}{0.55\textwidth}
\epsfig{file=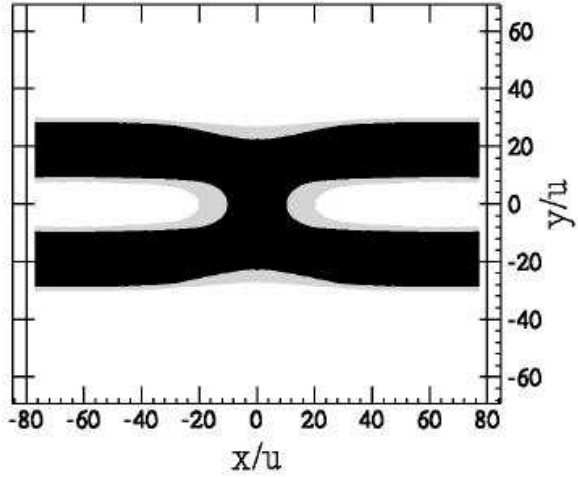,width=0.9\textwidth,clip=}
\end{minipage}
\begin{minipage}{0.45\textwidth}
\caption{Density distribution of segments in the stalk, averaged over all 32
simulation runs. At each point only the majority component is shown: solvent as
white, hydrophobic and hydrophilic segments of amphiphiles as black and gray
respectively. From Ref.~\protect\cite{Muller03}} \label{fig:stalk_profile}
\end{minipage}
\end{figure}

\begin{figure}[htbp]
\mbox{
\includegraphics[width=0.9\textwidth]{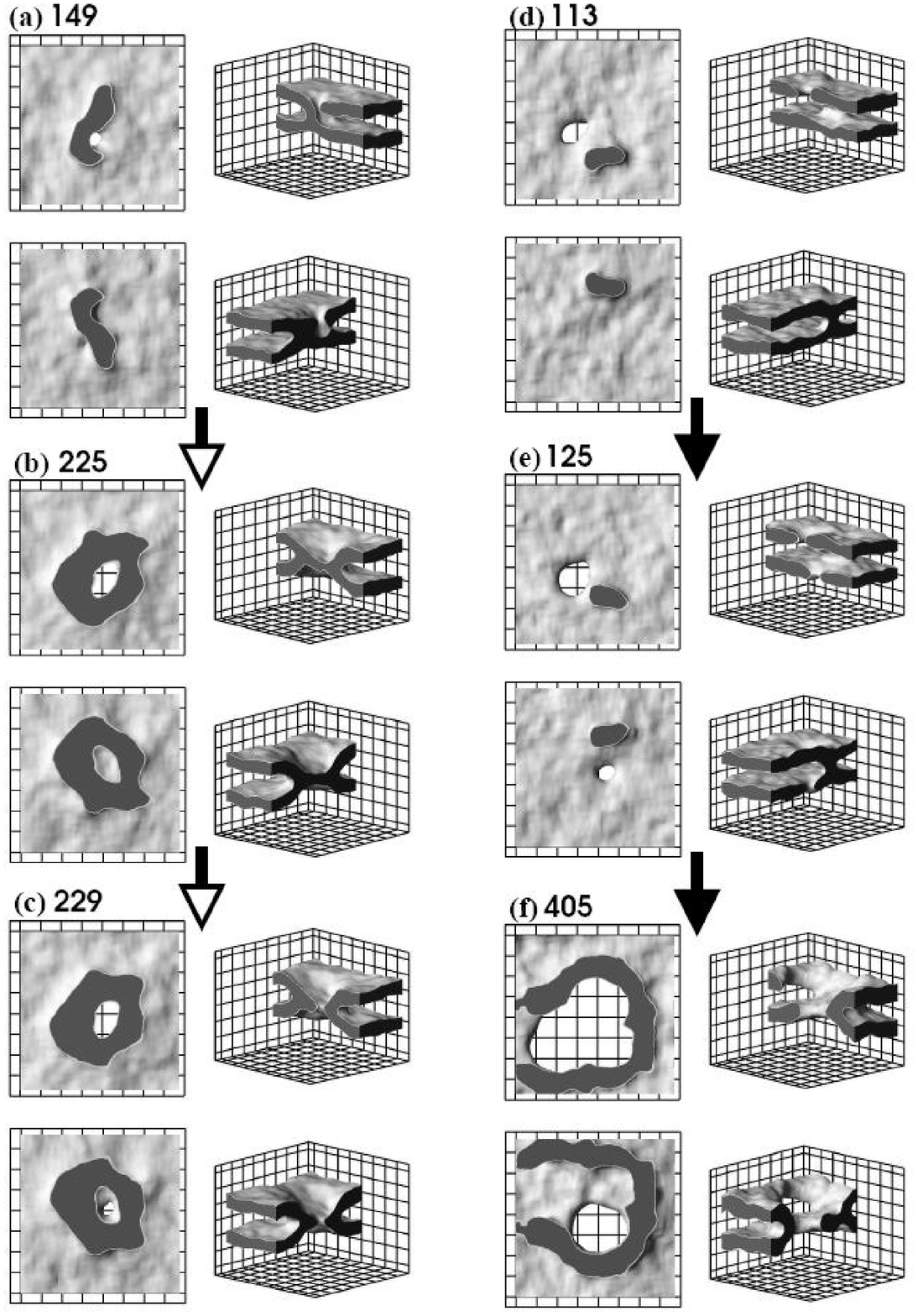}
}
\caption{Two observed pathways of fusion process. The snapshots are taken from
two representative simulation runs.  Each configuration is numbered by the time
(in multiples of $25,000$ MCS) at which it was observed.
For discussion of the mechanism see text. From Ref.~\protect\cite{Muller03b}} 
\label{fig:mechanism2}
\end{figure}

During the initial stage of simulations, interface fluctuations of the initially
flat bilayers are thermally excited and the bilayers collide with one another.
Sometimes these collisions give rise to small local interconnections. For the
most part, these contacts are fleeting.  Occasionally we observe sufficient
rearrangement of the amphiphiles in each bilayer to form a configuration, the
stalk, which connects the two bilayers. Once such a stalk has formed, it is
rather stable on the time scale $\tau_F$.  Density profiles of the hydrophilic
and hydrophobic parts of the amphiphiles in the presence of the stalk, 
obtained by averaging over configurations, are shown in
Fig.~\ref{fig:stalk_profile}. The dimples in the membranes at each end of the
stalk axis are notable.  What can barely be seen is a slight thinning of each
bilayer a short distance from the axis of the stalk.

Under the specific conditions of our simulations we believe that stalks are only
metastable. First, the observed stalks are isolated and their occurrence goes
along with a rise of the internal energy.  Second, occasionally we observe that
stalks disappear without proceeding further to a fusion pore. This behavior
indicates that the stalk represents a local (metastable) minimum along the fusion pathway.  
Once a stalk has formed, we observe that it elongates asymmetrically, and moves 
around in a worm-like manner. Evidently its elongation does not cost a great deal 
of free energy.

\begin{figure}[htbp]
\epsfig{file=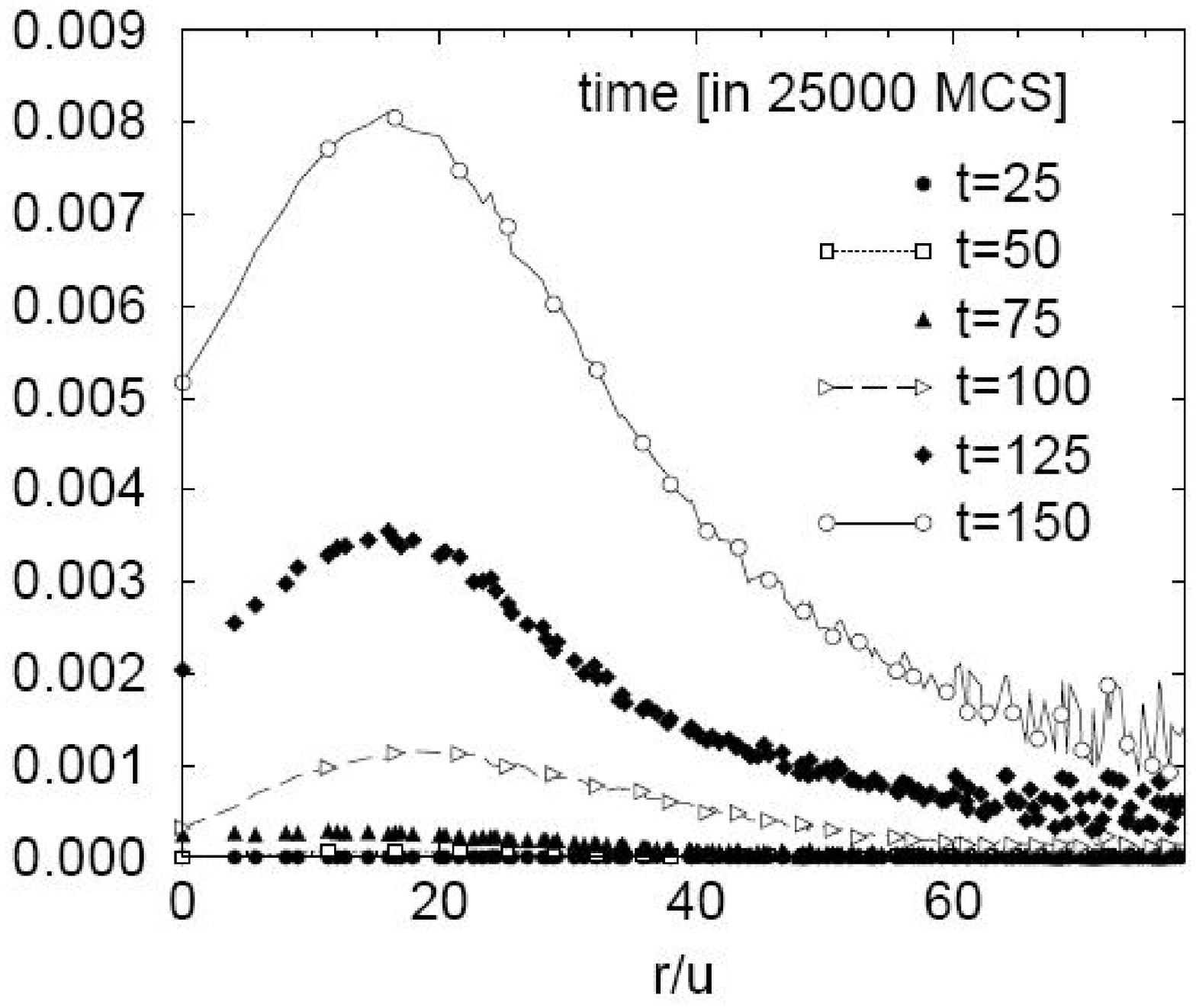,width=0.7\textwidth,clip=}
\caption{\label{fig:corr}
The hole--stalk correlation function, $g(r)$, at early times.
From Ref.~\cite{Muller03}.
}
\end{figure}

After stalks are formed, the rate of formation of holes in either of the
two bilayers increases. Stalk and hole formation are
not only correlated in time but also in space: Holes form close to the
stalks. This can be quantified by the hole-stalk correlation function:

\begin{equation}
g(r)\equiv \frac{\sum_{r_s,r_h}\delta(|{\bf r}_s-{\bf r}_h|-r) P_{sh}({\bf r}_s,{\bf r}_h)} {\sum_{r_s,r_h}\delta(|{\bf r}_s-{\bf r}_h|-r)}
\end{equation}
where $P_{sh}({\bf r}_s,{\bf r}_h)$ is the joint probability that the lateral
position ${\bf r}_s$ is part of a stalk and ${\bf r}_h$ is part of a hole, and
$\delta(r)$ is the Dirac delta function. The value of $g(r)$ at large distances
$r$ is proportional to the product of the areal fraction of holes and stalks.
This correlation function is shown in Fig.~\ref{fig:corr}.  The
scale of $g(r)$ increases with time indicating the simultaneous formation of
stalks and holes. The figure shows that the correlation peaks at a distance of
about 2/3 bilayer thickness, and rapidly decays at larger distances.

Now that a stalk connecting the two bilayers has appeared as well as a hole in
one of the bilayers, fusion proceeds along one of two closely related paths.
These are depicted in Fig.~\ref{fig:mechanism2}. The snapshots are taken from two
representative simulation run.  Configurations are numbered by the time (in
multiples of 25 000 MCS) at which it was observed.  Three-dimensional plots
of the local density are shown -- the hydrophobic core is depicted as dark gray, the
hydrophilic-hydrophobic interface (defined as a surface on which densities of
hydrophilic and hydrophobic segments are equal) is light gray. Hydrophilic
segments are not shown for clarity. Each configuration is shown
from four different viewpoints.  Top- and bottom- left sub-panels have been
generated by cutting the system along the middle plane parallel to the plane of
the bilayer. The top and bottom halves are viewed in the positive (up) and 
negative (down) direction correspondingly. In these panels one sees the cross-section
of connections between bilayers (hydrophobic core of stalks) as dark regions. Holes in the individual
bilayers appear as white regions in one of the panels. 
Top- and bottom- right sub-panels are side views with cuts
made by planes perpendicular to the bilayer. The grid spacing is $20u\approx
d_c$ (see Tab.~\ref{tab:bfm}).

\begin{enumerate}
\item A hole appears in one bilayer and the stalk completely surrounds it
rather rapidly.  The resulting configuration looks very much like a hemifusion
diaphragm which has been suggested by many authors as an intermediate stage in
fusion \cite{Markin83,Chernomordik85,Siegel93}.  However, this diaphragm is
quite different from the hemifusion diaphragm of the classical scenario that
consists of two {\em trans} monolayers of the fusing membranes.  In contrast, the
diaphragm we observe in our simulations is made of one of the pre-existing
bilayers; i.e., it is comprised of {\em cis and trans} leaves of the bilayer
that did not form a hole.  The appearance of a pore in this diaphragm and its
expansion completes the formation of the fusion pore.
\item A hole appears in one bilayer and, before the stalk completely surrounds it,
a second hole appears in the other bilayer. The stalk continues to surround them both,
and align them both during this process. Eventually, the stalk completely encircles
both holes and a complete and tight fusion pore is formed. This path is slower than
the previous one and the holes tend to expand more during this process.
\end{enumerate}

Once the fusion pore has formed, it expands because it reduces the bilayer area
and thereby relieves the tension. Note that in the canonical ensemble, the
total tensionless area is fixed from the beginning and the growth of the fusion
pore  ends when the pore reaches its optimum size determined by the finite size
of our simulation cell. Very similar finite-size related effects have been studied for the
hole formation in canonical simulations of isolated bilayers \cite{Tolpekina04}.

\begin{figure}[htbp]
\begin{minipage}{0.55\textwidth}
\epsfig{file=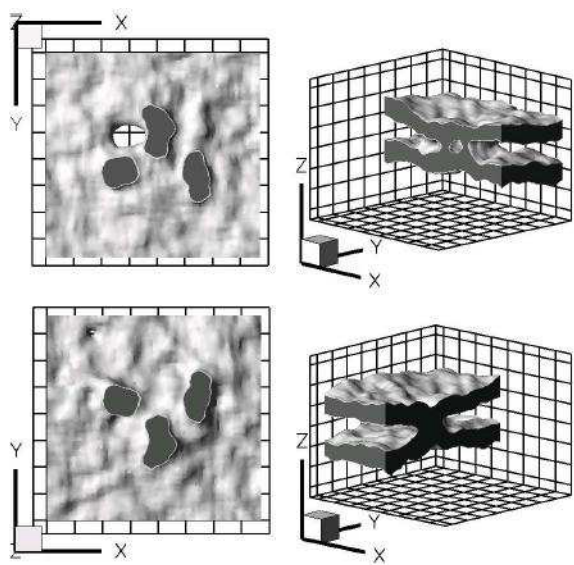,width=0.9\textwidth,clip=}
\end{minipage}
\begin{minipage}{0.45\textwidth}
\caption{\label{fig:multiple}
Snapshot from a simulation with small membrane tension and large bilayer distance (cf.~Fig.~\ref{fig:phase}) showing multiple stalks and a hole.
The configurations are depicted as in Fig.~\ref{fig:mechanism2}.
}
\end{minipage}
\end{figure}

Occasionally, more than one stalk forms in the simulated bilayer patch.
An example of configurations with multiple stalks is shown in
Fig.~\ref{fig:multiple}. The interactions between stalks have been considered
by Lukatsky and Frenkel within an effective interface model \cite{Lukatsky04}.
They argue that membrane-mediated interactions lead to an attraction of stalks, and 
that the collective clustering of stalks, in turn, aids  the fusion process.
This is compatible with the snapshots presented in Fig.~\ref{fig:multiple}
where there is apparently a higher probably of finding two stalks close to each
other.

\begin{figure}[htbp]
\begin{minipage}{8.5cm}
\epsfig{file=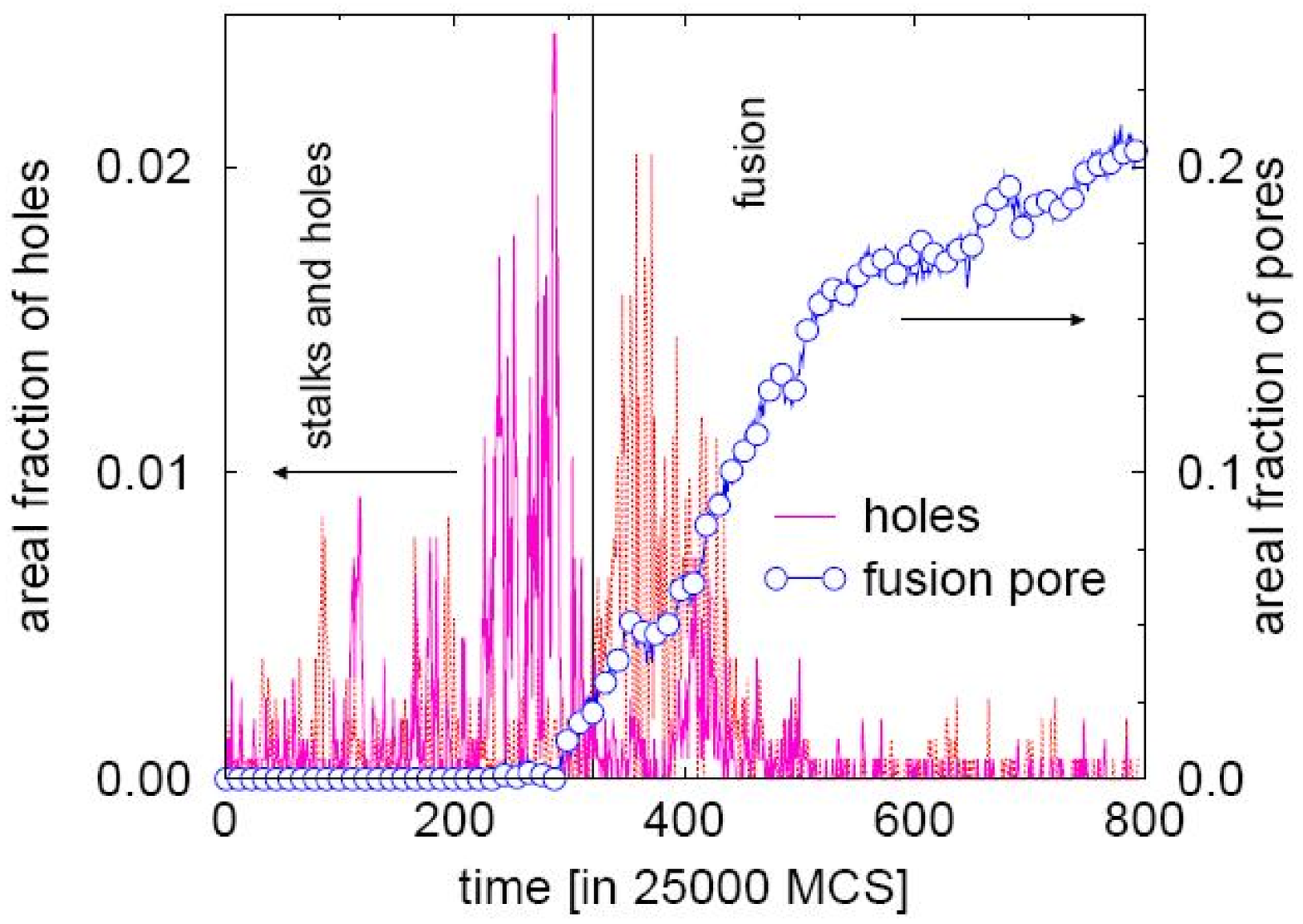,width=\textwidth,clip=}
\end{minipage}
\hfill
\begin{minipage}{6cm}
\epsfig{file=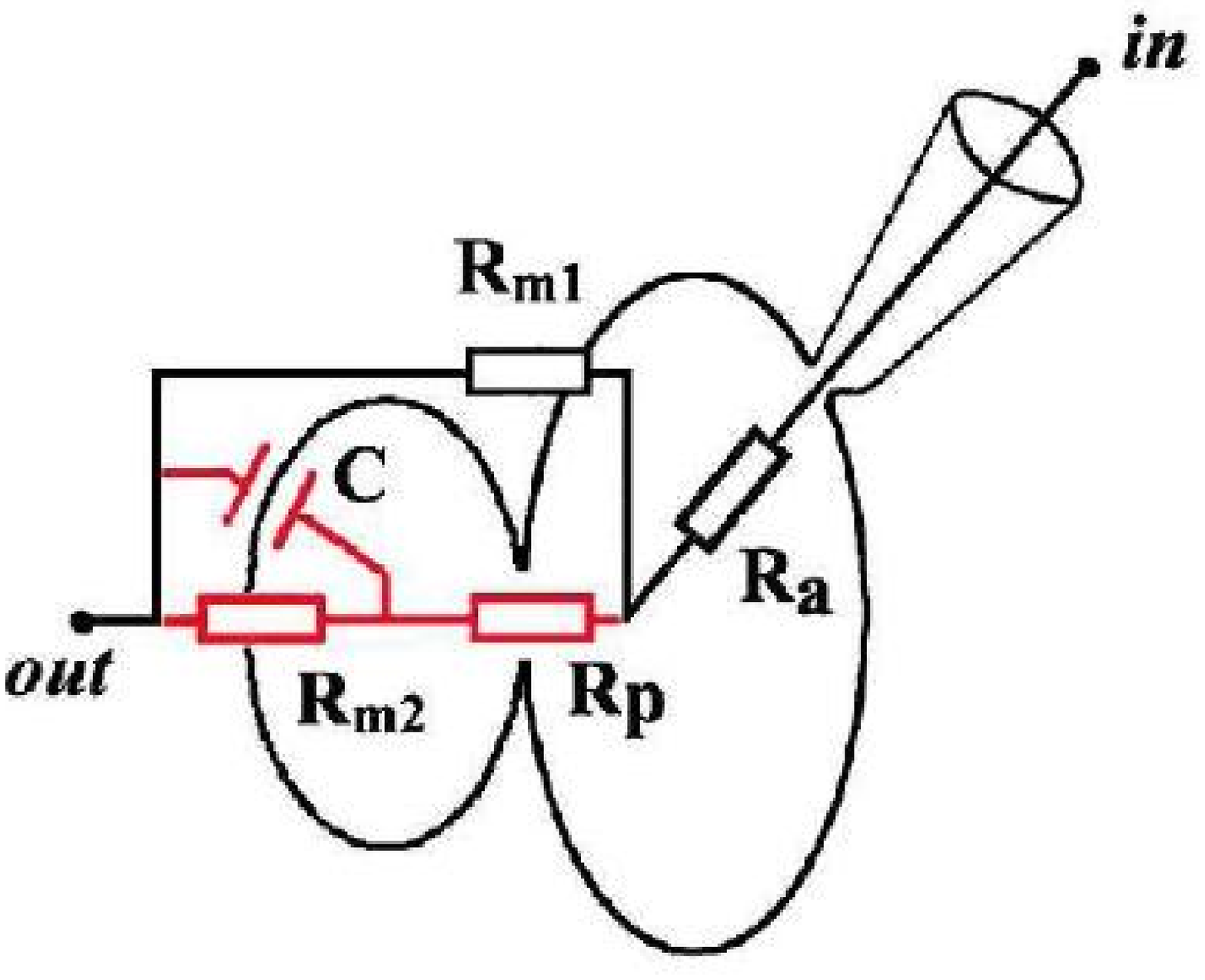,width=0.9\textwidth,clip=}
\epsfig{file=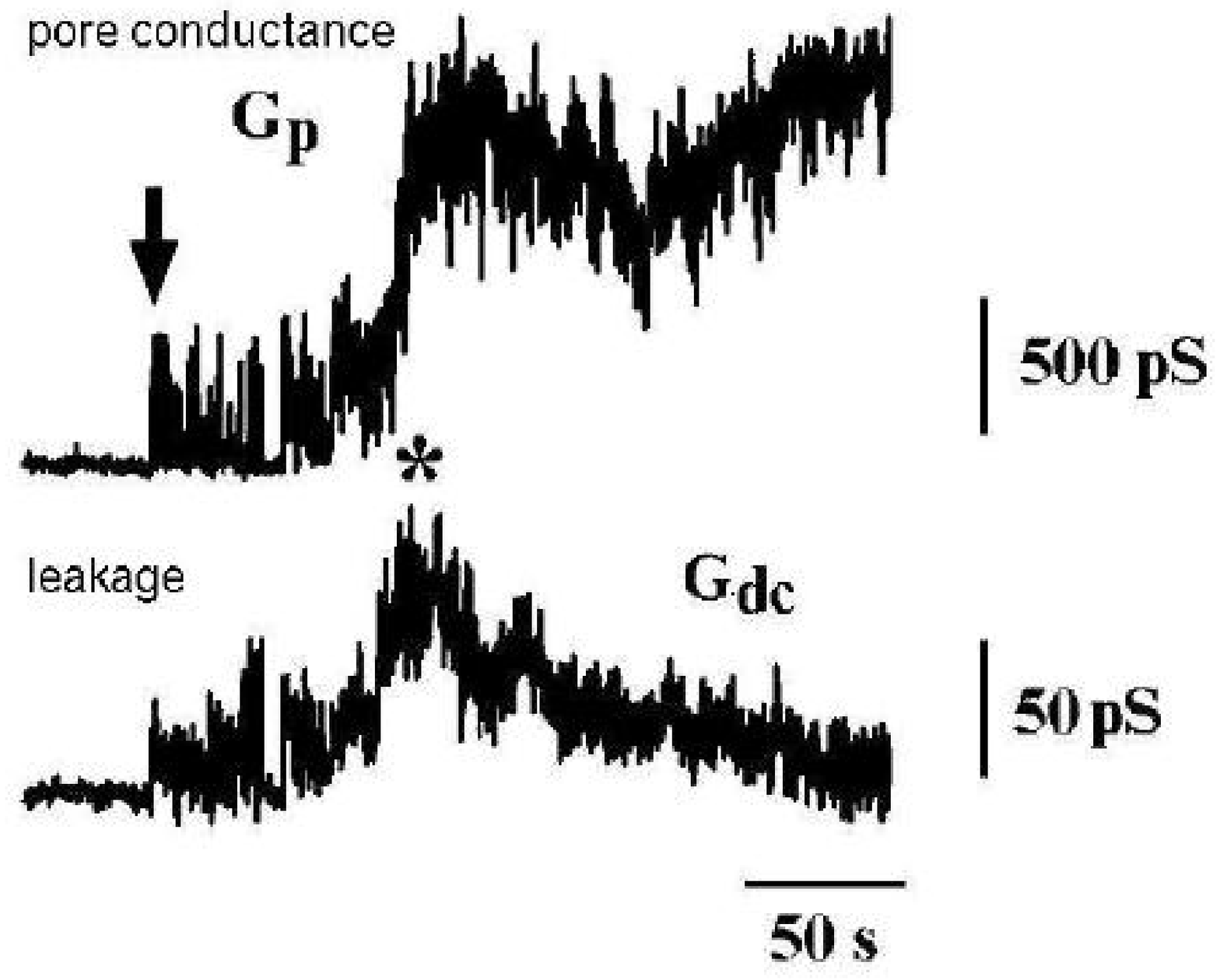,width=\textwidth,clip=}
\end{minipage}
\caption{\label{fig:exp}
(left) Time dependence of the area of holes (in single bilayers) and the fusion
pore in one (out of 32) representative simulation run. Note the different
scales for hole and pore areas.  From Ref.~\cite{Muller03} (right)
Electrophysiological experiment on influenza hemagglutinin-mediated fusion of
HAb2 and red blood cells.  upper panel: equivalent electrical circuit.  lower
panel: Fusion experiment showing leakage temporarily correlated with fusion.
The pore conductance $G_p=1/R_p$ marks the opening (arrow), flickering and
growth of the fusion pore. $G_{dc}$ is the conductance of the HAb2 cell when
the fusion pore is closed and is the sum of the conductances of the HAb2 and the red
blood cell when the fusion pore is open.  It is a measure for the leakiness of
the fusion event. Ten out of twelve experiments showed leakage.  From
Ref.~\cite{Frolov03}
} 
\end{figure}

The observed pathway of fusion  differs markedly from the classical hypothesis
in the following important aspects: First, the fusion intermediates we observe
break the axial symmetry which has been assumed in previous calculations.
Second, holes in the individual bilayers play an important role in the fusion
mechanism.  On the one hand, holes give rise to some degree of transient mixing
of the amphiphiles in the {\em trans and cis} leaves of the same membrane that is correlated
with the fusion event. Lipids can switch between the two leaves by diffusion
along the rim of these holes. On the other hand, the formation of holes in the
individual membranes implies a transient content leakage that is correlated in space and time with the fusion event.
Such leakage has been
observed in recent experiments \cite{Dunina00,Frolov03}.  In Fig.~\ref{fig:exp} we present the areal
fraction of holes in the individual membranes and the size of the fusion pore
from one simulation run. One clearly observes that hole formation precedes the
fusion pore. The extent of leakiness that can be observed in experiments
depends on the substance that passes through the holes of the vesicles and the
fraction of the rim of a pore that is sealed by the stalk. If the stalk is very
elongated and covers a substantial fraction of the incipient hole then leakage
will be very small. In the same figure, we present electrophysiological experiments
by Frolov and co-workers \cite{Dunina00,Frolov03} that show the conductance between two fusing
vesicles and the conductance between the individual vesicles and the
solution. The former quantifies the size of the fusion pore, while the latter
indicates the area of holes in the individual membranes. Consistent with the
Monte Carlo simulations there is no connectivity between the inside of the vesicle and the outside solution
before the fusion event, i.e., the vesicles are tight. Just before a current
between the vesicles is observed, however, the experiments reveal a substantial leakage
current.  Although the simulation and experiments
deal with quite different systems, they both observe leakage in contrast to
the classical hemifusion hypothesis. This exemplifies the insights into the mechanisms of collective phenomena that one can gain from simulations
of coarse-grained  membrane models.

\subsubsection{Comparison to other coarse-grained models}
While this qualitative agreement with experiment is very encouraging, the
coarse-grained simulation model differs in many aspects from experiment.
In Sec.~\ref{sec2.2} we  demonstrated that our model can
reproduce many properties of bilayer membranes, and in the previous section we
have argued the the time scale of fusion is clearly separated from the
relaxation time of local bilayer properties. Therefore we do not expect that
including details of the lipid architecture or details of the diffusive dynamics of the MC simulations will
qualitatively alter our conclusions.

Nevertheless to gauge the universality of the observation of this specific
coarse-grained model, it is very important to relate the findings to results of
alternative coarse-grained models. Since the various models include different
details, they emphasize different aspects of the fusion process. The
simulation studies of fusion differ both in the geometry of the initial state --
two apposed planar bilayers, or a planar bilayer in contact with a vesicle or
two small vesicles -- as well as in the representation of the amphiphiles and
the solvent.

\begin{figure}[htbp]
\begin{minipage}{0.55\textwidth}
\epsfig{file=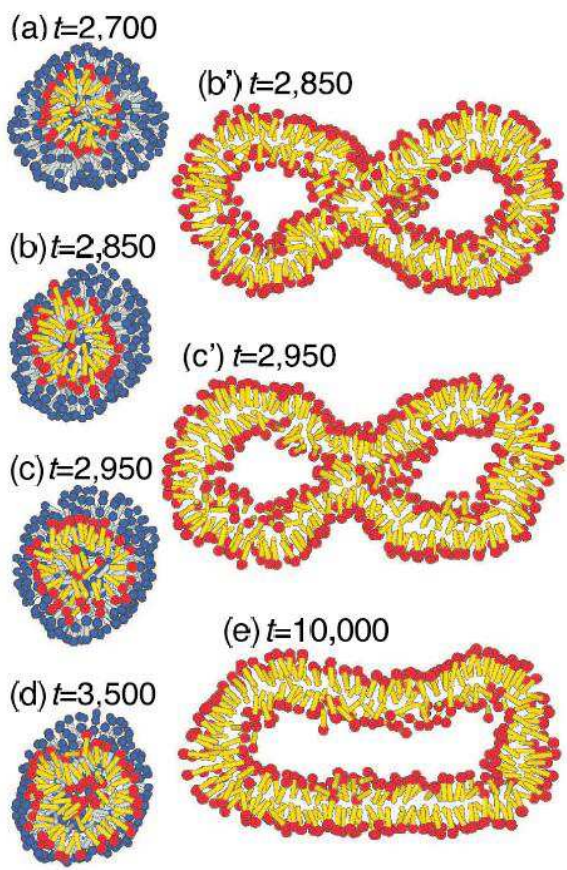,width=0.9\textwidth,clip=}
\end{minipage}
\begin{minipage}{0.45\textwidth}
\caption{\label{fig:Noguchi}
Sequential snapshots of the fusion of two vesicles within the solvent-free model of Noguchi and Takasu \cite{NOGUCHI01_2}.
Hydrophilic and hydrophobic segments are represented by spheres and cylinders, respectively. Panel (a)
presents a snapshot of the initial stalk between the two apposed vesicles. Panels (b), (c), (d)
show cuts perpendicular to the plane of contact while the images (b'), (e), (f), and (c') depict cuts
parallel to the plane of contact where the elongation of the stalk is clearly visible.
From Ref.~\cite{NOGUCHI01}.
}
\end{minipage}
\end{figure}

Chanturiya et al.~\cite{Chanturiya02} investigated the role of tension on fusion within a
two-dimensional model of a lipid bilayer. This limitation excluded the possibility of observing complex
fusion intermediates.

Noguchi and Takasu \cite{NOGUCHI01} studied the fusion of two small bilayer vesicles using a
solvent-free model \cite{NOGUCHI01_2} (see Sec.~\ref{sec2.2}). The amphiphiles were modeled as a rigid rod comprising one
hydrophilic and two hydrophobic beads.  Density dependent potentials were
tuned to bring about the self-assembly in the absence of solvent. One
important difference to the fusion of two planar bilayer membranes consists in
the rather small contact zone of the two vesicles and the high curvature of the
bilayer membranes.  They observed two distinct fusion mechanisms: One mechanism
resembled the classical hemifusion mechanism starting from the formation of a
stalk and subsequent transmembrane contact.  The transmembrane contact, however, did not
significantly expand but remained confined to the rather small contact area
of the two vesicles. A pore in the transmembrane contact completed the fusion.
The other mechanism they saw was the stalk-hole mechanism which we had observed. Their work was  
completely independent of ours, and was carried out at about the same time. In particular,
they observed the elongation of the stalk along the edge of the contact zone
between the two vesicles. These observations are visualized in Fig.~\ref{fig:Noguchi}.

The self-assembly of vesicles from a disordered solution and their fusion has been studied 
using MESODYN simulations \cite{Altevogt99,Vlimmeren99} of diblock copolymers \cite{Sevink05}. The model is similar to the one used in 
Sec.~\ref{sec:FSCF} but the use of a very coarse discretization of
the molecular architecture permits the study of large system sizes in three dimensions. Interestingly, these 
calculations also observe the formation of stalks between two apposed vesicles and the formation 
of holes in the vicinity of stalks in both bilayers (stalk-hole mechanism 2). The subsequent 
encircling of both holes by the stalk completes the fusion process.

\begin{figure}[htbp]
\begin{minipage}{0.55\textwidth}
\epsfig{file=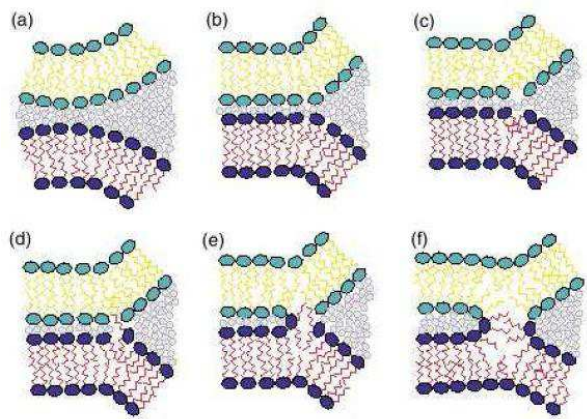,width=0.9\textwidth,clip=}
\end{minipage}
\begin{minipage}{0.45\textwidth}
\caption{\label{fig:stevens}
Sketch of stalk formation due to association of splayed lipids suggested by
coarse-grained simulations. (a) Two vesicles are brought together and (b) a
flat contact forms where, at the edge, the area per lipid in the outer leaflet
increases as the membrane is strained. (c) Lipids tilt at the contact and (d)
the hydrophobic tails of some amphiphiles begin to splay.  (e) Splayed
molecules then associate by their tails to form a new hydrophobic core, (f)
which expands as the tails extend to form a classical stalk-like structure.
From Ref.~\cite{stevens03}.  }
\end{minipage}
\end{figure}

Stevens, Hoh and Woolf \cite{stevens03} also studied the fusion of two small
vesicles that had been pushed together via an external force. The double-tailed
amphiphiles were described by a bead-spring model of eleven particles, and the
solvent as consisting of single particles.  Their simulation showed a single
pathway to fusion,  the stalk-hole mechanism. As in Noguchi and Takasu's second
mechanism, they found a highly asymmetric expansion of the stalk along the edge
of the contact zone.  Most notably, the simulations provided a detailed
description of stalk formation. Since the hydrocarbon tails were modeled as
semi-flexible chains, they were found to tilt at the rim of the contact zone
between the vesicles. The model also afforded the possibility of double-tailed
lipids bridging between the {\em cis}-layers of the two apposed vesicles.  The
authors argue that this effect is important for nucleating the initial stalk.
The details of the stalk formation inferred from the simulations are sketched
in Fig.~\ref{fig:stevens}.

\begin{figure}[htbp]
\epsfig{file=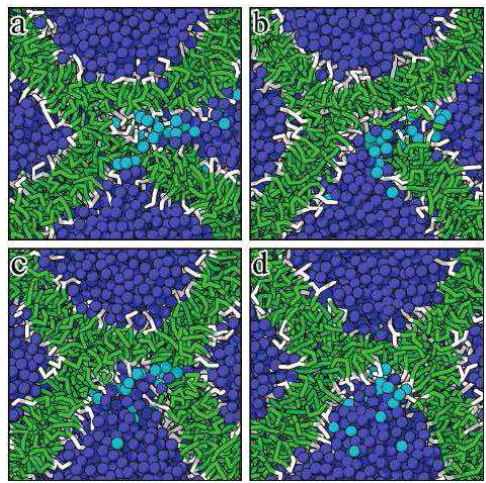,width=0.4\textwidth,clip=}
\caption{\label{fig:smeijers}
Fusion of two tensionless vesicles observed in a coarse-grained bead-spring
model \cite{Marvoort05}.  Shown is a cross-section along the vesicle-vesicle
axis. Solvent is shown as dark blue spheres, and the solvent particles that
enter the bottom vesicle highlighted light blue. (a) Some solvent enters the
vesicle and initializes a hole in the lower vesicle (26.5 ns). (b) The hole is
evident (26.8 ns). (c and d) As the stalk encircles the hole and the last
solvent particles enter the vesicle, a hemifusion diaphragm is formed by the
two monolayers of the upper vesicle (27.0 ns and 27.2 ns). 
From Ref.~\cite{Smeijers06}
}
\end{figure}

Utilizing a similar coarse-grained bead-spring model \cite{Marvoort05}, Smeijers et al
\cite{Smeijers06} also provide a detailed description of stalk formation and
emphasize the role of fluctuations that give rise to microscopic
hydrophobic patches of the bilayers. In contrast to the simulations of Stevens
et al \cite{stevens03}, however, stalks do not necessarily form at the edge of the
contact zone. One important difference between Smeijers' simulations and others
is the absence of tension in the vesicles that have been prepared by
spontaneous self-assembly from a disordered solution. For the small vesicles
considered in the simulations, the curvature suffices to induce fusion.  For
two planar bilayers they observed the formation of stalks but no fusion \cite{Hilbers06}. In these
simulations the stalk-hole mechanism was observed (see Fig.~\ref{fig:smeijers}) significantly more often
than the radial expansion of the stalk.

\begin{figure}[htbp]
\begin{minipage}{0.4\textwidth}
\epsfig{file=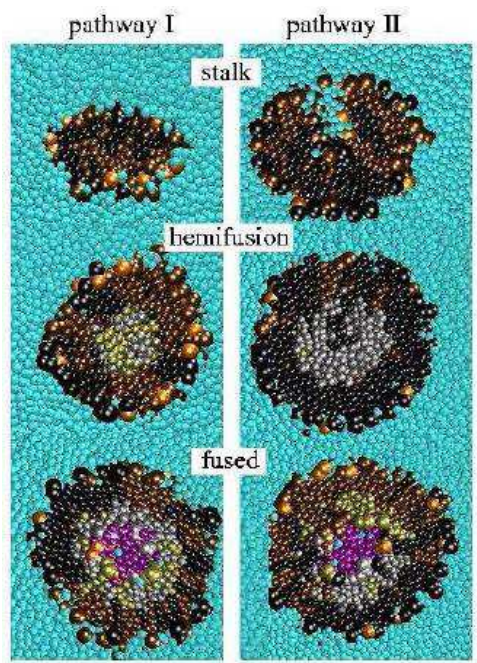,width=\textwidth,clip=}
\end{minipage}
\begin{minipage}{0.55\textwidth}
\caption{\label{fig:Marrink}
Comparison of the transition from stalk to opening of the fusion pore in both
pathways observed for the fusion of mixed DPPC/DPPE vesicles. Slabs
perpendicular to the fusion axis are shown, cutting through the stalk or
hemifusion diaphragm. Lipid headgroups are represented by large spheres.
Different colors distinguish between lipids in the inner (yellow/silver) and
outer monolayer (brown/black) and between the two vesicles (brown and yellow
vs.~black and silver). Orange spheres denote the ethanolamine site of PE, blue
spheres denote exterior water, purple spheres interior water. Note the
differences in stalk structure (bent in pathway II) and in the composition of
the HD (mixed in pathway I, almost entirely from a single vesicle in pathway
II). From Ref.~\cite{Marrink03} (suppl. information).  }
\end{minipage}
\end{figure}

Even more details of the lipid architecture are incorporated in the
systematically coarse-grained representation of Marrink and Mark
\cite{MARRINK04} who studied fusion of two very small vesicles
\cite{Marrink03}. This model corroborates the tilting of the amphiphilic tails
in the stalk. Marrink and Mark observe two pathways: the classical hemifusion
mechanism (pathway I) and the stalk-hole mechanism, with an elongated stalk
(pathway II). Importantly, transient pores in one of the bilayers were observed
in the vicinity of the stalk in the second mechanism. The two observed pathways
are depicted in Fig.~\ref{fig:Marrink}.

The simulation model of Shillcock and Lipowsky takes a step towards an even more
coarse-grained approach utilizing a DPD-model \cite{SHILLCOCK02}.  This allows
them to study a large number of fusion events between a tense planar bilayer
and a tense vesicle \cite{SHILLCOCK05}. The larger degree of coarse-graining
also allows them to systematically vary the tension of the two membranes. While they
do not discuss the detailed mechanism of the fusion, they find that successful
fusion is limited to a narrow range of rather high tensions.  If the tension is
too high, however, the membranes rupture instead of fusing.  If the tension is
below some threshold, the two apposed membranes do not fuse but rather adhere
on the time scale of the simulations. This strong tendency to adhere has not
been reported in other simulation models. Thus it does not appear to be a
universal characteristic, and it remains unclear which specific property of the
model is responsible for this feature.

Utilizing a DPD-model, Li and Liu investigated the structure of the fusion
intermediates \cite{LI05}.  They found an asymmetric elongation of stalks
similar to our observation. For rather symmetric amphiphiles, however, the
elongated stalks expanded into an axially symmetric transmembrane contact,
while for more asymmetric lipids, holes formed in the individual membranes in
four out of five simulation runs.

Taken together, the different coarse-grained simulation studies provide a
consistent picture of the microscopic details of membrane fusion. In
particular, the stalk-hole mechanism appears to be a viable alternative to the
classical hemifusion mechanism. Although the different lattice, bead-spring, and soft DPD models differ substantially in
their microscopic interactions, and in the fusion geometry,  planar bilayer or
highly curved vesicle, they all observe the non-axially symmetric elongation of
the stalk and transient holes in the individual membranes in the vicinity of a
stalk. This is also compatible with the experimental observation of transient
leakage in some experiments.

It is difficult to determine from the above results the conditions under which
the classical fusion mechanism or the stalk-hole  mechanism is the favored one. 
Field-theoretic methods, however,
are well suited to explore model parameters and to answer this, and other, questions.

\subsection{SCF calculations}
\label{sec:FSCF}
One advantage of our coarse-grained model is the fact that it can be mapped
onto the standard Gaussian chain model that is routinely utilized in SCF
calculations \cite{matsen02b,Matsen06}. The length scale, $R_e$, and the incompatibility $\chi N$, can be extracted from the simulations. Further, simulation results of the
bond fluctuation model have quantitatively been compared to SCF calculations
for many properties \cite{Muller95b,Schmid95,Muller97,Muller98b,Muller99,Werner99b,SCFT2}.  
The degree of the quantitative agreement without any adjustable parameter can be 
gauged from Fig.~\ref{fig:tension}.

The SCF theory approach allows us to calculate the free energy of axially symmetric
intermediates utilizing the radius of the structures as a reaction coordinate.
First, we examine the dependence of the intermediates along the classical hemifusion path
as a function of the bilayer tension and lipid architecture. Then, we proceed
to estimate the free energy of the alternative fusion intermediates observed in the
simulations by patching together radially symmetric structures. These SCF
calculations go beyond the phenomenological approaches that utilize effective interface Hamiltonians in two crucial aspects:
(1) They retain the notion of amphiphilic molecules and calculate the changes
of the molecular conformations in the complex, spatially inhomogeneous
environment. No assumptions about the chain packing or stretching have to be
made. (2) Only the radius (reaction coordinate) and the topology of the intermediate is dictated. The
detailed geometry of the intermediate structure is optimized as to minimize the
free energy that comprises contributions both from the interface between
hydrophobic and hydrophilic components, and from the loss of configurational entropy
due to changes of the chain conformations.

\subsubsection{Axially symmetric configurations along the classical fusion pathway}
\begin{figure}[htbp]
\epsfig{file=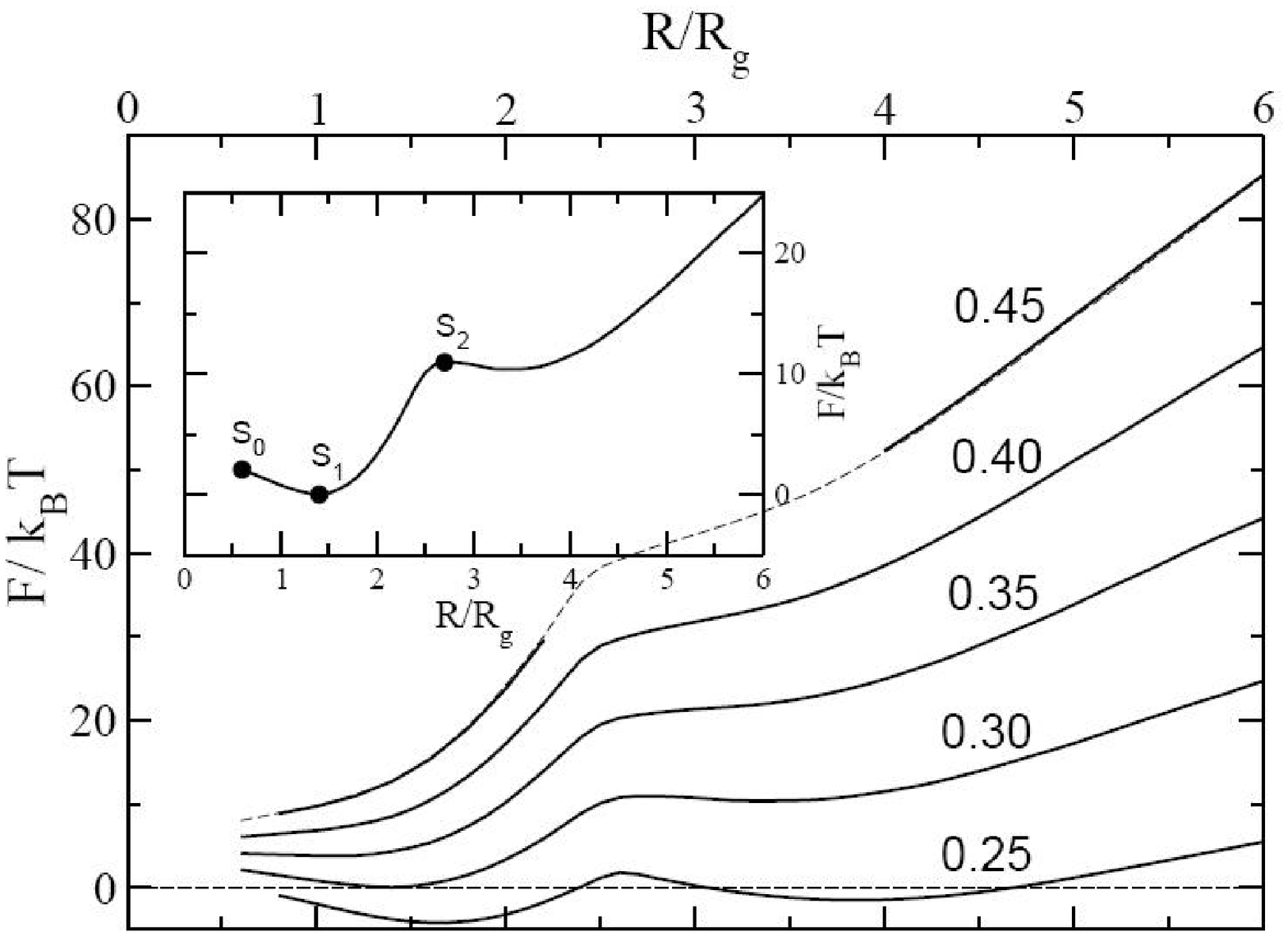,width=0.6\textwidth,clip=}
\\[4mm]
\hspace*{0.65cm}
\epsfig{file=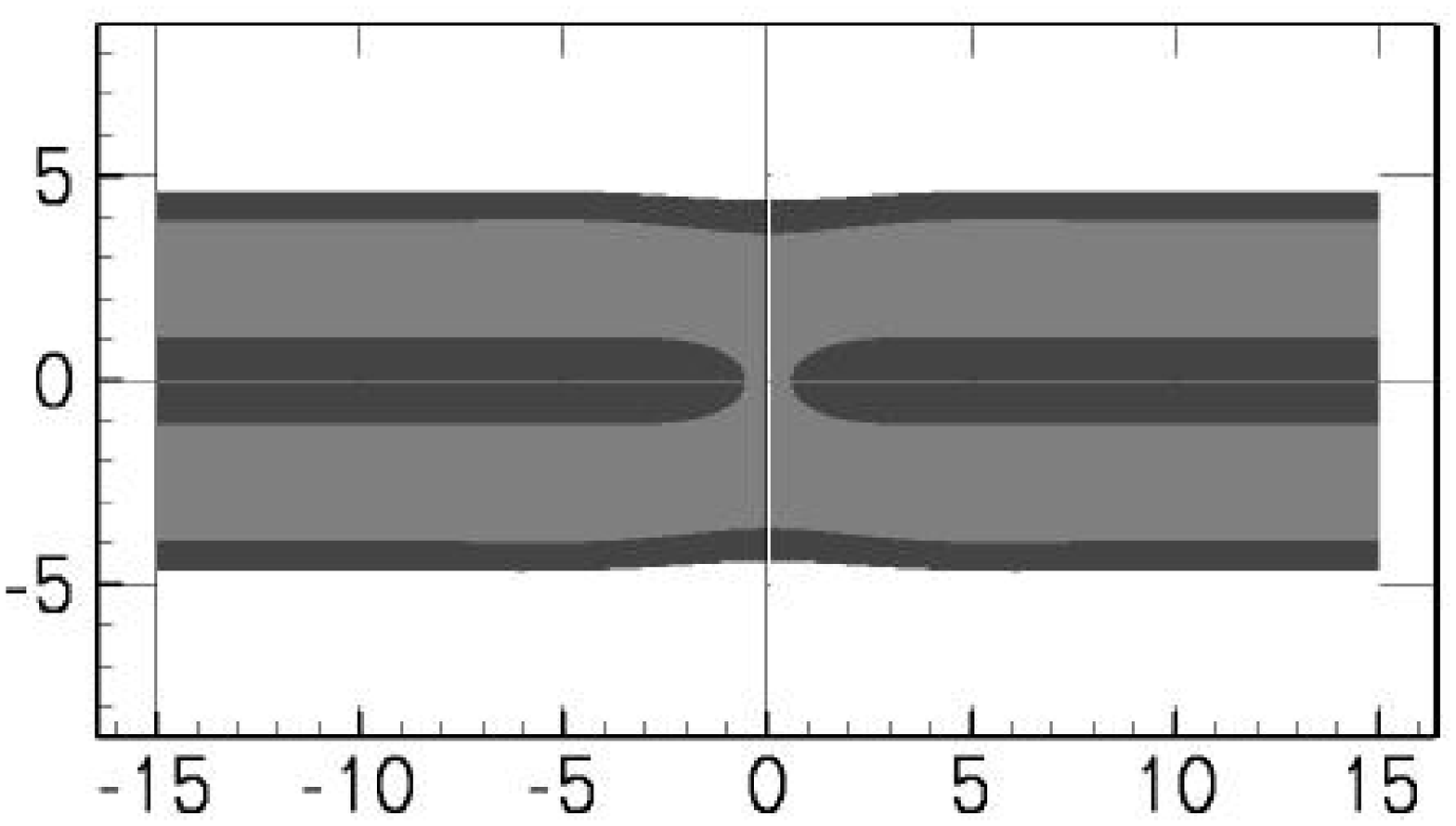,width=0.18\textwidth,clip=}
\epsfig{file=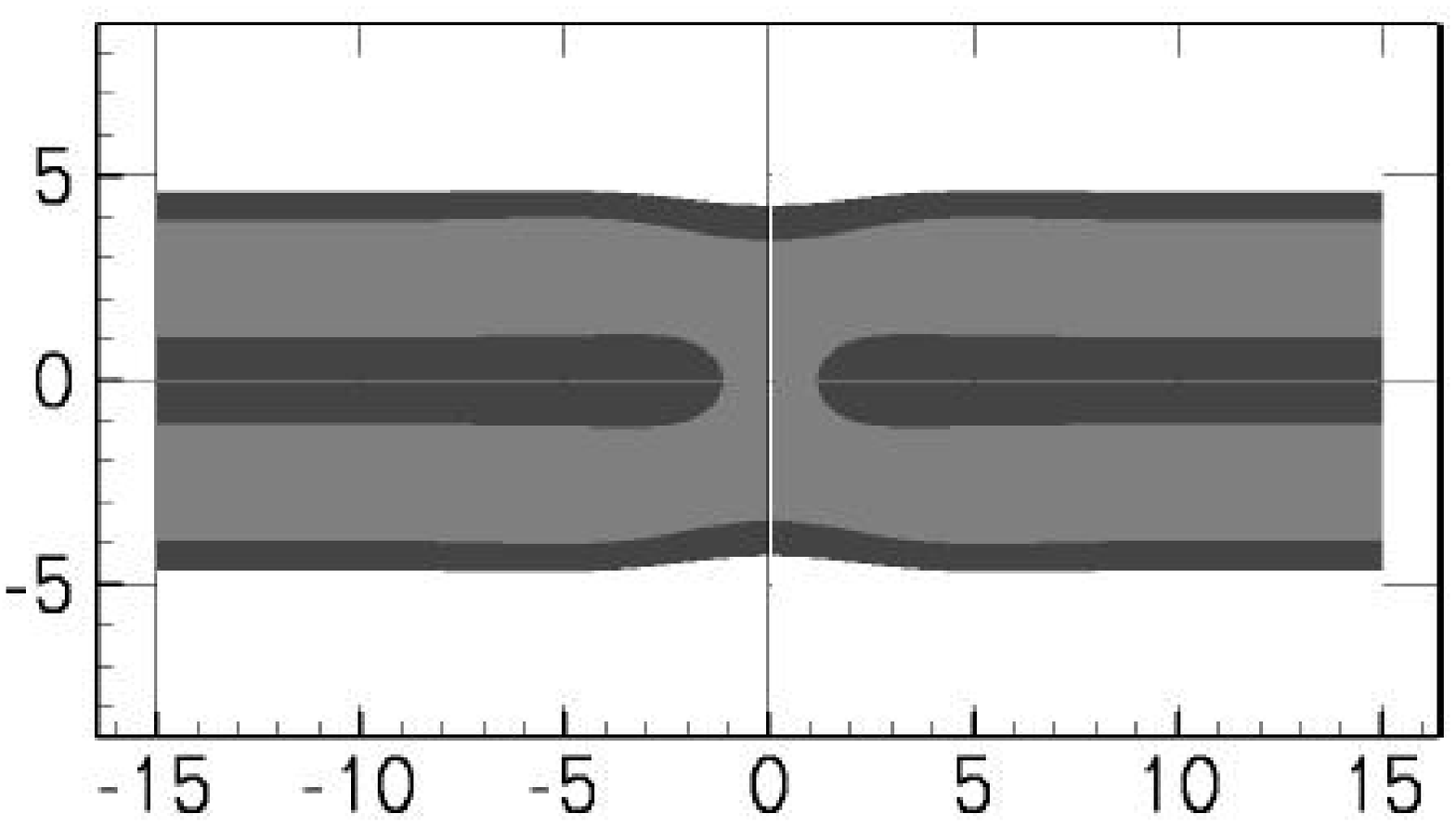,width=0.18\textwidth,clip=}
\epsfig{file=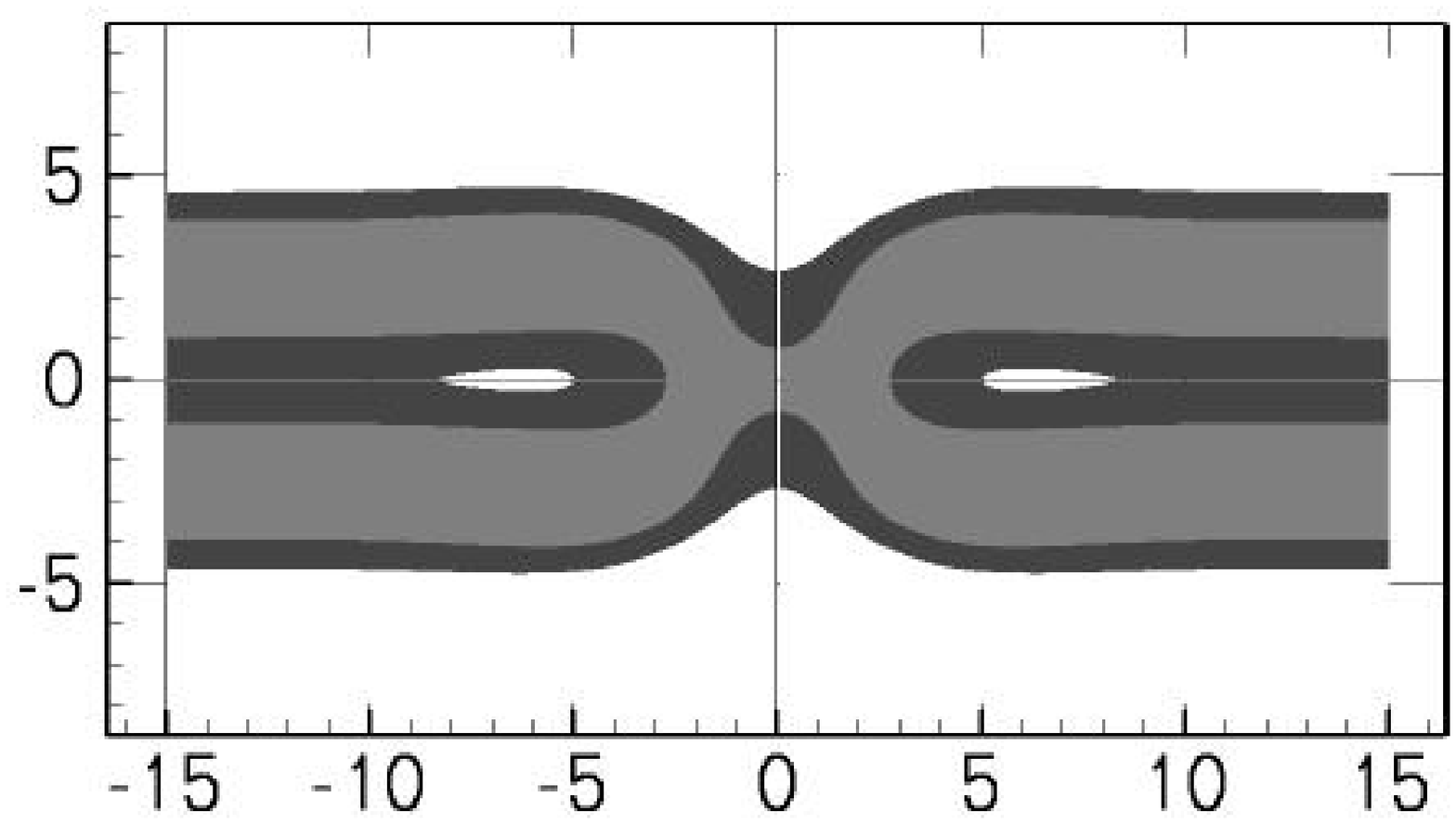,width=0.18\textwidth,clip=}
\\[4mm]
\epsfig{file=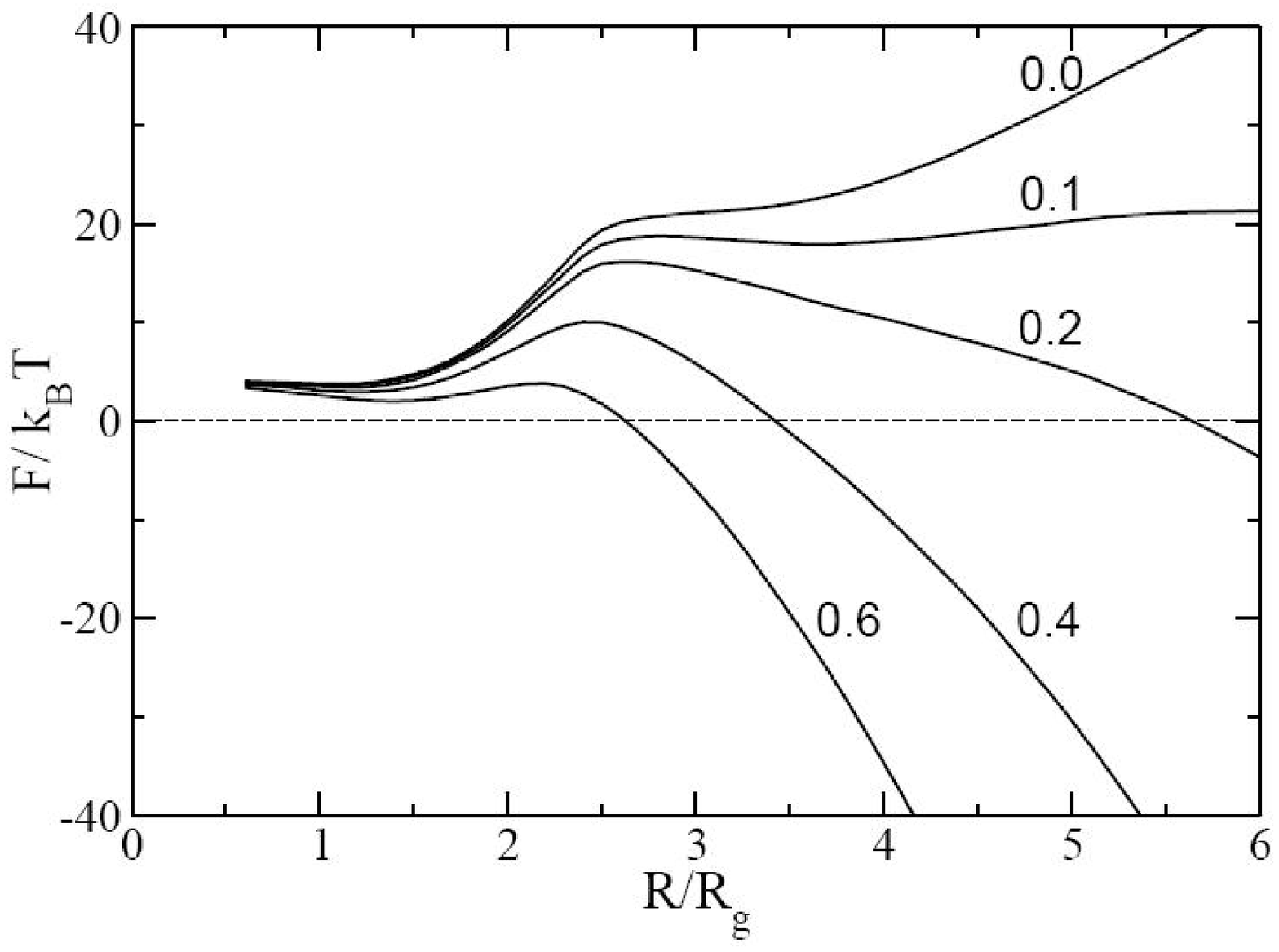,width=0.6\textwidth,clip=}
\caption{\label{fig:stalk}
(a) The free energy, $F$, of the stalk-like structure connecting
bilayers of fixed, zero, tension is shown for several different values of
the amphiphile's hydrophilic fraction $f$. In the inset we identify the
metastable stalk, $S_1$, the transition state, $S_0$, between the system
with no stalk at all and with this metastable stalk, and the transition
state, $S_2$, between the metastable stalk and a hemifusion diaphragm. The
architectural parameter is $f=0.30$ for this inset.  No stable stalk
solutions were found for $f=0.45$ in the region shown with dashed lines.
They were unstable to pore formation. Profiles showing the majority component
of the two barriers $S_0$ and $S_2$, and the metastable stalk, $S_2$, are shown below the main panel.
(b) The free energy of the expanding
stalk-like structure connecting bilayers of amphiphiles with fixed
architectural parameter $f=0.35$ is shown for several different bilayer
tensions. These tensions, $\gamma/\gamma_{\rm int}$, are shown next to
each curve.
From Ref.~\cite{Katsov04}.
}
\end{figure}

The free energy of an axially symmetric stalk that connects the two apposed
bilayer membranes as a function of its radius is shown in Fig.~\ref{fig:stalk}. The
free energy profiles exhibit two maxima -- the first one, $S_0$, corresponds
to the formation of an initial connection between the bilayers (stalk) and the second one, $S_2$, corresponds
to the expansion of the stalk to a hemifusion diaphragm. These two maxima are
separated by a minimum, $S_1$, that marks the radius of a metastable stalk. The
radius of this metastable configuration is set by the bilayer thickness.
Density plots of the majority component of the three extremal states are shown
in Fig.~\ref{fig:stalk}. While the phenomenological calculations focused much effort
on calculating the free energy cost of forming a stalk, $S_0$, the SCF calculations
show that the main barrier along the path towards the expanded hemifusion
diaphragm is not the formation of the initial stalk, $S_0$, but it is 
determined by the expansion, $S_2$, of the metastable stalk to the diaphragm.

\begin{figure}[htbp]
\begin{minipage}{0.55\textwidth}
\epsfig{file=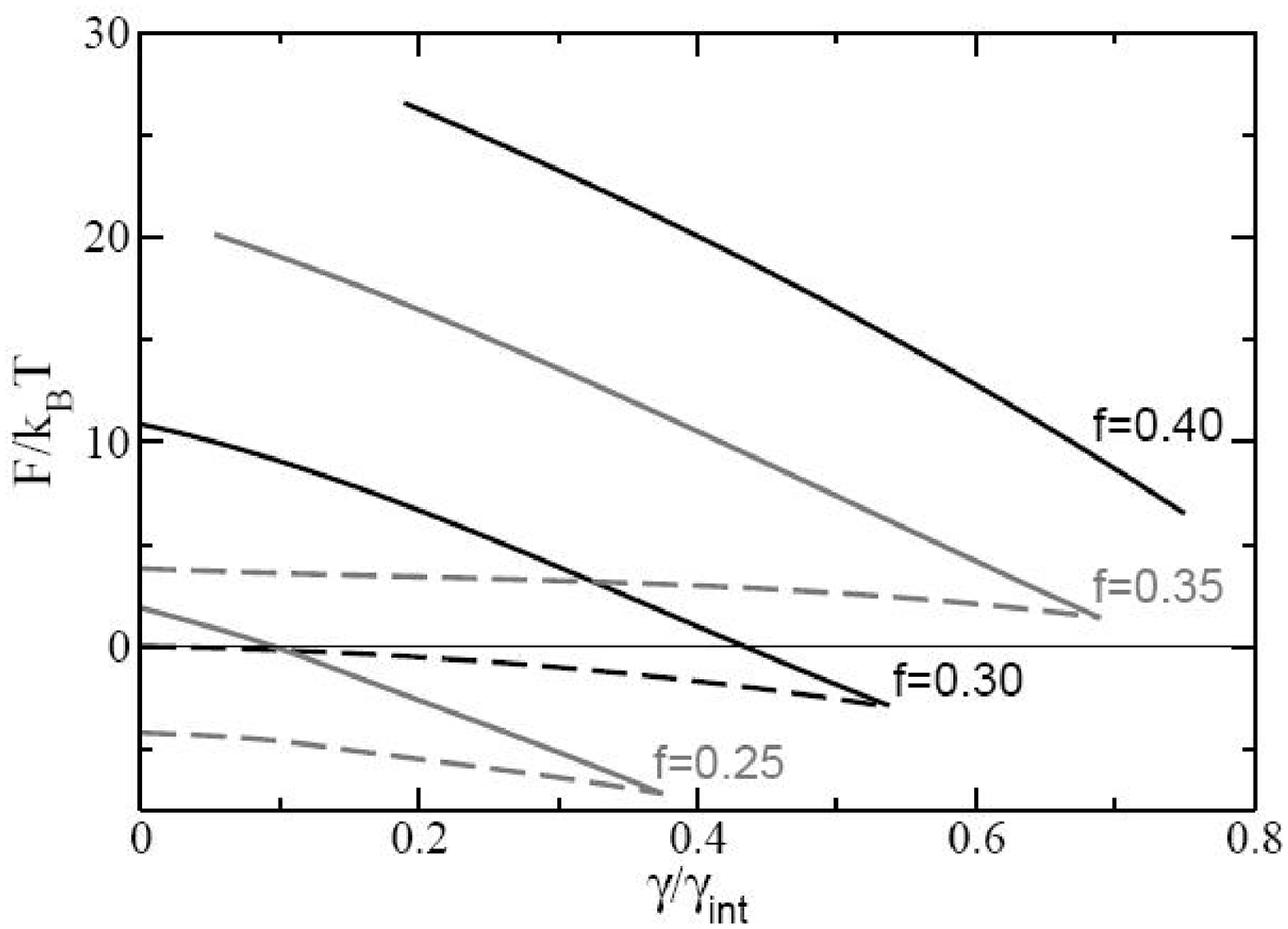,width=0.9\textwidth,clip=}
\end{minipage}
\begin{minipage}{0.45\textwidth}
\caption{\label{fig:stalk2}
The free energy, $F$, of the metastable stalks $S_1$ (dashed lines) and the
transition states $S_2$ (full lines) along the path towards the hemifusion diaphragm are shown as a
function of the tension for different architectures $f=0.25,\ 0.30,\ 0.35,\
0.40$. Notice that there is no metastable stalk (solution $S_1$) for $f=0.4$ at zero 
tension. From Ref.~\cite{Katsov04}.
}
\end{minipage}
\end{figure}

The free energies of the metastable stalk, $S_1$, and the saddle point, $S_2$
along the path
towards the hemifusion diaphragm are shown in Fig.~\ref{fig:stalk2} as a function of
the amphiphilic architecture, $f$, and the bilayer tension, $\gamma$.  The free
energy barrier, $S_2$, associated with the stalk expansion strongly decreases
with tension. Within the classical model, this explains why the fusion rate
increases with tension. The free energy of the metastable stalk hardly depends
on tension but it decreases substantially as we lower the fraction $f$ of
hydrophilic segments and thereby decrease the spontaneous curvature of a
monolayer to more negative values. For very asymmetric amphiphiles, those that
form the inverted hexagonal phase in the bulk, the free
energy of a stalk actually becomes negative. In this case, many stalks are
expected to form  and to condense on an hexagonal lattice thereby creating a ``stalk phase".
Such hexagonally perforated lamellar phases have been observed both in diblock
copolymers \cite{Hajduk97,Loo05b} and in lipid/water mixtures \cite{Yang02}. Making the amphiphiles even more
asymmetric, we observe that stalks spontaneously elongate and the inverted
hexagonal phase forms. On the other hand, if we make the amphiphiles more symmetric, the
metastable stalk configuration eventually disappears. In this case, the
fusion rate is not determined by the free energy difference between the saddle
point $S_2$ and the metastable stalk $S_1$ but by the large barrier $S_2$ only.
Thus, the absence of a metastable stalk will strongly suppress fusion.

\begin{figure}[htbp]
\epsfig{file=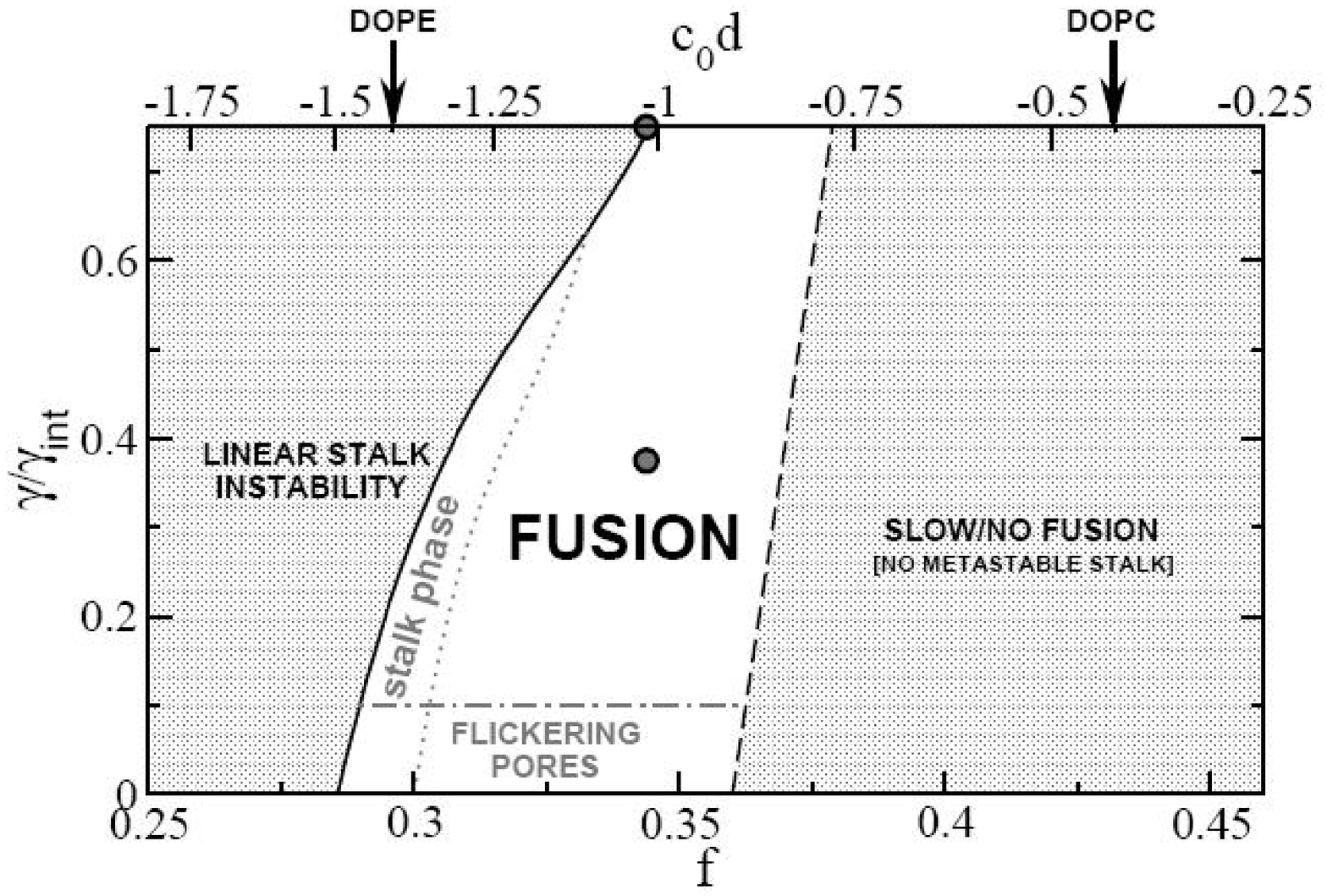,width=0.7\textwidth,clip=}
\caption{\label{fig:phase}
A ``phase diagram'' of the fusion process in the hydrophilic
fraction-tension, ($f,\gamma$), plane. Circles show points at which
simulations were performed by us. Successful fusion
only can occur within the unshaded region. As the tension, $\gamma$, decreases
to zero, the barrier to expansion of the pore increases without limit as
does the time for fusion. As the right-hand boundary is approached, the
stalk loses its metastability causing fusion to be extremely slow. As the
left-hand boundary is approached, the boundaries to fusion are reduced, as
is the time for fusion, but the process is eventually pre-empted due to
the stability either of radial stalks, forming the stalk phase, or linear
stalks, forming the inverted hexagonal phase.
From Ref.~\cite{Katsov04}.
}
\end{figure}

Both in the classical mechanism, as well as in the alternate stalk-hole fusion mechanism observed in the
simulations, stalks play a pivotal role. The limits of metastability of stalks
explored by the SCF calculations are compiled in Fig.~\ref{fig:phase}. Most notably, the free
energy calculations demonstrate that fusion is restricted to a rather narrow
window of amphiphilic architecture characterized by the ratio of spontaneous
monolayer curvature and bilayer thickness. This narrow range of spontaneous
curvatures extracted from a coarse-grained model is bound by the spontaneous
curvatures of two relevant lipids, DOPE and DOPC.

\begin{figure}[htbp]
\begin{minipage}{0.55\textwidth}
\epsfig{file=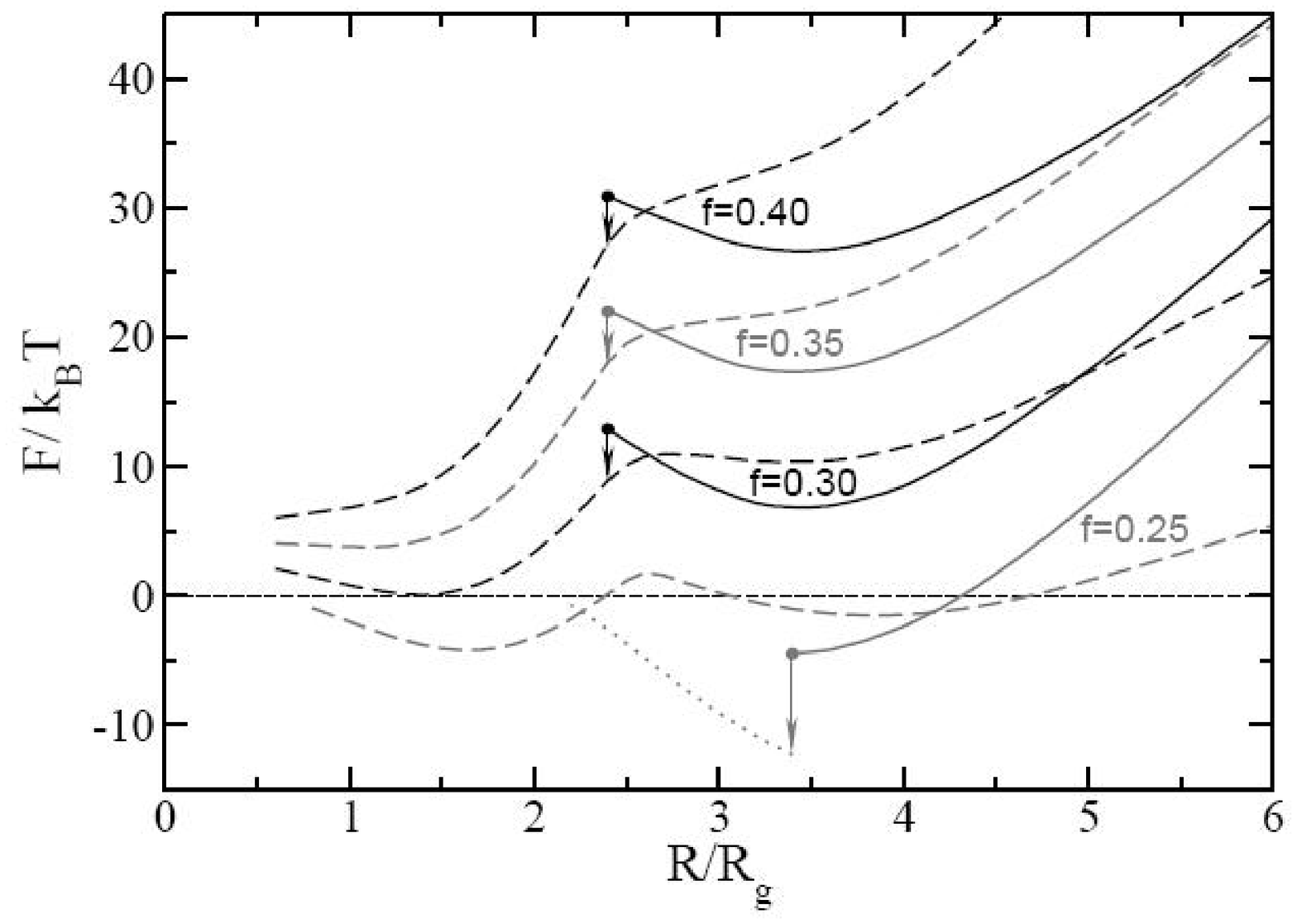,width=0.9\textwidth,clip=}
\end{minipage}
\begin{minipage}{0.45\textwidth}
\caption{\label{fig:pore}
The free energies, $F$, of a fusion pore (solid lines) and of a stalk (dashed
lines) of radius $R$ are shown. The membranes are tensionless and the
architecture $f$ of the amphiphiles is indicated in the key.  The instability of
the fusion pores at small radius is indicated by arrows.  For $f=0.3$, 0.35,
and 0.4, the stalk-like structure converts into a pore when it expands to a
radius $R\approx 2.4R_g$ at which the free energies of stalk-like structure and
pore are equal. For the system composed of amphiphiles of $f=0.25$, however,
the stalk-like structure converts at $R\approx 2 R_g$ into an inverted micellar
intermediate, IMI, whose free energy is shown by the dotted line. The fusion
pore is unstable to this IMI intermediate when its radius decreases to
$R\approx 3.4R_g.$ Thus the IMI is the most stable structure under these
conditions.
From Ref.~\cite{Katsov04}.  }
\end{minipage}
\end{figure}

The rupture of the hemifusion diaphragm completes the fusion process. The free
energy of a fusion pore in tensionless bilayers as a function of its radius, $R$, is shown in
Fig.~\ref{fig:pore}. For large radii the free energy linearly increases with  $R$,
the slope being proportional to the effective line tension of the pore's rim. For very
small radii, however, the free energy also increases as we decrease the size of
the fusion pore because the head groups begin to repel each other across the
pore. Thus, the SCF calculations predict a barrier for the closing of a fusion
pore. This prediction qualitatively agrees with the experimental observation of
flickering fusion pores at very low tension. In this case, fusion pores once
formed do not expand due to the lack of tension, but they do not close either
because of the barrier just mentioned. Thus they remain in a metastable state
and their radius fluctuates around its preferred value. Experimentally this is
detected by a flickering of the current between the two fusion vesicles \cite{Fernandez84,Spruce90,Chanturiya97}. The
region where flickering pores are expected is also indicated in the fusion diagram (cf.~Fig.~\ref{fig:phase}).

\subsubsection{Barriers along the stalk-hole path observed in the simulations}
The fusion path observed in the simulation differs from the classical hypothesis
by the occurrence of non-axially symmetric intermediates and the formation of
holes in the individual bilayers.

\begin{figure}[htbp]
\epsfig{file=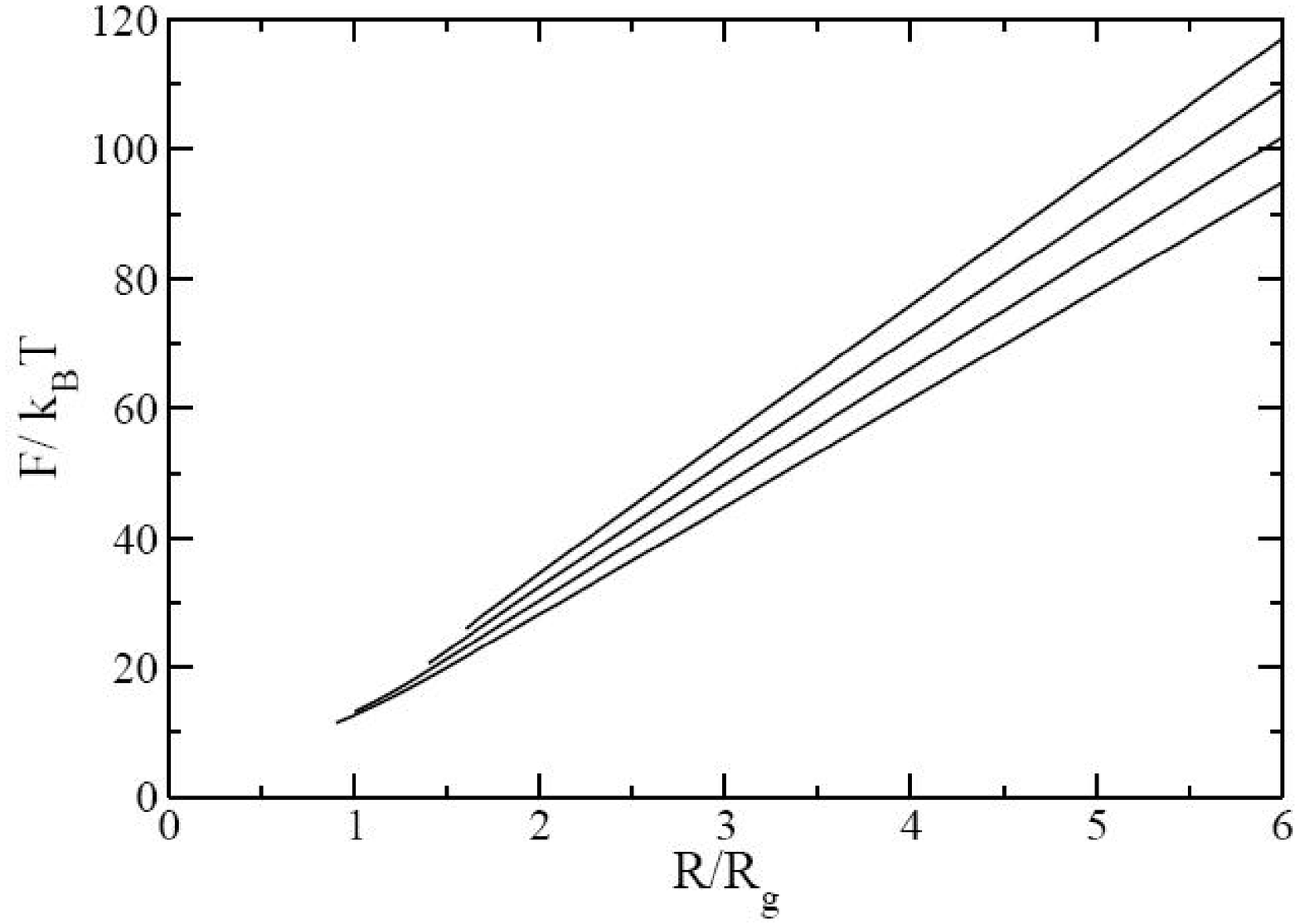,width=0.7\textwidth,clip=}\\[4mm]
\hspace*{1.cm}
\epsfig{file=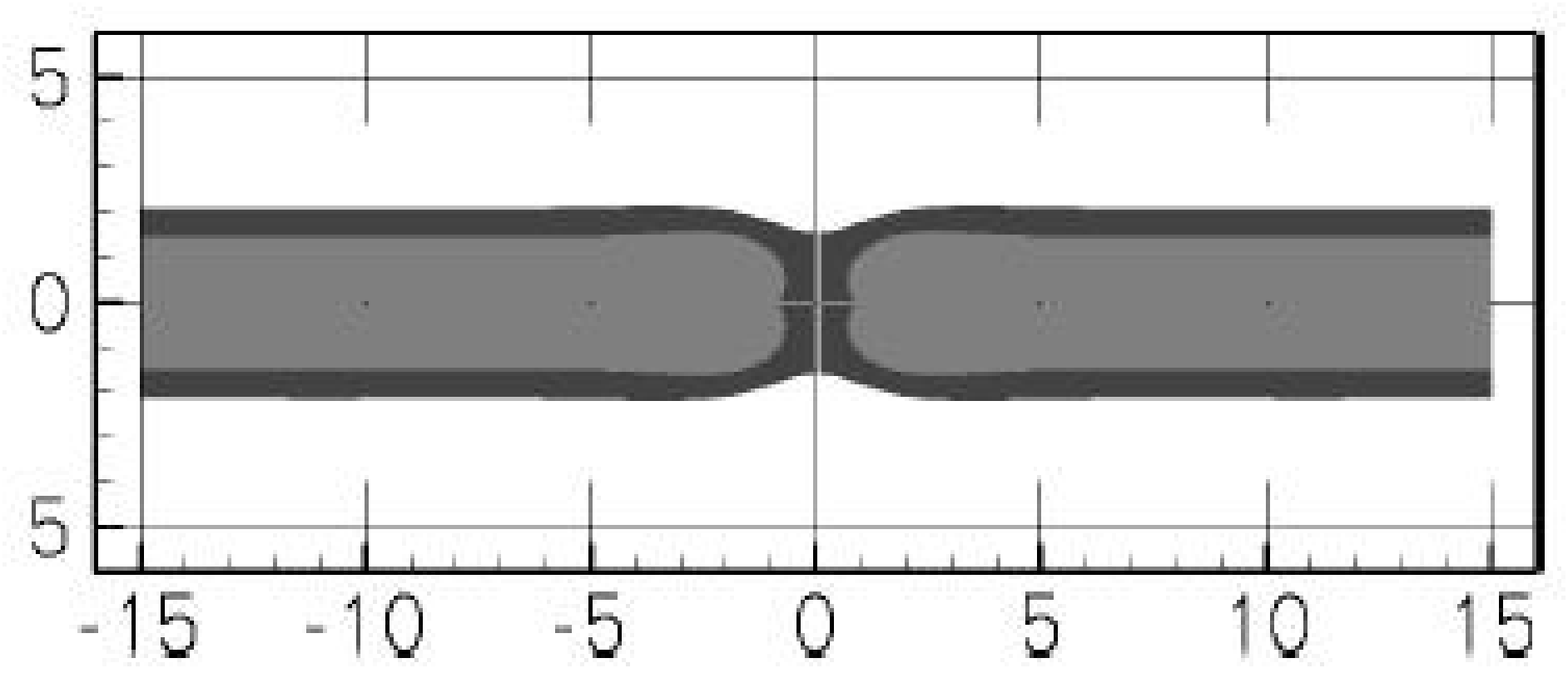,width=0.2\textwidth,clip=}
\epsfig{file=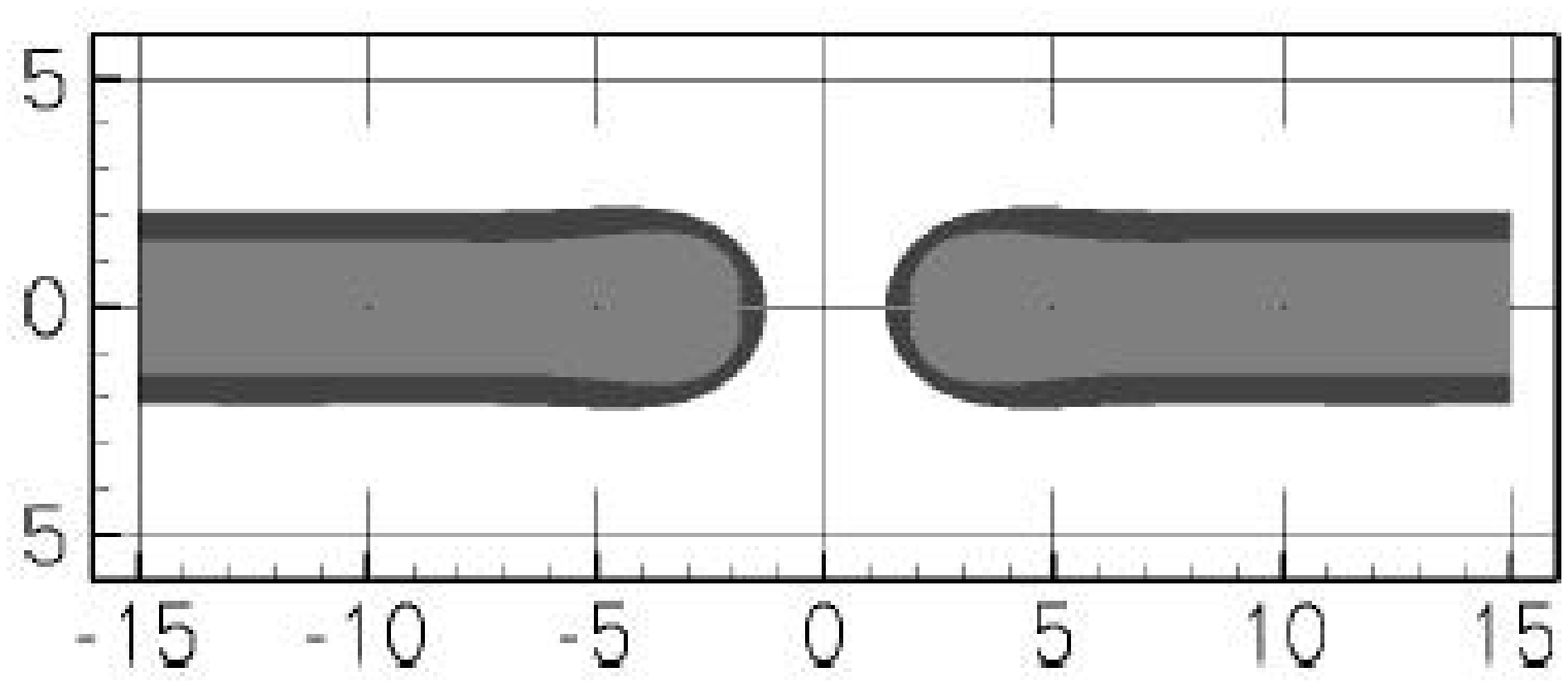,width=0.2\textwidth,clip=}
\epsfig{file=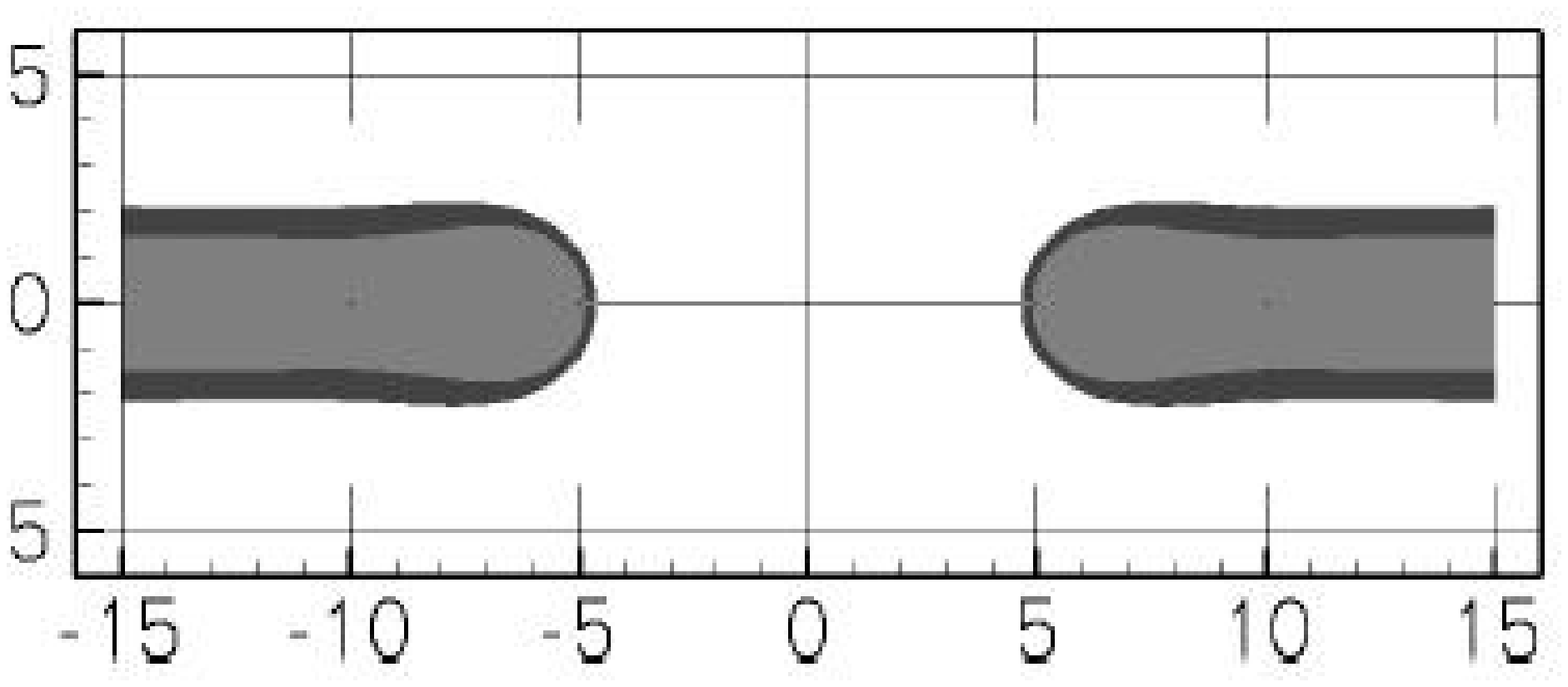,width=0.2\textwidth,clip=}
\\[4mm]
\hspace*{-0.3cm}
\epsfig{file=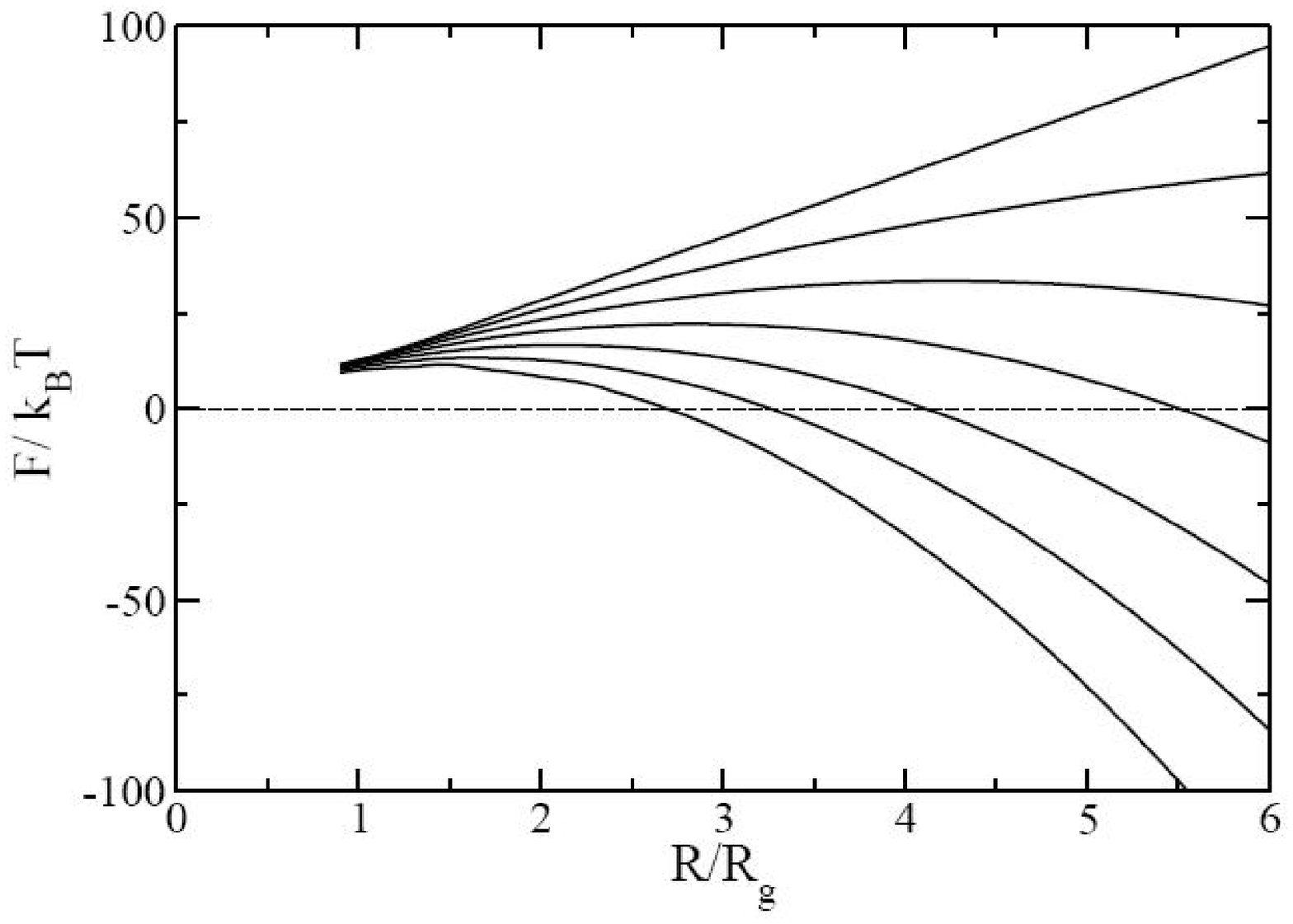,width=0.7\textwidth,clip=}
\caption{\label{fig:hole}
(a) Free energy of a hole in an isolated bilayer as a function of $R/R_g$ at
zero tension for various amphiphile architectures, $f$. From top to bottom the
values of $f$ are 0.29, 0.31, 0.33, and 0.35. (b) Same as above, but at fixed
$f=0.35$ and various tensions $\gamma/\gamma_{\rm int}$. From top to bottom,
$\gamma/\gamma_{\rm int}$ varies from 0.0 to 0.6 in increments of 0.1.
From Ref.~\cite{Katsov06}.
}
\end{figure}

First, we discuss the free energy of an isolated hole in a single membrane. 
Hole formation has been investigated by previous simulations
\cite{Shillcock96,Mueller96,KNECHT05,GROOT01,Tolpekina04,TOLPEKINA04b,LEONTIADOU04,FARAGO05,WANG05_2,LOISON04,HU05,Cooke05,Wohlert06}.
These studies have focused on the fluctuations of the pores, their shape and detailed structure, e.g., the bulging at the
rim (see Fig.~\ref{fig:hole} middle panel). The results indicate that hole formation is well described by classical
nucleation theory \cite{Taupin75,Litster75}. Fig.~\ref{fig:hole} presents the free energy of a hole as a
function of its radius R measured in units of the radius of gyration $R_g$ of
the amphiphilic molecules. While previous SCF calculations \cite{NETZ96} conceived
holes as weakly segregated equilibrium structures (the hexagonally perforated phase) of tense bilayers
the holes that form at large incompatibility must be stabilized by an external constraint (see Sec.~\ref{sec:SCFT})
and correspond to unstable structures, in agreement with simulation and experiment.
For a tensionless membrane, $\gamma=0$, the free
energy linearly increases with its radius, and from the asymptotic behavior one
can identify the line tension, $\lambda$ of the hole's rim. Upon increasing the
membrane tension, the free energy curves adopt a parabolic shape expected from
classical nucleation theory, $F_{\rm hole}(R) = 2\pi R \lambda - \pi R^2
\gamma$ \cite{Taupin75,Litster75}. Note that the free energy barrier, $\Delta F^* = \pi
\lambda^2/\gamma$, is on the order of a few tens of the thermal energy scale,
$k_BT$, for the parameters utilized in the MC simulations of the bond fluctuation model.
Although tense membranes are metastable and eventually will rupture,
the high barrier makes the homogeneous nucleation of a hole in a bilayer an
unlikely event on the time scale on which fusion occurs. Thus, isolated bilayer
membranes are stable (cf.~also Fig.~\ref{fig:holes_2vs1})

In contrast to the above calculation of the free energy of a hole in a bilayer, the direct calculation of the free energy of the stalk-hole intermediate is much more difficult. The reason is that whereas the former structure is axially symmetric, the latter is not. As noted, the axially symmetry is broken explicitly. This means that one has to calculate the free energy of an intrinsically three-dimensional structure. This is computationally very demanding. Moreover,
the choice of a suitable reaction coordinate is less obvious for the complex,
non-axially-symmetric intermediates of the stalk-hole mechanism. For these practical reasons we estimate
the free energy of the intermediates observed in the simulations by patching
together axially symmetric configurations calculated within SCF theory. Two
structures, the elongated stalk and the stalk-hole complex, are considered.

\begin{figure}[htbp]
\epsfig{file=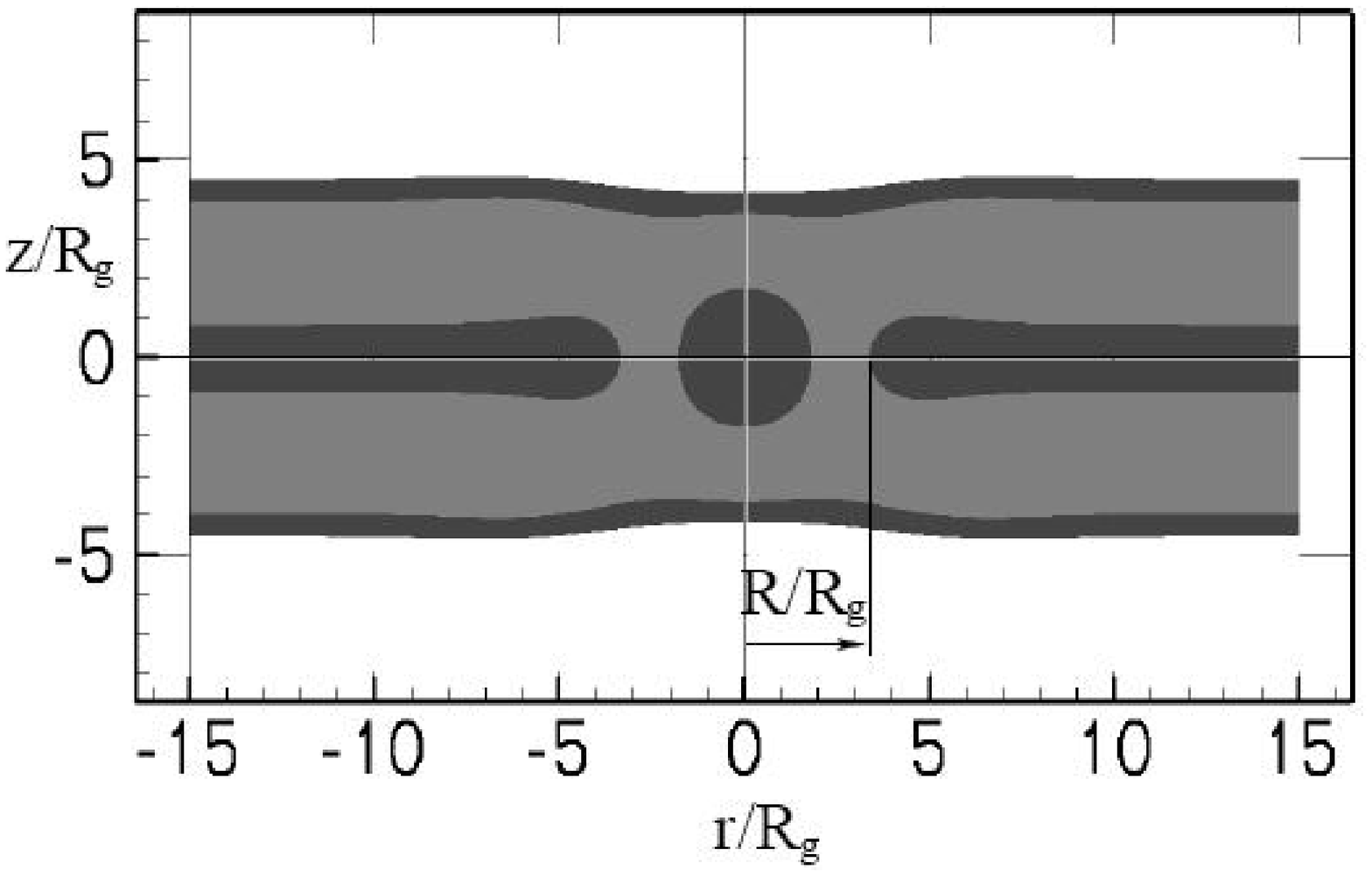,width=0.7\textwidth,clip=}
\epsfig{file=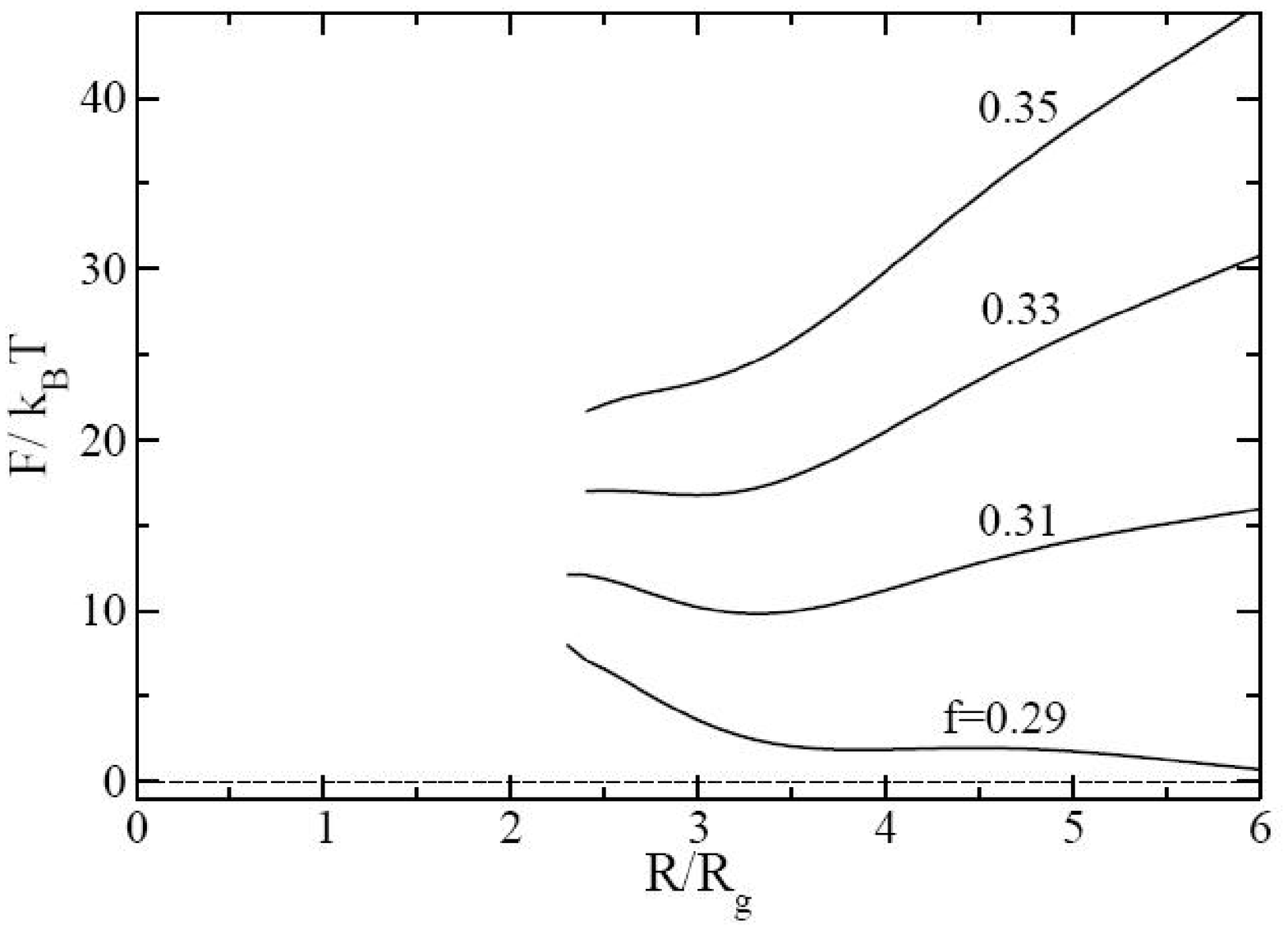,width=0.7\textwidth,clip=}
\caption{\label{fig:IMI}
(a) Density profile of an IMI -- an elongated stalk that closes in itself forming a circular structure.
The amphiphiles are characterized by $f=0.3$.
The radius of the IMI, in units of the  radius of gyration, $R/R_g$ is 3.4. (b)
Free energy of an IMI as a function of $R/R_g$ at zero tension for various
amphiphile architectures, $f$ From Ref.~\cite{Katsov06}.
}
\end{figure}

The geometry of the elongated stalk, which is observed before a hole forms next to it,  is characterized by the arc of a circle of radius,
$R$, and length $2\pi R \alpha$, with $0 \leq \alpha \leq 1$. We decompose
the free energy of this structure into the contribution of its two end caps,
which each resemble  half of a metastable, circular stalk, and a fraction, $\alpha$, of a ring-shaped stalk, or inverted micellar intermediate (IMI) (see
Fig.\ref{fig:IMI}),

\begin{equation}
F_{\rm es} = F_{\rm stalk}+\alpha F_{\rm IMI}(R).
\label{justbefore}
\end{equation}
By constructing the free energy via the simple addition of the free energy of axially
symmetric structures and not allowing for additional optimization of the shape,
we will overestimate the free energy of the intermediate.

After a hole forms next to the stalk, the geometry of the stalk-hole complex can be characterized by the radius of the
hole, $R-\delta$, where $\delta$ is radius of the metastable stalk, and the
fraction $\alpha$ of the hole's rim that is covered by the stalk. The geometry
of the latter configuration resembles a hemifusion diaphragm. Thus we can
approximate the free energy of the stalk-hole complex by 
\begin{equation} 
F_{\rm sh} = \alpha F_{\rm HD}(R) + (1-\alpha) F_{\rm hole}(R-\delta) 
\end{equation} 
where we neglect the free energy costs of the two end points of the stalk. The
free energy of these two saddle-shaped point defects presumably is small.

\begin{figure}[htbp]
\epsfig{file=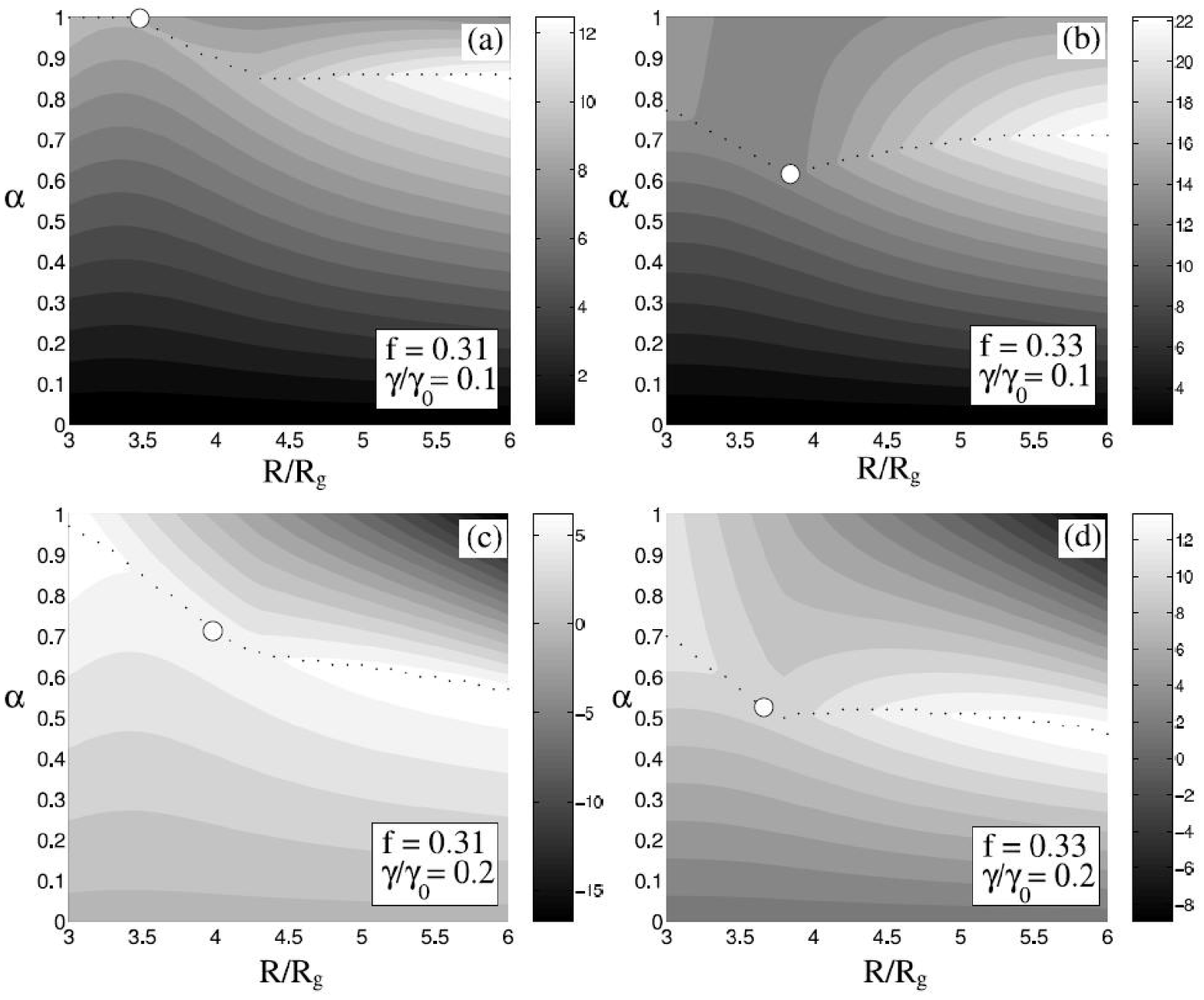,width=0.8\textwidth,clip=}
\caption{\label{fig:landscape}
Four free energy landscapes (in units of $k_BT$) of the fusion
process, plotted as a function of the radius, $R$ (in units of
$R_g$) and circumference fraction $\alpha$. The architecture of the
amphiphiles, $f$, and the value of the tension, $\gamma/\gamma_{\rm int}$, are given.
The dotted line shows a ridge of possible transition states, separating
two valleys. The region close to the $\alpha=0$ line corresponds to
a barely elongated stalk intermediate (see Eq.~(\ref{justbefore})).
The other valley, close to $\alpha=1$ states, corresponds to a hole
almost completely surrounded by an elongated stalk. The saddle point on the ridge,
denoted by a white dot, corresponds to the optimal, lowest free energy,
transition state.
From Ref.~\cite{Katsov06}.
}
\end{figure}

These free energy estimates allow us to explore the free energy landscape,
$F(R,\alpha) = \min\left[F_{\rm es},F_{\rm sh} \right]$ of the fusion process
as a function of the two reaction coordinates, the radius of the elongated
stalk, $R$ and the fraction $\alpha$ of the hole's rim covered by the stalk.
Here, we additionally have assumed that a pore will form with a negligible
barrier when the stalk-hole free energy, $F_{\rm sh}$, is lower than the free energy of an
elongated stalk, $F_{\rm es}$.  The formation of a hole in the vicinity of a stalk produces a
ridge in the free energy landscape.  This is, of course, a simplification, but we
expect the corresponding free energy barrier to be small. The main effect of
this simplification is that the locus of transitions between the extended stalk and the
stalk-hole complex is a sharp ridge, while in a more complete description it would occur over a range of values over which the
free energy of the two competing structures differs by order $k_BT$.  Once the
hole has formed, the free energy decreases as the stalk surrounds the hole
(increasing $\alpha$), and the stalk-hole intermediate expands (increasing $R$).

\begin{figure}[htbp]
\epsfig{file=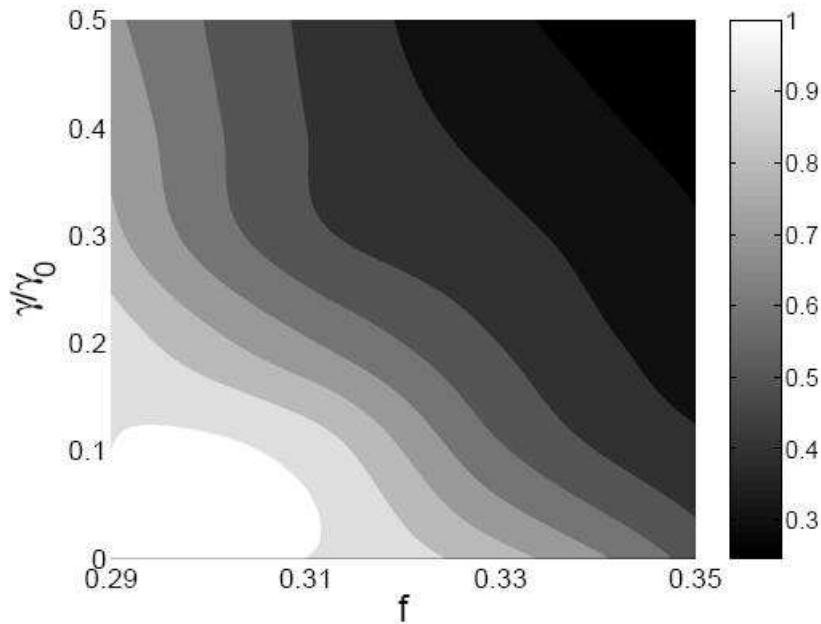,width=0.7\textwidth,clip=}
\caption{\label{fig:alpha}
Plot of $\alpha^*$, which corresponds to the optimal transition
state in the stalk-hole mechanism, as a function of architecture of the
amphiphiles and the tension of the membrane. Small values of $\alpha^*$
(dark) correspond to leaky fusion events.
From Ref.~\cite{Katsov06}.
}
\end{figure}

The minimum of the free energy along the ridge $F_{\rm es}(R,\alpha)=F_{\rm
sh}(R,\alpha)$ defines the transition state characterized by $R^*$ and
$\alpha^*$. The value, $\alpha^*$, is an important characteristic of the fusion
intermediate. The larger $\alpha^*$, the larger is the fraction of the hole's
rim that is surrounded by the stalk when the hole is formed. Thus, larger values of
$\alpha^*$ correspond to the tighter fusion events while small values of $\alpha^*$
suggest that fusion is leaky and possibly competes with rupture. The value of
$\alpha^*$ as a function of tension, $\gamma$, and amphiphilic architecture, $f$,
is shown in Fig.~\ref{fig:alpha}. Small values of the tension and small values of
$f$, i.e., large negative curvatures, favor tight fusion. Stalks expand more
readily because the line tension of the stalk decreases with $f$ (cf.~Fig.~\ref{fig:hole}); 
the line tension, $\lambda_H$, of a hole increases for small $f$; and
holes expand less quickly for smaller tension.

\begin{figure}[htbp]
\epsfig{file=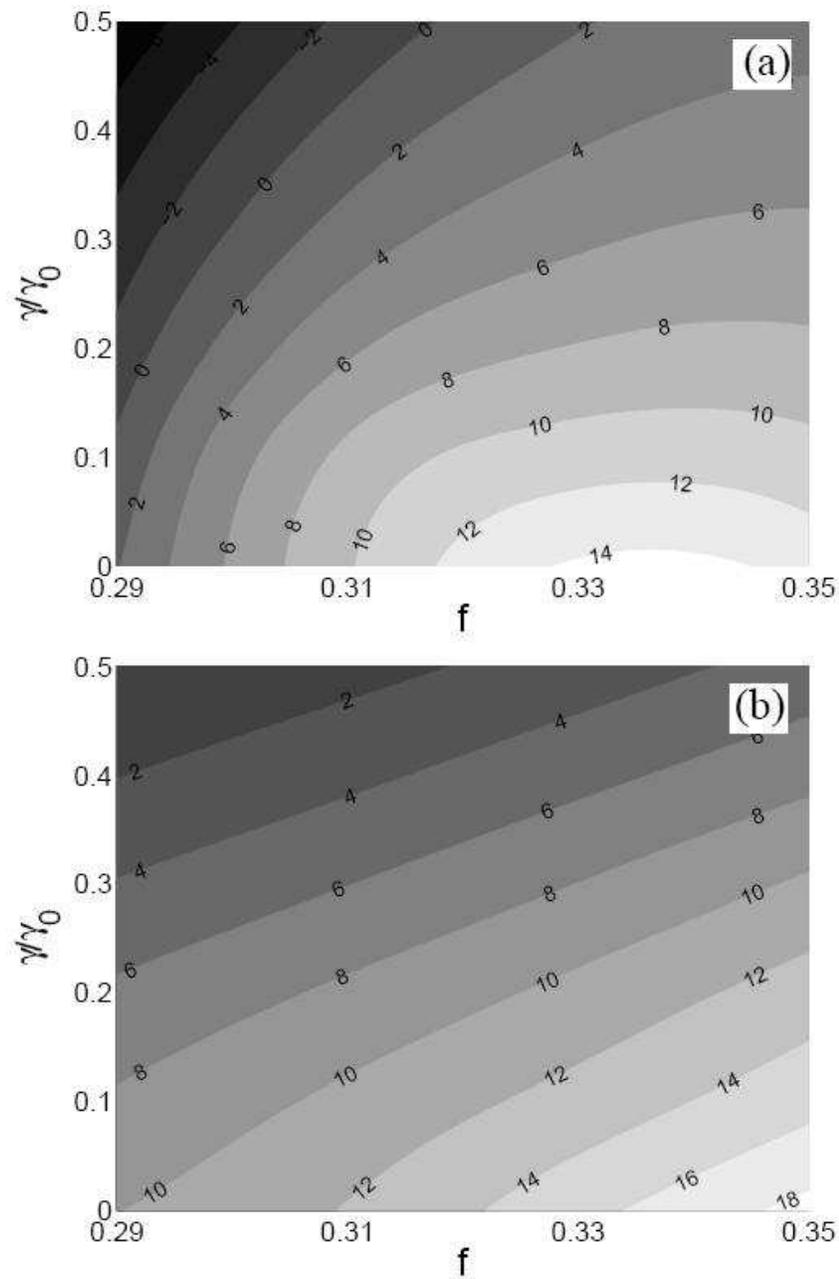,width=0.7\textwidth,clip=}
\caption{\label{fig:comp} Free energy barriers measured relative to the initial
metastable stalk, in units of $k_BT$, in (a) the new stalk-hole complex
mechanism, and (b) the standard hemifusion mechanism.  From
Ref.~\cite{Katsov06}.  }
\end{figure}

In Fig.~\ref{fig:comp} we compare the free energy of the fusion intermediates along the classical
hemifusion path with that of  the stalk-hole mechanism as a function of tension and architecture.
In agreement with experimental observation, the fusion barrier in both scenarios is
the lower the more negative is the spontaneous curvature of the monolayers and the larger
the membrane tension is. For the parameters studied, the stalk-hole mechanism has a slightly
smaller free energy barrier than that of the classical hemifusion path. The free energy difference, however,
amounts only to a few $k_BT$ indicating that the stalk-hole mechanism is a viable alternative
to the classical hemifusion path, but that the choice of the pathway will depend on  specific
details of the system. This is in agreement with the simulation data.

\section{Conclusion and outlook}
\label{sec7}
We hope that we have demonstrated that coarse-grained models are a valuable
tool for investigating universal collective phenomena in bilayer membranes.
These models bridge the gap between atomistic simulations that are limited to
rather small time and length scales, and phenomenological  models
that ignore much of the internal structure of the bilayer. They allow for
direct insights into processes that involve many lipids and take place on time
scales of milliseconds and length scales of micrometers.

With a specific example, the fusion of membranes, we have illustrated what can
be learned from studying coarse-grained models. The apparent universality of
membrane fusion that is observed experimentally and the separation of time
scales between the local structural relaxation in an individual membrane and
the fusion process make it a suitable subject of such models. A rather simple
model already yields direct information about the fusion pathway that cannot
easily be obtained otherwise, and leads to experimentally verifiable
predictions. For high membrane tension and asymmetric lipid architecture, the
SCF calculations predict transient leakage that is correlated in time and space
with the fusion event. This is observed in simulation of coarse-grained models
\cite{NOGUCHI01,Mueller02,Muller03,Marrink03,LI05,Smeijers06} that
substantially differ in their interactions and is corroborated by
electrophysiological experiments \cite{Dunina00,Frolov03}. Moreover, the
coarse-grained models resolve the overt contradiction between the stability of
isolated membranes or vesicles that is necessary for their function and the
fusion process that involves hole formation in the two apposed membranes. This
can be understood as follows. The line tension of holes in the membranes is
reduced by the presence of stalks. Therefore, the rate of hole formation is
dictated by heterogeneous nucleation at stalks, which is much greater than the
rate of homogeneous nucleation in an unperturbed bilayer.  The study of fusion
also illustrates the fruitful interplay between particle-based simulations and
field-theoretic techniques.  The latter technique provides quantitative free
energy estimates of the fusion barrier and permits the exploration of a wide
range of molecular architectures and membrane tensions. It reveals that fusion
is strongly suppressed outside a narrow range of monolayer curvatures (see
``phase diagram'' in Fig.~\ref{fig:phase}).  This result suggests ways to
control fusion by tuning the stability of stalks and the formation of holes.

The increase of computing resources and algorithmic advances will allow
atomistic simulations to explore phenomena at ever larger time and length
scales. The role of coarse-grained models, however, will not merely consist in
studying problems that are not yet accessible by atomistic simulations. Rather
the reduction of the complexity of the system will highlight the relevant
ingredients and allow for a systematic quantitative analysis. Thus,
coarse-grained models will provide insight into specific mechanisms and
principles as well as the degree of universality. Coarse-grained models can
explore a variety of conformational properties, thermodynamic quantities (such
as phase behavior, bending elastic constants, tension and areal
compressibility) and dynamic characteristics (e.g., structural relaxation of
the bilayer, dynamics of undulations, lateral diffusion of lipids)
simultaneously, and the parameters of the model system can be varied
independently and over a wide range.  The wealth of information these models
provide concerning equilibrium and dynamic properties can be
compared both to experiment and to simpler phenomenological approaches in order
to assess the validity of the model description.  In this way, coarse-grained
models identify interesting parameter regimes, test phenomenological concepts,
and permit systematic exploration of  collective phenomena in biological
membranes.

A variety of coarse-grained models has been devised in recent years which
differ in the degree of molecular detail, the type of interactions, and the
representation of the solvent.  In view of the richness and complexity of
collective phenomena in membrane physics, no single model will emerge that
captures all aspects. ``Systematic'' coarse-graining procedures
\cite{SHELLEY01_1,AYTON02,MARRINK04,NIELSEN04_1,IZVEKOV05,BOEK05} or tightly
coupled multi-scale simulation techniques
\cite{AYTON02,Hummer03b,Chang05b,Praprotnik05,Praprotnik06} will provide
information about the specific interactions that have to be incorporated into
coarse-grained models to describe the phenomena of interest.

There are a host of biological problems that can be studied by the kind of
coarse-grained models we have discussed, and we expect their use in biological
physics to increase greatly in the near future. The formation of bilayers and
vesicles \cite{GOETZ98,NOGUCHI01_2,Yamamoto02,Marrink03b,Sevink05,Cooke05,Ortiz06} has
been investigated.  The studies of different bilayer phases
\cite{KRANENBURG04_1,Brannigan04b,Lenz05,Kranenburg05,Cooke05}, of the kinetics
of phase transition \cite{Marrink04b,Marrink05},  of self-assembly
\cite{NOGUCHI01_2,Yamamoto02,Marrink03b,Sevink05}, and of phase separation in
membrane consisting of different components
\cite{AYTON05,KNECHT05,Faller04,IMPARATO03,FORET05,VEATCH05,ELLIOT05,KHELSHVILI05,LIU05,Shi05,Izvekov06}
have already yielded important information about these fascinating systems.
Other important questions that can be tackled by coarse-grained models include
the interaction between membranes,  between membranes and solid substrates, and
the interplay between membrane collective phenomena  and peptides, proteins and
polymers
\cite{Srinivas04b,Venturoli05,Srinivas05,Lopez06,Sperotto06,Brannigan06b,Farago06},
or active components (e.g., ion pumps
\cite{Manneville99,Gov04,Girard05,Lin06,Brannigan06b}, or fusion proteins
\cite{Hilbers06,Shillcock06}).

We close by considering a simple, but important, application to an area which
has received much study recently; the possibility of phase separation in the
plasma membrane. The basic idea is that the plasma membrane enclosing the cell,
rather than being a homogeneous mixture of lipids and cholesterol, may in fact
be rather inhomogeneous, with regions of saturated lipids, like sphingomyelin,
and cholesterol, aggregated and floating, like ``rafts", in  a sea of
unsaturated lipids, like most phosphatidylcholines. The field grew quickly when
it was observed that some signaling proteins preferred the raft environment. As
a consequence, these proteins were not distributed randomly throughout the
plasma membrane, but were concentrated in the rafts and could therefore
function more efficiently.

This seems to be a system which could be studied by coarse-grained
models \cite{scott}. One  object would be to capture the relevant properties of
cholesterol which appears to order the saturated lipids in its
vicinity \cite{vist90}. With a suitable model, one can then calculate phase
diagrams and compare them to the several experimental ones.

As interesting as this would be, and it would be very interesting indeed, one
could then try to understand the observation that rafts act as a nucleation
site for the fusion process. One sees how ``...way leads onto way" and to the
study of ever more complex phenomena, those which are characteristic of
biological systems. 

\subsection*{Acknowledgment}
It is a great pleasure to acknowledge stimulating discussions with V.~Frolov.
We also acknowledge K. Ch. Daoulas, S.~J.~Marrink and H. Noguchi for a critical reading of the manuscript.
Financial support was provided by the Volkswagen Foundation and the NFS under grant DMR-0140500 and 0503752.
The simulations were performed at the von Neumann Institute for Computing at J{\"u}lich, Germany.

\bibliography{\bibdata}
\end{document}